\documentclass[a4,fleqn,usenatbib,useAMS]{mn2e}
\usepackage{color}
\usepackage{amsmath} \usepackage{amssymb} \usepackage{multicol} \usepackage{bm} 
\usepackage{textcomp,amssymb}
\usepackage{leftidx}
\usepackage{hyperref} 
\hypersetup{dvips, colorlinks=true, linkcolor=black, citecolor=black, filecolor=blue, urlcolor=black}
\usepackage{float}
\usepackage{epsfig}
\usepackage{array}
\usepackage{supertabular}
\usepackage{hhline}
\usepackage{multirow}
\usepackage{aas_macros}
\usepackage{balance}
\usepackage{times}
\usepackage{graphicx}
\usepackage{times}
\usepackage{soul}
\usepackage{color}
\usepackage{bm}\usepackage[caption=false]{subfig}
\usepackage{times}
\allowdisplaybreaks

\newcommand{\ri}{\mathrm{i}}
\newcommand{\re}{\mathrm{e}}

\newcommand{\rf}{\mathrm{f}}

\newcommand{\Rb}{R_{\mathrm{b}}}
\newcommand{\pb}{p_{\mathrm{b}}}

\newcommand{\rd}{\mathrm{d}}

\newcommand{\vtheta}{\vartheta}
\newcommand{\Mtot}{M_{\mathrm{tot}}}
\newcommand{\bT}{\bm{\theta}}
\newcommand{\bJ}{\mathbf{J}}
\newcommand{\bO}{\mathbf{\Omega}}

\newcommand{\bS}{\mathbf{S}}
\newcommand{\bs}{\mathbf{s}}
\newcommand{\tbJ}{\widetilde{\mathbf{J}}}
\newcommand{\tbO}{\widetilde{\mathbf{\Omega}}}

\newcommand{\Sign}{\mathrm{Sign}}

\newcommand{\wbM}{\widehat{\mathbf{M}}}
\newcommand{\wM}{\widehat{M}}
\newcommand{\bn}{\mathbf{n}}
\newcommand{\tbn}{\widetilde{\mathbf{n}}}
\newcommand{\bx}{\mathbf{x}}
\newcommand{\bxp}{\mathbf{x}^{\prime}}
\newcommand{\rpp}{r^{\prime}}
\newcommand{\upp}{u^{\prime}}
\newcommand{\mC}{\mathcal{C}}
\newcommand{\rperi}{r_{\mathrm{p}}}
\newcommand{\rapo}{r_{\mathrm{a}}}
\newcommand{\deltaD}{\delta_{\mathrm{D}}}

\newcommand{\bI}{\mathbf{I}}
\newcommand{\veps}{\varepsilon}

\newcommand{\Rs}{R_{\mathrm{s}}}

\newcommand{\nmax}{n_{\mathrm{max}}}

\newcommand{\RePart}{\mathrm{Re}}

\newcommand{\Arg}{\mathrm{Arg}}

\usepackage{acronym}
\newacro{DF}{distribution function}
\newacroplural{DFs}{distribution functions}
\newcommand{\DF}{\ac{DF}}
\newcommand{\DFs}{\acp{DF}}

\begin{document}

\setcounter{tocdepth}{3}

\title[Stability of stellar rotating spheres via linear response theory]
{Mapping the stability of  stellar rotating spheres\\via linear response theory}

\author[S. Rozier, J-B. Fouvry, P.G. Breen, A.L. Varri, C. Pichon, D.C. Heggie]
{S. Rozier$^{1}$, J.-B. Fouvry$^{2,1}$, P.~G. Breen$^{3}$, A.~L. Varri$^{4}$, C. Pichon$^{1,4,5}$ and D.~C. Heggie$^{3}$
\vspace*{6pt}\\
\noindent
$^{1}$ Institut d'Astrophysique de Paris, and UPMC Univ. Paris 06, (UMR7095), 98 bis Boulevard Arago, 75014 Paris, France\\
$^{2}$ Institute for Advanced Study, Einstein Drive, Princeton, New Jersey 08540, United States of America\\
$^{3}$ School of Mathematics and Maxwell Institute for Mathematical Sciences, University of Edinburgh, Kings Buildings, Edinburgh EH9 3FD, UK\\
$^{4}$ Institute for Astronomy, University of Edinburgh, Royal Observatory, Blackford Hill, Edinburgh EH9 3HJ, UK\\
$^{5}$ Korea Institute for Advanced Study (KIAS), 85 Hoegiro, Dongdaemun-gu, Seoul, 02455, Republic of Korea
}

\date{\today}

\maketitle

\begin{abstract}
Rotation is ubiquitous in the Universe, and recent kinematic surveys have shown that early type galaxies and globular clusters are no exception. 
Yet the linear response of spheroidal rotating stellar systems has seldom been studied. 
This paper takes a step in this direction by considering the behaviour of spherically symmetric systems with differential rotation. Specifically, the stability of several sequences of Plummer spheres is investigated, in which the total angular momentum, 
 as well as the degree and flavour of anisotropy in the velocity space are varied.
To that end, the response matrix method is customised to spherical rotating equilibria.
 The shapes, pattern speeds and growth rates
of the systems' unstable modes are computed.
Detailed comparisons to appropriate $N$-body measurements are also presented.
 The marginal stability boundary is charted in the parameter space of velocity anisotropy and rotation rate.
When rotation is introduced, two sequences of growing modes are identified corresponding to radially and tangentially-biased
anisotropic spheres respectively.
For radially anisotropic spheres,
growing modes occur on two intersecting surfaces (in the parameter space of anisotropy and rotation), 
which correspond to fast and slow modes, 
depending on the net rotation rate.
Generalised, approximate stability criteria are finally presented.
\end{abstract}


\section{Introduction}
\label{sec:introduction}

 Most stellar systems are rotating. 
 While this  has been clear for spiral galaxies since the beginning of extra-galactic spectroscopy, 
 recent kinematic surveys have shown that 
    early type galaxies and globular clusters can also be partially centrifugally-supported. 
    
Globular clusters have offered an essential empirical
probe to develop and test theories of stellar dynamics.
The long-term evolution of such self-gravitating systems generically comprises two stages.
First, on a few dynamical times, as a result of strong potential fluctuations, the cluster can undergo a violent (collisionless) relaxation~\citep{LyndenBell1967}, allowing it to reach a steady state.
Then, on longer timescales, the cluster will slowly explore
subsequent thermodynamical equilibria, as a result of two-body relaxation sourced by finite-$N$ fluctuations~\citep{BinneyTremaine2008}.
During the first stage of evolution, the globular cluster is insensitive to its finite number of constituents, and can therefore be treated as a collisionless system.
It is also during this stage that the system can develop linear instabilities,
 that deeply rearrange the system's orbital structure, should the violent relaxation stage
 have  left it in an unlikely (low entropy) configuration.
This is where lies the interest of this paper.

The analysis of the linear stability of spherically symmetric stellar systems
has already been the subject of numerous investigations.
For example, the Doremus-Feix-Baumann theorem~\citep{BinneyTremaine2008} states that radial modes of an ergodic spherical model are all stable if the gradient of the system's distribution function w.r.t.\ the energy is negative, while systems with a positive gradient w.r.t.\ the angular momentum tend to be unstable~\citep{Tremaine2005}.
Similarly, systems that support too many radial orbits tend to undergo the so-called radial orbit instability~\citep[see, e.g.\@,][]{Polyachenko1981,Saha1991,Weinberg1991}.
However, all these investigations were limited to configurations having a spherically symmetric velocity distribution
(i.e.\@, a distribution function which is invariant under rotation in configuration space).
This study sets out to lift this restriction and investigate the linear stability of rotating, anisotropic equilibria, i.e.\ systems with a non-zero total angular momentum.

This theoretical endeavour is particularly timely, for a number of reasons.
The interest in the role of angular momentum in the evolution of globular clusters is now peaking, as a result of a variety of new empirical evidence. This includes recent astrometric measurements by HST~\citep{Bellini2017}, Gaia DR2~\citep{Bianchini2018,Sollima2019}, and spectroscopic surveys, such as the MIKiS survey~\citep{Ferraro2018} and dedicated programmes with MUSE~\citep{Kamann2018}.
Similarly, line-of-sight measurement of nuclear star clusters are also revealing them as fast rotators~\citep{Seth2008,FeldmeierKrause2017}.
Finally, on larger scales, the classical ${V/\sigma}$ vs. ellipticity diagram~\citep{Davies1983} is now rich in spheroidal, rotation-supported early-type galaxies (see e.g.~\cite{Roger2011} for a summary), following major surveys such as SAURON~\citep{Bacon2001} and ATLAS3D~\citep{Cappellari2011}.
While these systems are often modelled with Jeans and Schwarzschild techniques, it is important to 
explore the stability properties of the equilibria identified in these studies.

The main goal of this work is to eventually extend to stellar systems the classical knowledge of 
the stability of uniformly rotating fluid dynamical systems (for a summary of the theory of the ellipsoidal figures of equilibrium, see~\citealt{Chandrasekhar1969}),
and its subsequent developments into the realm of differentially rotating polytropes, which have been found to exhibit instability at surprisingly low values of the total angular momentum
~(e.g., see~\citealt{Pickett1996,New2001,Centrella2001,Shibata2002,Ou2006} and, more recently, \citealt{Galeazzi2012}; suggestions for a possible dynamical interpretation in which corotation points play a primary role have been offered, among others, by~\citealt{Watts2005,Passamonti2015,Yoshida2017}).

The present study is therefore part of a concerted effort to systematically explore the theoretical dimension of the newly appreciated ``kinematic richness'' (i.e., any deviations from the traditional assumptions of isotropy in the velocity space and absence of internal rotation), which is empirically emerging in a variety of spheroidal stellar systems. Such a richness has implications on (i) the construction of appropriate self-consistent equilibria (e.g., see the fresh efforts by~\citealt{Binney2014,Evans2015,Sanders2015,Posti2015} and, for a summary of recent developments with applications to globular clusters, \citealt{Varri2016} and references therein), (ii) their dynamical and secular (collisionless) behaviour, as well as (iii) the long-term (collisional) evolution of stellar systems (for an analysis of the effects of ``primordial'' velocity anisotropy and differential rotation on spherically symmetric collisional systems, see~\citealt{Breen+2017,Breen+2019a}, respectively).

With the exception of some pioneering studies on the counterparts of uniformly rotating polytropes (see especially \citealt{Vandervoort1980}), very few investigations of the stability properties of spheroidal, differentially rotating stellar systems have been conducted so far,
and
mostly by means of numerical approaches~\citep[see, e.g.\@, see][]{Barnes1986,Kuijken1994,vanderMarel1997,Varri2012}, partly with the hope of exploring the validity of the classic~\citet{OstrikerPeebles1973} stability criterion in that context.
It is therefore opportune to explore the role of angular momentum in the early evolution of spheroidal stellar systems
and investigate the possible interplay between velocity anisotropies and internal rotation.

The present paper is organised as follows.
Section~\ref{sec:DF} presents the equilibrium model which we consider as initial state,
in particular its \DF\ with velocity anisotropies, and how solid rotation is introduced (through the ``Lynden-Bell trick'').
With the aim of characterising the linear stability of these systems, Section~\ref{sec:OneMode} presents an extension of the matrix method for rotating spherical systems,
applies this theory to rotating, anisotropic Plummer spheres,
and provides also a direct comparison with $N$-body simulations.
Section~\ref{sec:Charting} extends this analysis to the whole two-dimensional parameter space of velocity anisotropy and rotation, and investigates their respective roles in sourcing dynamical instabilities.
We finally discuss our results in Section~\ref{sec:Discussion},
and conclude in Section~\ref{sec:Conclusion}.

\section{Describing rotating spheres}
\label{sec:DF}

Let us start by describing in detail 
the equilibrium model adopted for the initial state of the system,
focusing on its two-parameter distribution function
that encodes both the velocity anisotropies
and the system's total rotation.

\subsection{Angle-action coordinates}
\label{sec:AA}

The mean stellar system is always assumed to be spherically symmetric,
and characterised by a mean radial potential ${ \psi (r) }$,
with ${ (r , \vtheta , \phi) }$ the spherical coordinates.
Throughout this study, we take the mean potential to be a Plummer potential~\citep{Plummer1911},
\begin{equation}
 \psi (r) = - \frac{G M_{\mathrm{tot}}}{ \Rs} \frac{1}{\sqrt{1+({r}/{\Rs})^{2}}} ,
 \end{equation}
and work in the dimensionless units ${ G = \Rs = M_{\mathrm{tot}} = 1 }$,
with $M_{\mathrm{tot}}$ the system's total mass.
A spherically symmetric potential guarantees the existence of angle-action
coordinates, ${ (\bT , \bJ) }$,~\citep{BinneyTremaine2008}, with
\begin{equation}
\bJ = (J_{1} , J_{2} , J_{3}) = (J_{r} , L , L_{z}) ,
\label{def_actions}
\end{equation}
standing respectively for the radial action, the norm of the angular momentum, and its projection along the vertical axis.
The orbit's inclination is naturally defined as ${ \cos (i) = L_{z} / L }$, with ${ 0 \leq i \leq \pi }$.
According to the Jeans theorem~\citep{Jeans1915}, the mean-field distribution of the system is fully characterised by its \DF\@, ${ F (\bJ) }$, normalised as ${ \!\int\! \rd \bT \rd \bJ \, F = \Mtot }$.
Because the system is spherically symmetric,  the orbital frequencies, ${ \bO (\bJ) = \partial H / \partial \bJ }$, satisfy the degeneracy condition ${ \Omega_{3} (\bJ) = 0 }$,
i.e. stars orbit in a fixed orbital plane.
Similarly to~\cite{Hamilton2018}, this already entices us to define the reduced actions
${ \tbJ = (J_{r} , L) }$, and the associated reduced frequencies ${ \tbO = \partial H (\tbJ) / \partial \tbJ }$.
Note that spherical systems with a spherically symmetric velocity distribution satisfy ${ F = F (\tbJ) }$.

\subsection{Introducing velocity anisotropies}
\label{sec:DejongheFamily}

\begin{figure}
\center\includegraphics[width=0.45\textwidth]{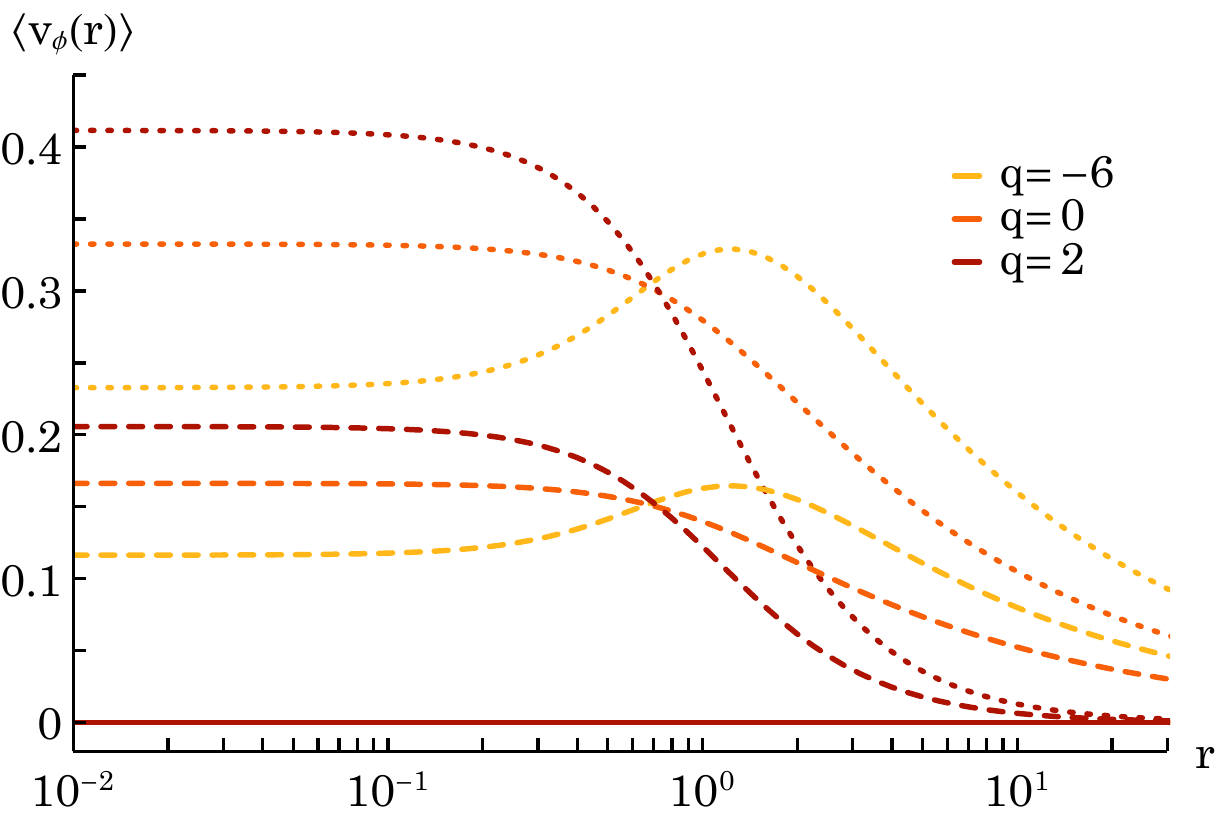}
\center\includegraphics[width=0.45\textwidth]{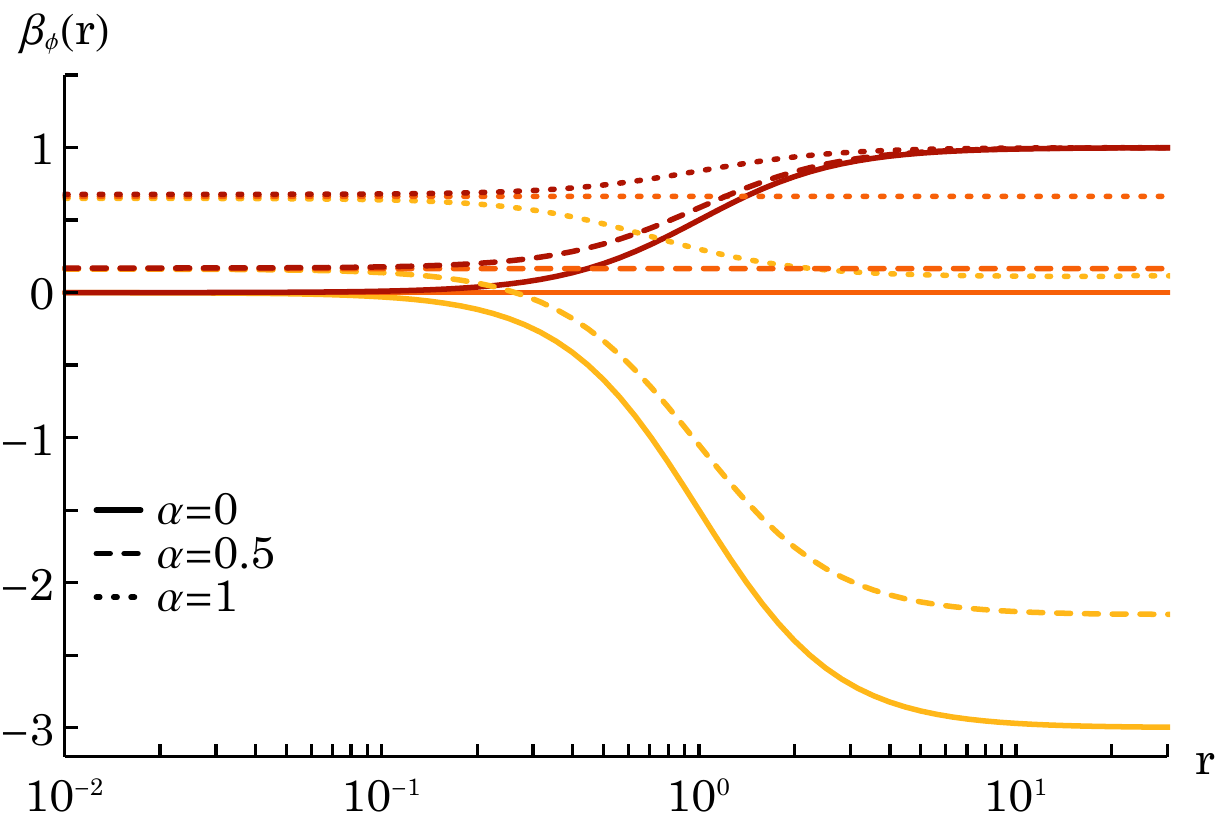}
\caption{ Illustration of the radial dependence of ${ \langle v_{\phi} (r) \rangle }$
(top panel) and ${ \beta_{\phi}(r) }$ (bottom panel)
for various velocity anisotropies (via $q$, indicated by the shade of colour) and total angular momentum (via $\alpha$, indicated by the line style).
Here, $\alpha$ controls ${ \langle v_{\phi} (r) \rangle }$, while $q$ controls $\beta_\phi$,
noticeably in the outer region of the cluster.
We note that for ${ (\alpha,q) =(1,-6) }$ a significant fraction of the radial support is centrifugal,
whereas $\sigma_\phi$ remains comparatively small.
The sphere has therefore a large reservoir of kinetic energy,
which can be converted into thermal energy through an instability.
}
\label{fig:KinematicMoments}
\end{figure}

As in the investigation conducted by~\cite{Breen+2017},
in order to study possible effects associated with velocity anisotropies,
we consider the family of \DFs\ for the Plummer sphere proposed by~\cite{Dejonghe1987}
\begin{equation}
F_0(q,E,L)\! =\! \frac{ 3 \Gamma( 6\! -\! q )}{ 2 (2\pi)^{5/2} } (\!-\!E)^{\frac{7}{2} - q} \mathcal{F}\! \left(\!0, q/2, \frac{9}{2}\! -\! q, 1; \!-\frac{L^2}{2 E} \!\right)\!\!,
\label{def_F0}
\end{equation}
with $q$ controlling the flavour and degree of velocity anisotropy of the system,
and $\Gamma$ the gamma function.
Here, $\mathcal{F}$ is expressed in terms\footnote{This formulation slightly simplifies Dejonghe's by suppressing vanishing ${ \Gamma(q/2) }$ terms, which are indeterminate when $q$ is an even, negative integer.} of the hypergeometric function $\!\ _2 F_1$
\begin{equation}
\mathcal{F} (a,\! b,\! c,\! d; x) \!\! = \!\! \left\{ \!\!
\begin{aligned}
& x^{a} \frac{_2 F_1(a\!+\!b, \!1\!+\!a\!-\!c; a\!+\!d; x) }{ \Gamma(c\!-\!a) \Gamma(a\!+\!d) } ,
\!\!\! & \text{if } x\leq 1;
\\
& \frac{1}{x^{b}}\frac{\!\ _2 F_1(a\!+\!b,\! 1\!+\!b\!-\!d; b\!+\!c; \frac{1}{x}) }{ \Gamma(d\!-\!b) \Gamma(b\!+\!c) } ,
\!\!\! & \text{if } x\geq 1.
\end{aligned}
\right.
\end{equation}
As illustrated in Figure~\ref{fig:KinematicMoments}, the parameter $q$ controls the flavour of the velocity anisotropy in the system, with ${ q < 0 }$ (resp. ${ q > 0 }$) associated with tangential (resp. radial) velocity anisotropies.
One of the advantages of this \DF\ is the simple forms taken by its velocity dispersions in all directions. Indeed, one has
\begin{align}
\sigma_r^2(r) &\,= \frac{(1+r^2)^{-1/2}}{6-q},
\nonumber
\\
\sigma_{\theta}^2(r) &\,= \sigma_{\phi}^2(r) = \sigma_{r}^{2} (r) \left( 1-\frac{q}{2} \frac{r^2}{1+r^2} \right)\!\!,
\end{align}
leading to the anisotropy parameter:
\begin{equation}
\beta_\theta(r)=1-\frac{\sigma_{\theta}^2(r)}{\sigma_{r}^2(r)}=\frac{q}{2}\frac{r^2}{1+r^2}.
\label{eq:defbeta}
\end{equation}
We note that after introducing rotation as below,
this relation still holds, as long as one uses equation~\eqref{eq:defbeta}
to define $\beta_\theta$, ignoring the dispersion in $\phi$.

\subsection{Introducing rotation}
\label{sec:LDBTrick}

\begin{figure}
\center\includegraphics[width=0.45\textwidth]{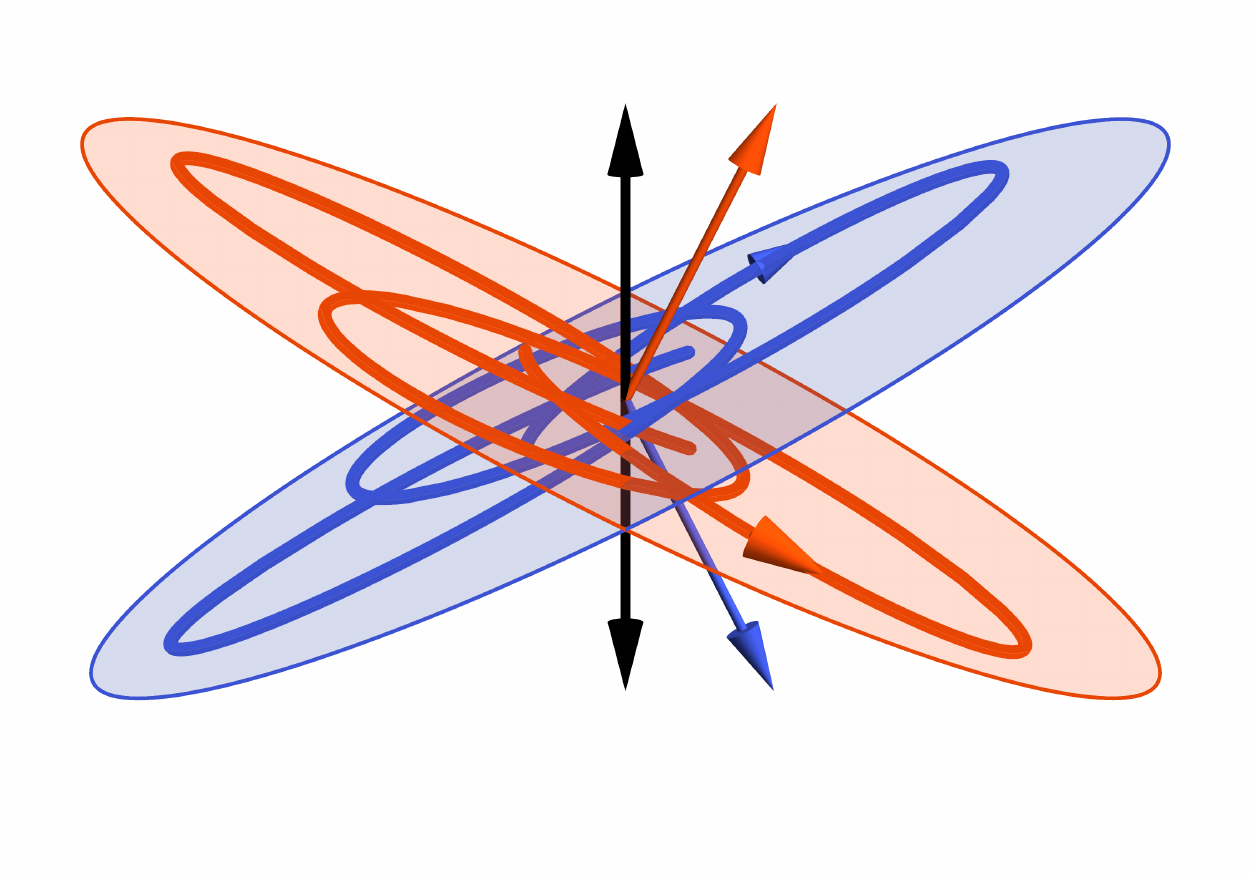}
\caption{Illustration of Lynden-Bell's trick with two orbits, red and blue,
having the same total angular momentum modulus
(the red and blue arrows are perpendicular to the orbital planes),
but opposite $L_z$ (black arrows).
The $\alpha$ parameter of equation~\eqref{solid_DF}
defines the relative number of such orbits (which only differ by the orientation of their orbital plane).
When ${ \alpha = 1 }$, no blue orbits are left.}
\label{fig:lb-trick}
\end{figure}

Following~\cite{LyndenBell1962}, a simple way of introducing rotation
in a spherically symmetric system is to modify its \DF\ as\footnote{We note that any odd fonction of $L_z$ could be used to generalise our choice of ${ \Sign (L_{z}) }$ to introduce rotation.}
\begin{equation}
F (\alpha , q , \bJ) = F_{0} (q , \tbJ) + \alpha F_{0} (q , \tbJ) \, \Sign (L_{z}) \,,
\label{solid_DF}
\end{equation}
with the sign function ${ \Sign (x) }$ equal to ${ -1 }$ for ${ x < 0 }$, then $0$ for ${ x=0 }$ and $1$ for ${ x > 0 }$.
In equation~\eqref{solid_DF}, ${ |\alpha| \leq 1 }$ is a dimensionless parameter
that controls the amount of rotation in the system.
We note that ${ F (\alpha = 0 , q , \bJ) = F_{0} (q , \tbJ) }$, corresponding to the case without rotation,
and that the normalization of $F$ is guaranteed by that of $F_0$, since ${ \!\int_{-L}^{L} \!\! \rd L_{z} \, \Sign (L_{z}) = 0 }$.
The \DF\ of equation~\eqref{solid_DF} does not change the distribution of the orbits' shapes,
but rather the distribution of their orientations, as illustrated in Figure~\ref{fig:lb-trick}.
For each orbit (i.e.\ each $\tbJ$), the probability to have ${ L_z > 0 }$ is multiplied by ${1+\alpha}$, compared to the non-rotating \DF\@. In the limiting case ${ \alpha = 1}$, all particles have ${ L_{z} > 0 }$.

Since a more extensive characterisation of the properties of these configurations will be provided in~\citet{Breen+2019a}, here we just wish to highlight a few selected features\footnote{An implementation of the construction of such anisotropic, differentially rotating spherical models is already freely available at \url{https://github.com/pgbreen/PlummerPlus}.}.
In Figure~\ref{fig:KinematicMoments}, we illustrate how $\alpha$ creates some mean azimuthal motion
in an otherwise non-rotating system.
We also note that $\alpha$ induces an anisotropy in the azimuthal motions,
by breaking the symmetry in the two tangential directions,
as given by the azimuthal anisotropy parameter
\begin{equation}
\beta_{\phi} (r) = 1 - \frac{\sigma_{\phi,\alpha}^2(r)}{\sigma_{r}^2(r)}.
\end{equation}
The bottom panel of Figure~\ref{fig:KinematicMoments} also shows
how $\alpha$ biases the azimuthal anisotropy towards radial motions
by increasing the mean azimuthal velocity.
Similarly, increasing $q$ increases the mean azimuthal velocity
in the central region of the cluster and decreases it in the outer region.
For ${ q \neq 0 }$, a radial variation of $\beta_\phi$ is also introduced in the outer region of the cluster.
Moreover, increasing $q$ decreases ${\beta_\phi}$.
We note that by construction  ${ \langle v_{\phi} (r) \rangle \propto \alpha }$.
Finally, we note that even for $q=-6$, when $\alpha=1$,
${ \sigma_r>\sigma_\phi }$.

\begin{figure}
\includegraphics[width=0.47\textwidth]{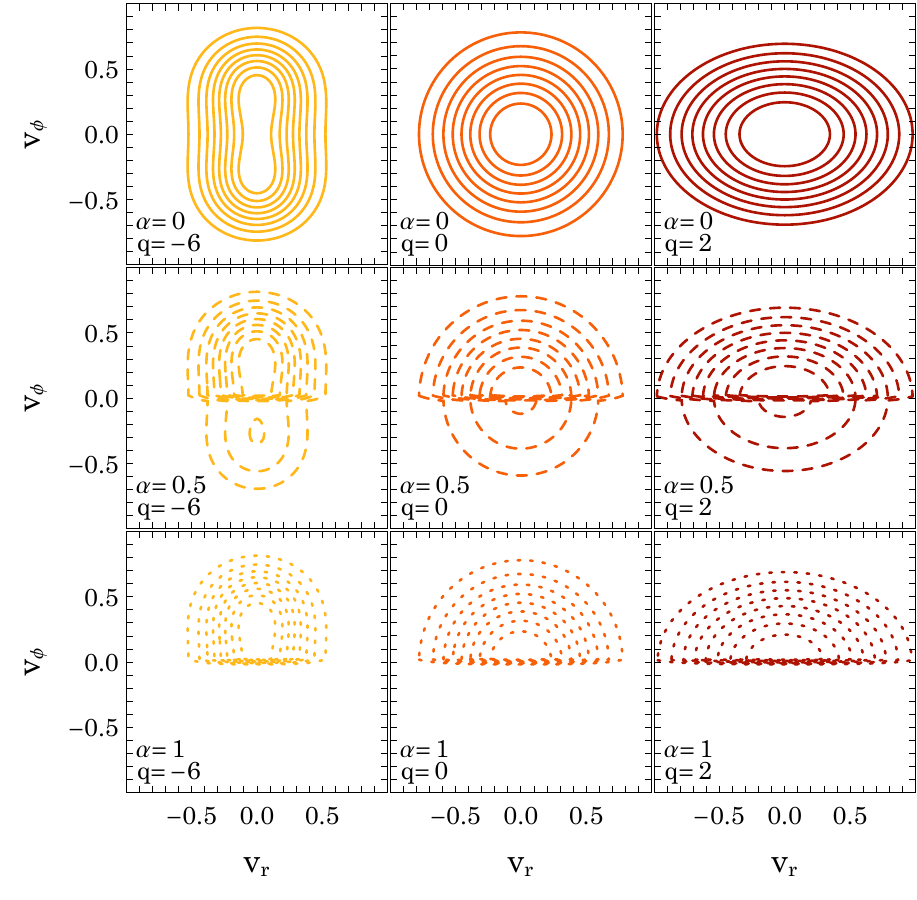}
\caption{Number density of stars in the ${ (v_r,v_{\phi})}$-space
at the constant radius ${r=1}$.
Whenever ${ \alpha \neq 0 }$, the PDF is discontinuous on the line ${ v_{\phi} = 0 }$.
As $q$ increases, the spread in $v_{\phi}$ decreases, and the system becomes more radially biased.
As $\alpha$ increases, stars are swapped towards positive $v_\phi$.}
\label{fig:PDFvrvphi}
\end{figure}

In Figure~\ref{fig:PDFvrvphi}, we illustrate the PDF of ${ (v_r , v_{\phi}) }$ for various values 
of $\alpha$ and $q$.
We note that the ansatz given by equation~\eqref{solid_DF} introduces a sharp discontinuity
in the distribution of stars in the $v_{\phi}$-direction as soon as ${ \alpha \neq 0 }$.
It is possible that this discontinuity plays a role in driving the instabilities,
as discussed further in Section~\ref{sec:discontinuity}.

\section{Characterising unstable modes}
\label{sec:OneMode}

\subsection{Matrix method and rotating spheres}
\label{sec:matrix}

The stability properties of anisotropic, spherical equilibria
have already been the object of many studies~\citep{Polyachenko1981,Saha1991,Weinberg1991,Polyachenko:2015ii}.
Below, we follow the notations from~\cite{Hamilton2018}.

The linear stability of a long-range interacting integrable system
can generically be characterised by its response matrix ${ \wbM (\omega) }$~\citep{BinneyTremaine2008},
which in the case of ${3D}$ spherical systems is given by
\begin{equation}
\!\wM_{pq} (\omega) \!=\! (2 \pi)^{3} \!\! \sum_{\bn \in \mathbb{Z}^{3}\,} \!\! \int \!\! \rd \bJ \, \frac{\bn \cdot \partial F / \partial \bJ}{\omega - \bn \cdot \bO (\bJ)} \big[ \psi^{(p)}_{\bn} (\bJ) \big]^{*}\! \psi_{\bn}^{(q)} (\bJ) .
\label{Fourier_M}
\end{equation}
A system is said to be linearly unstable if there exists a frequency ${ \omega = \omega_{0} + \ri \eta }$ (with ${ \eta > 0 }$), such that ${ \wbM (\omega) }$ admits an eigenvalue equal to $1$. The instability criterion for the  sphere therefore reads
\begin{equation}
\det [ \bI - \wbM (\omega_{0} + \ri \eta)  ] ={0}\,.
\label{eq:det0}
\end{equation} In that case, the system supports an unstable mode of oscillation frequency $\omega_{0}$ and growth rate $\eta$.
To capture the system's self-gravitating response, equation~\eqref{Fourier_M} involves a biorthogonal basis of potentials and densities, ${ \psi^{(p)} (\bx) }$ and ${ \rho^{(p)} (\bx) }$, whose conventions are given explicitly in Appendix~\ref{appendix:MatrixMethod}.
The same Appendix details how the response matrix associated with the rotating \DF\ from equation~\eqref{solid_DF} is computed, and shows that
\begin{equation}
\wbM (\alpha , \omega) = \wbM_{0} (\omega) + \alpha \, \wbM_{1} (\omega) ,
\label{final_Matrix}
\end{equation}
where ${ \wbM_{0} (\omega) }$ corresponds to the system's response in the absence of rotation, while ${ \alpha \, \wbM_{1} (\omega) }$ captures the effects associated with rotation.
Equation~\eqref{final_Matrix} is the master equation characterising its linear stability\footnote{We note that in the slow rotating rate regime, i.e. ${ \alpha \ll 1 }$,
seeking eigenvalues for equation~\eqref{final_Matrix}
is formally identical to the corresponding problem addressed by
perturbation theory in quantum mechanics.
In particular, given the knowledge of 
the full spectrum of ${ \wbM_{0} (\omega) }$, the shift in eigenvalues
and eigenvectors associated with the introduction of ${ \alpha \ll 1 }$
can be computed perturbatively.}. 
Because of its reduced number of symmetries, we also note that the matrix ${ \wbM_{1} (\omega) }$
is less sparse than the zero-rotation matrix ${ \wbM_{0} (\omega) }$,
and hence the numerical task is more cumbersome.
In Appendix~\ref{appendix:FastWMat}, we detail our numerical implementation
of the computation of ${ \wbM (\alpha , \omega) }$.

Before applying the previous machinery to the study of the linear stability
of a wide range of  equilibria (i.e. different ${ (\alpha , q) }$),
we will start by considering one particular case.
Following equation~\eqref{solid_DF}, we focus on the fiducial model
${ (\alpha , q) = (1 , -6) }$, i.e.\ a model with the maximum rotation
and with tangential velocity anisotropies,
as already illustrated in Figure~\ref{fig:KinematicMoments}. 
From now on, our search is restricted to unstable modes with the pattern number ${ m = 2 }$,
i.e. bars and two-armed spirals.
This is motivated by previous results from $N$-body simulations,
which were conducted in continuation of~\cite{Breen+2017}
and will appear in a separate study~\citep{Breen+2019b}, which
suggested that bi-symmetric modes dominate.

\subsection{Matrix method methodology}
\label{sec:MatrixMethodology}

\begin{figure}
\centering
\includegraphics[width=0.45\textwidth]{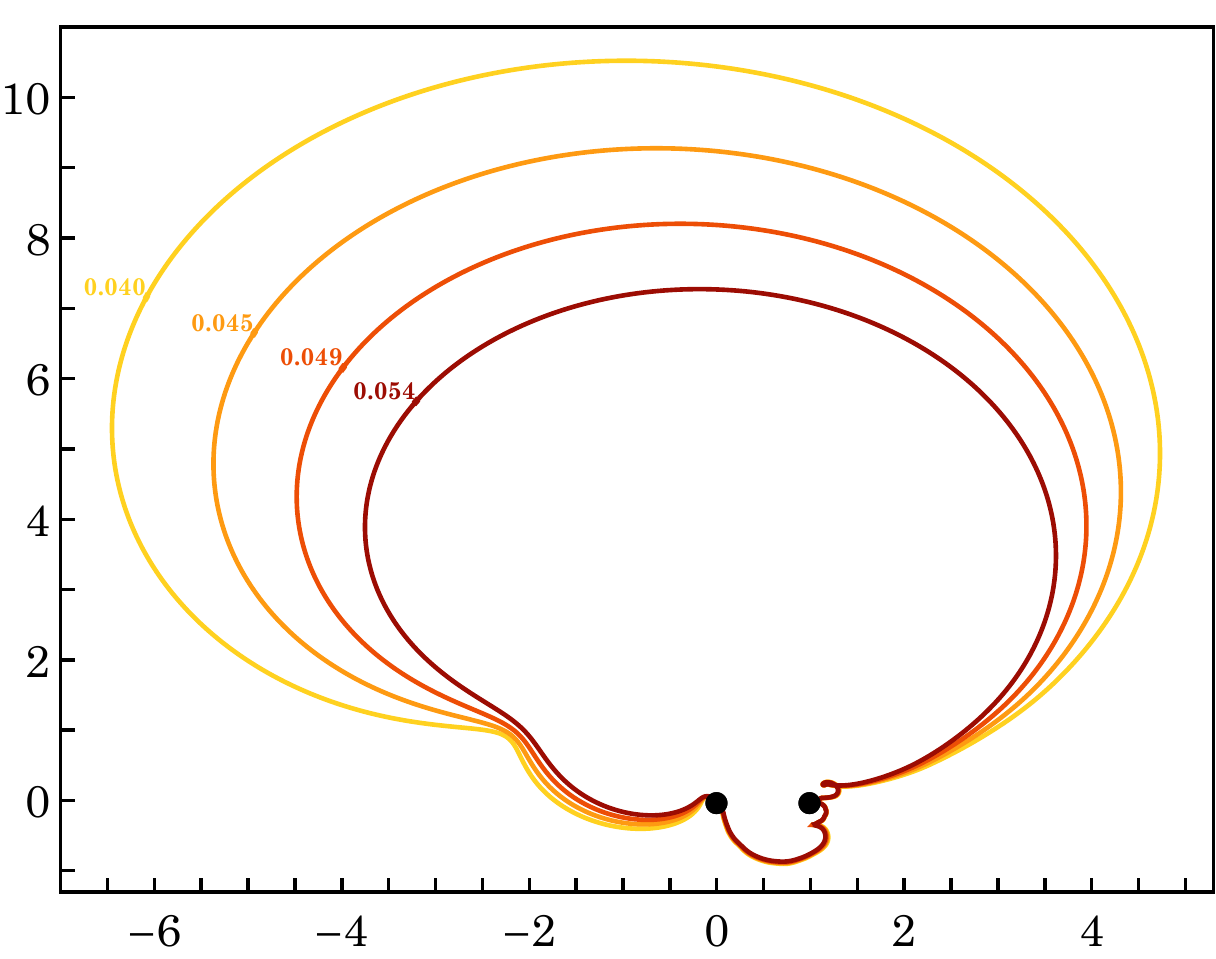}
\includegraphics[width=0.45\textwidth]{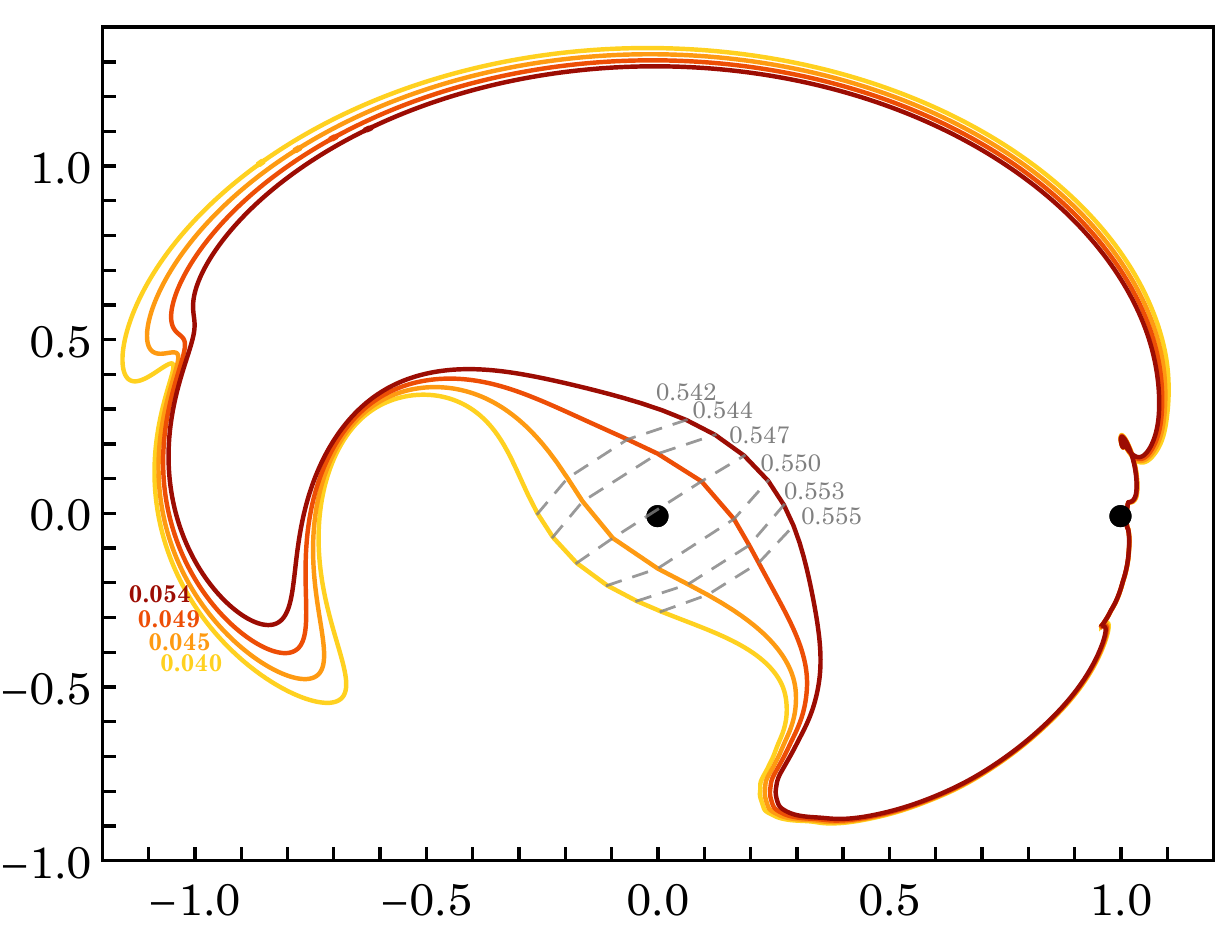}
\caption{Nyquist diagram of the fastest growing mode of the model
${ (\alpha , q) = (1 , -6) }$, as given by the
curves ${ \omega_{0} \mapsto \det [\bI - \wbM (\omega_{0} + \ri \eta)] }$.
The top panel represents the Nyquist diagram without renormalization.
Following~\protect\cite{PichonCannon1997}, in the bottom panel,
a logarithmic scaling is applied,
${ r \re^{\ri \theta} \mapsto \tfrac{1}{3} \log (1 \!+\! 10^{3} r) \, \re^{\ri \theta} }$,
to highlight the structure of the contour around the origin
and make sure that the first growing mode is indeed selected.
Different colours are associated with different growth rates,
while the changes in $\omega_{0}$ along each curve are indicated with the gray dashed lines.
This diagram shows that the cluster supports an unstable mode
with a growth rate, ${ 0.045 \leq \eta \leq 0.049 }$,
and oscillation frequency, ${ 0.547 \leq \omega_{0} \leq 0.550 }$.}
\label{fig:NyquistDiagram}
\end{figure}

In order to investigate the stability properties of the fiducial model,
the numerical methods detailed in Appendices~\ref{appendix:MatrixMethod}
and~\ref{appendix:FastWMat} are implemented.
Following~\cite{PichonCannon1997}, unstable modes are identified using Nyquist contours.
To do so, for a given value of ${ \eta > 0 }$, one constructs the curve
${ \omega_{0} \mapsto \det [ \bI - \wbM (\omega_{0} + \ri \eta) ] }$.
If ever this contour encloses the origin, then the cluster supports
an unstable mode with a growth rate larger than $\eta$. 
Figure~\ref{fig:NyquistDiagram} illustrates the Nyquist contours associated with our fiducial model.
The set of numerical parameters of this computation is detailed in Appendix~\ref{appendix:ConvergenceMatrix},
where we also explore the convergence of the matrix method
w.r.t.\ these various numerical parameters.
In order to find the instability with the largest growth rate,
we draw Nyquist contours for progressively decreasing values of $\eta$,
from a value at which the Nyquist contour does not enclose the origin, down to the first crossing point.
For the fiducial model, we identify an unstable mode with frequency
${ (\omega_{0} , \eta) \simeq (0.547,0.047) }$.
We note that as one keeps decreasing the value of $\eta$,
additional sub-dominant modes may be found.
This will be of particular interest in Section~\ref{sec:q>0space},
when considering radially-biased systems.

\begin{figure}
    \centering
    \subfloat[Linear theory]{        \includegraphics[width=0.93\columnwidth]{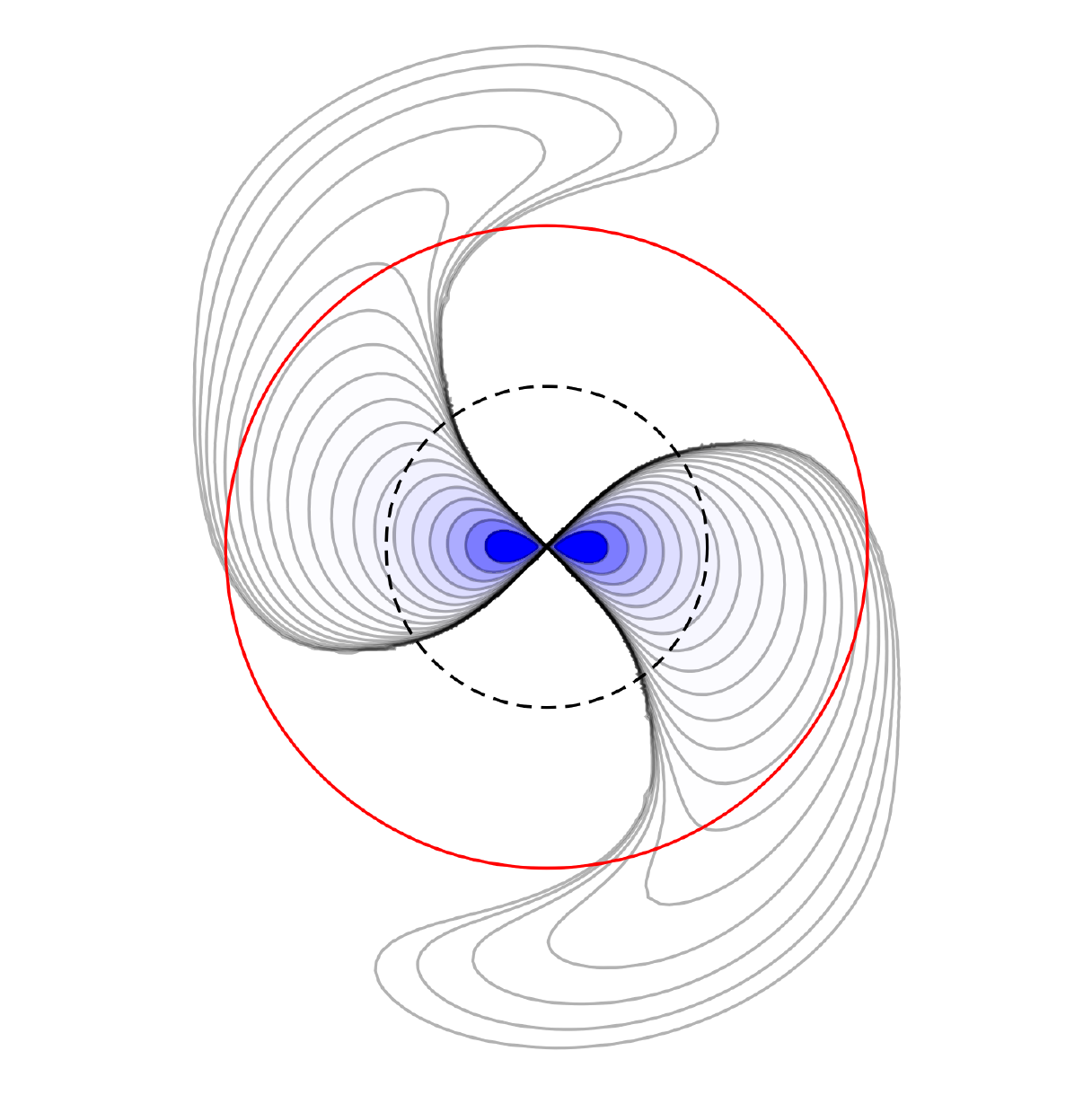}
                \label{subfig:ShapeModeMatrix}}\qquad
      \subfloat[$N$-body simulation]{        \includegraphics[width=0.93\columnwidth]{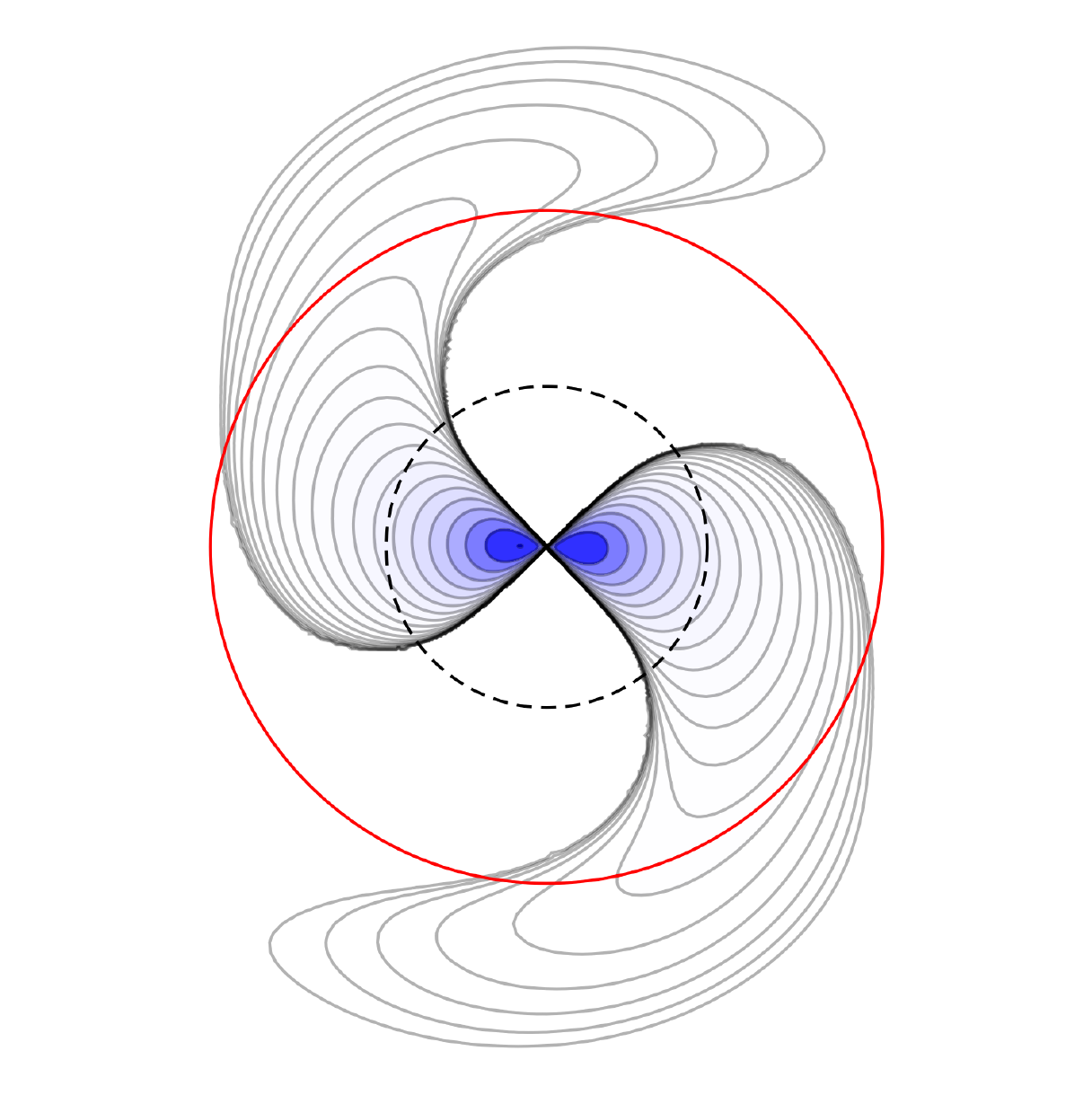}
               \label{subfig:ShapeModeNbody}}
      \caption{Radial dependence in the equatorial plane
   of the density of the unstable mode recovered in
   Figures~\ref{fig:NyquistDiagram} and~\ref{fig:GrowthRateNbody}.
   The top panel is the linear theory's prediction,
   and the bottom panel the measurement from $N$-body simulations.
   The black dashed circle represents the scale radius of the background Plummer potential,
   and the red circle the mode's corotation radius.
   Isocontours are logarithmically sampled, with a factor ${\sim\!1.6}$ between two consecutive lines, so that the outermost line has a density ${1.5 \!\times\! 10^{4} }$ times smaller than the peak value.
   The similarity between the two results is striking.}
   \label{fig:ShapeMode}
\end{figure}

\begin{figure}
    \centering
   \includegraphics[width=0.45 \textwidth]{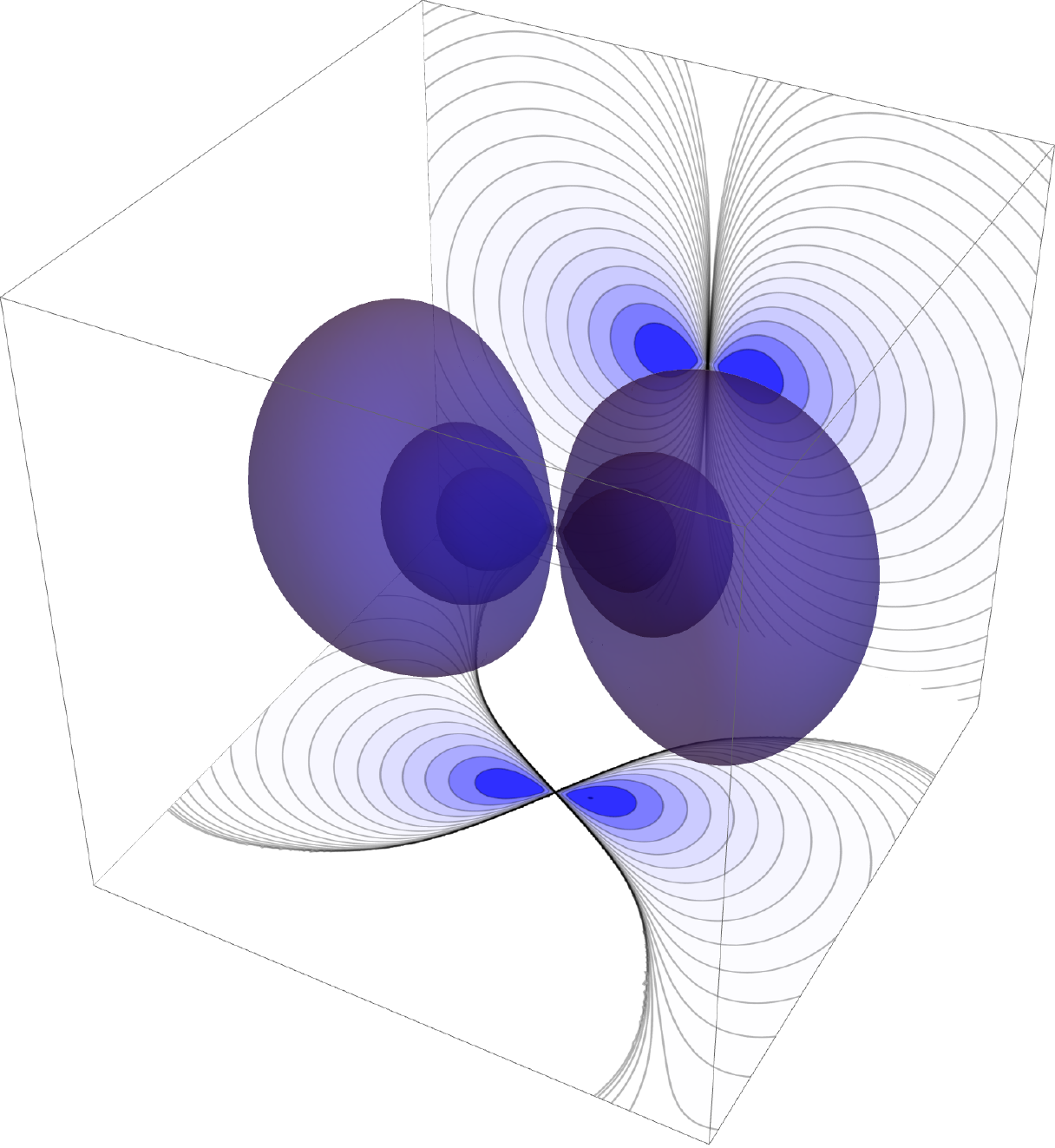}
   \caption{${3D}$ isosurfaces of the density of the unstable mode recovered in Figure~\ref{fig:NyquistDiagram}, and their intersection with the equatorial and meridional planes.}
   \label{fig:ShapeMode3D}
\end{figure}

Once the unstable mode has been identified,
the response matrix's eigenvector associated with the eigenvalue nearly equal to $1$
can be determined, which allows us to obtain the mode's shape via
\begin{equation}
\rho(r,\vtheta,\phi) = \RePart\!\left( \! \sum_p X_p \, \rho^{(p)}(r,\vtheta,\phi) \! \right),
\label{modereconstruction}
\end{equation}
with ${X_p}$ the eigenvector.
In Figure~\ref{fig:ShapeMode},
we represent in the equatorial plane
the density of the mode of the fiducial model.
In Figure~\ref{fig:ShapeMode3D}, we illustrate the ${3D}$ density isocontours
of the same unstable mode.
We note that the density response exhibits a large scale
spiral pattern extending beyond corotation,
and that the mode's thickness in the vertical direction is significant.

\subsection{$N$-body methodology}
\label{sec:NbodyMethodology}

In order to validate the linear stability analysis further,
$N$-body simulations were carried out for the same fiducial model,
with ${ N = 2^{17} }$ particles, integrated for a time interval
of $100$ H\'enon Units~\citep{Henon1971}.
The simulations were performed
using \texttt{gyrfalcON}~\citep{Dehnen2002}.
To characterise the frequencies of the unstable modes,
we Fourier-analysed the space density as follows.
For a given pattern number ${ m = 2 }$, we define ${ A_{m} (t) \!=\! C_{m} (t) / C_{0} (t) }$,
where
\begin{equation}
C_{m} (t) = \!\! \int_{0}^{+ \infty} \!\!\!\!\!\!\! \rd r \, r \!\! \int_{- \infty}^{+ \infty} \!\!\!\!\!\! \rd z \, \!\! \int_{0}^{2 \pi} \!\! \frac{\rd \phi}{2 \pi} \, \rho (r , z , \phi , t) \, \re^{- \ri m \phi} ,
\label{def_Cm}
\end{equation}
with ${ \rho (R , z , \phi , t) }$ the simulation's instantaneous density
(i.e. a sum of $N$ delta functions).
Should the cluster support an unstable mode,
one can in principle extract its frequencies
from the numerically-measured function ${ t \mapsto A_{m} (t) }$,
as one has
\begin{equation}
\frac{\rd \, \log (|A_{m} (t)|)}{\rd t} = \eta ; \;\;\; \frac{\rd \Arg (A_{m} (t))}{\rd t} = \omega_{0} \,.
\label{ModeFromAm}
\end{equation}

In practice, measuring the growth rate of an instability in $N$-body data has two major difficulties. The first is noise due to the discrete nature of the system. The second is that the $N$-body system cannot sustain an exponential growth indefinitely, and the mode must saturate at some point. The transition from exponential growth to saturation is not easy to model,  and may have some dependence on the phase space distribution function.  The result is that this transition can possibly result in a bias such that growth rates based on an exponential fit may systematically underestimate the true growth actually experienced by the mode in its early linear phase.

\begin{figure}
\includegraphics[width=0.24\textwidth]{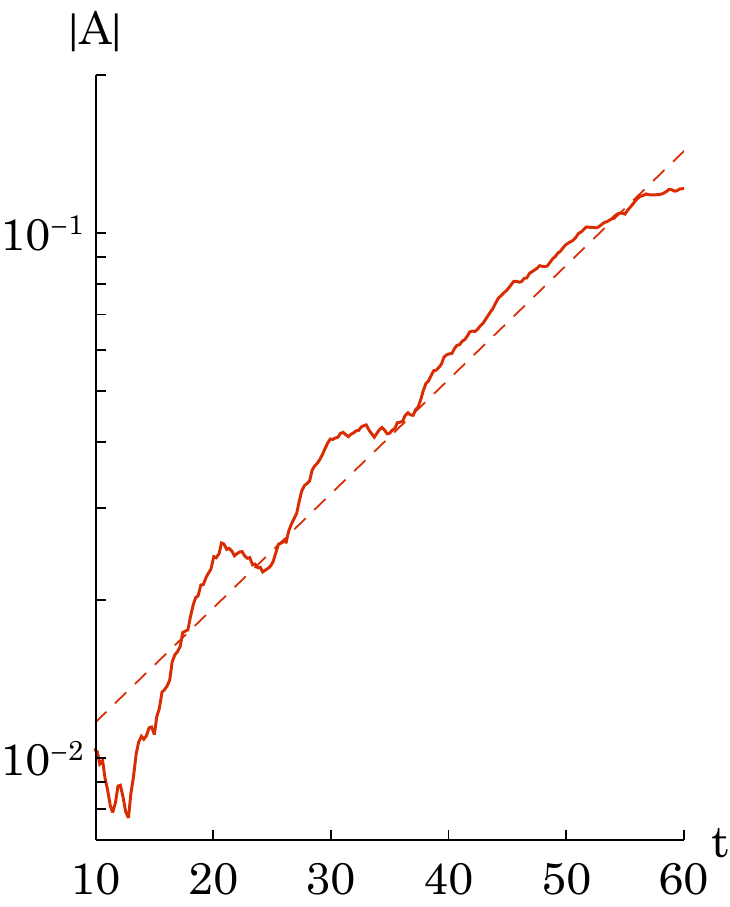}
\includegraphics[width=0.24\textwidth]{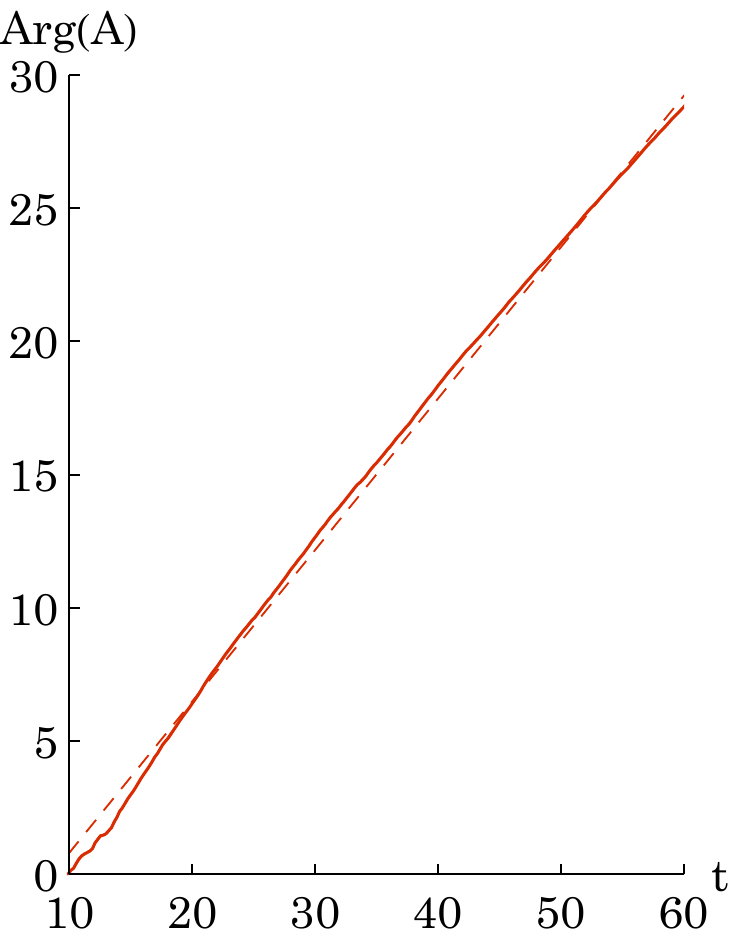}
\caption{Illustration of the time-dependence of ${ |A_{m} (t)| }$ (left panel)
and ${ \Arg (A_{m} (t)) }$ (right panel) for the same cluster
model as in Figure~\ref{fig:NyquistDiagram}.
Following equation~\eqref{ModeFromAm}, linear fits are also represented.
We recovered ${(\omega_{0} , \eta) =  (0.57, 0.022)}$ that should
be compared with the measurements of Figure~\ref{fig:NyquistDiagram}.}
\label{fig:GrowthRateNbody}
\end{figure}

In order to automatically detect and measure growth rates in a grid of $N$-body simulations, we use an iterative scheme which involves fitting an exponential curve to the amplitude ${ |A_{m} (t)| }$ of the mode, first over the full duration of the simulation, but then reducing the time interval of the fit, until the maximum absolute difference between the fit and data (the normalised amplitude) falls below a predetermined value of tolerance. The choice of tolerance should be small enough such that the time interval is reduced when there is a poor fit to a mode  (because of its saturation) but large enough that the interval is not reduced merely by the presence of noise. Based on a grid of 300 different realisations of the reference model (${ N = 2^{17}, q = -6, \alpha = 1 }$), a value of 0.009 was found to produce robust measurements of the growth rate $\eta$.
Similarly, $\omega_0$ is determined by a linear fit to ${ {\rm{Arg}}(A_m(t)) }$ (\!\!\!\!$\mod{2\pi}$ to account for complete revolutions) over the same time interval. 
Figure~\ref{fig:GrowthRateNbody} illustrates such measurements of $\eta$ and $\omega_{0}$
for the fiducial model.

Having characterised the frequencies of the unstable modes,
one can use the same numerical simulations to extract the mode's shape.
To do so,  the simulation's particles are projected onto the same
${ \rho^{(p)}(r,\vtheta,\phi)}$ basis as for the matrix method,
to reconstruct the mode using equation~\eqref{modereconstruction}. 
This is illustrated in the bottom panel of Figure~\ref{fig:ShapeMode}.
In Appendix~\ref{appendix:ConvergenceNbody}, we briefly check the convergence
of the $N$-body runs w.r.t.\ their various control parameters.

The comparison between linear theory and $N$-body simulations
in the fiducial model shows a ${ \sim 15\% }$ discrepancy in the measured
value of $\eta$, and a ${ \sim 3\% }$ discrepancy in $\omega_0$.
Moreover, as highlighted in Figure~\ref{fig:ShapeMode},
the modes' shapes are in close agreement.
This allows us to quantitatively validate our implementation of the matrix method
as well as our extension of the linear stability analysis to rotating systems.

\section{Charting the instabilities}
\label{sec:Charting}

Having found good agreement between the matrix method
and the direct $N$-body approach,
we now set out to investigate the properties of the unstable modes
in the whole ${ (\alpha , q)}$-space, to account for the contributions
from both velocity anisotropy and  total angular momentum.
As in the previous sections, we will focus on the three characteristics of unstable modes,
namely the growth rate $\eta$,
the oscillation frequency $\omega_{0}$,
and the mode's shape.

\subsection{Instability in the ${(\alpha, q)}$-space}
\label{sec:entirespace}

\begin{figure*}
    \centering
    \subfloat[Growth rate, matrix method.]{        \centering
        \includegraphics[width=0.44\textwidth]{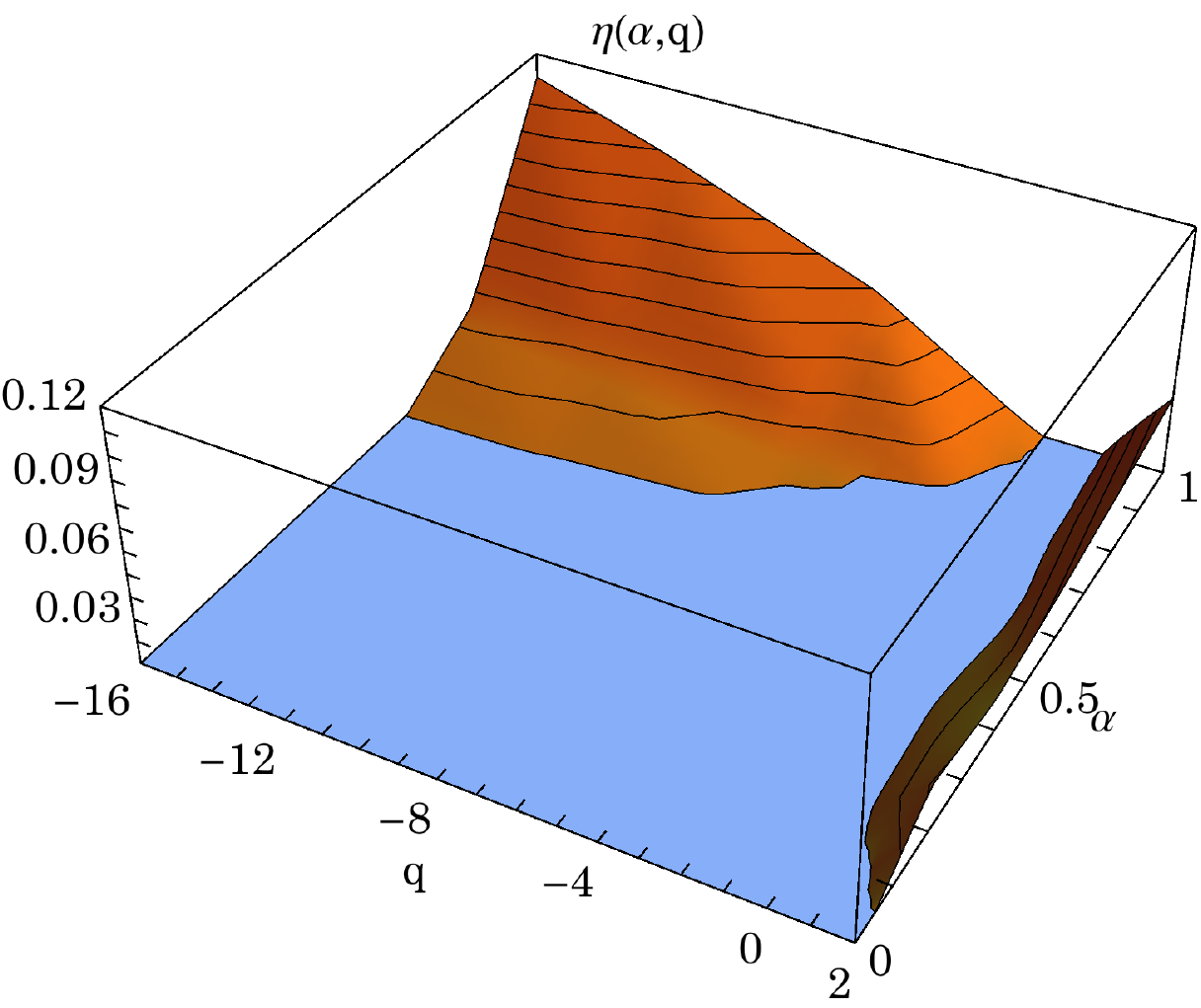} 
               \label{subfig:EtaSurfaceMatrix}}
        \hspace*{1cm}
    \subfloat[Growth rate, $N$-body measurements.]{        \centering
        \includegraphics[width=0.44\textwidth]{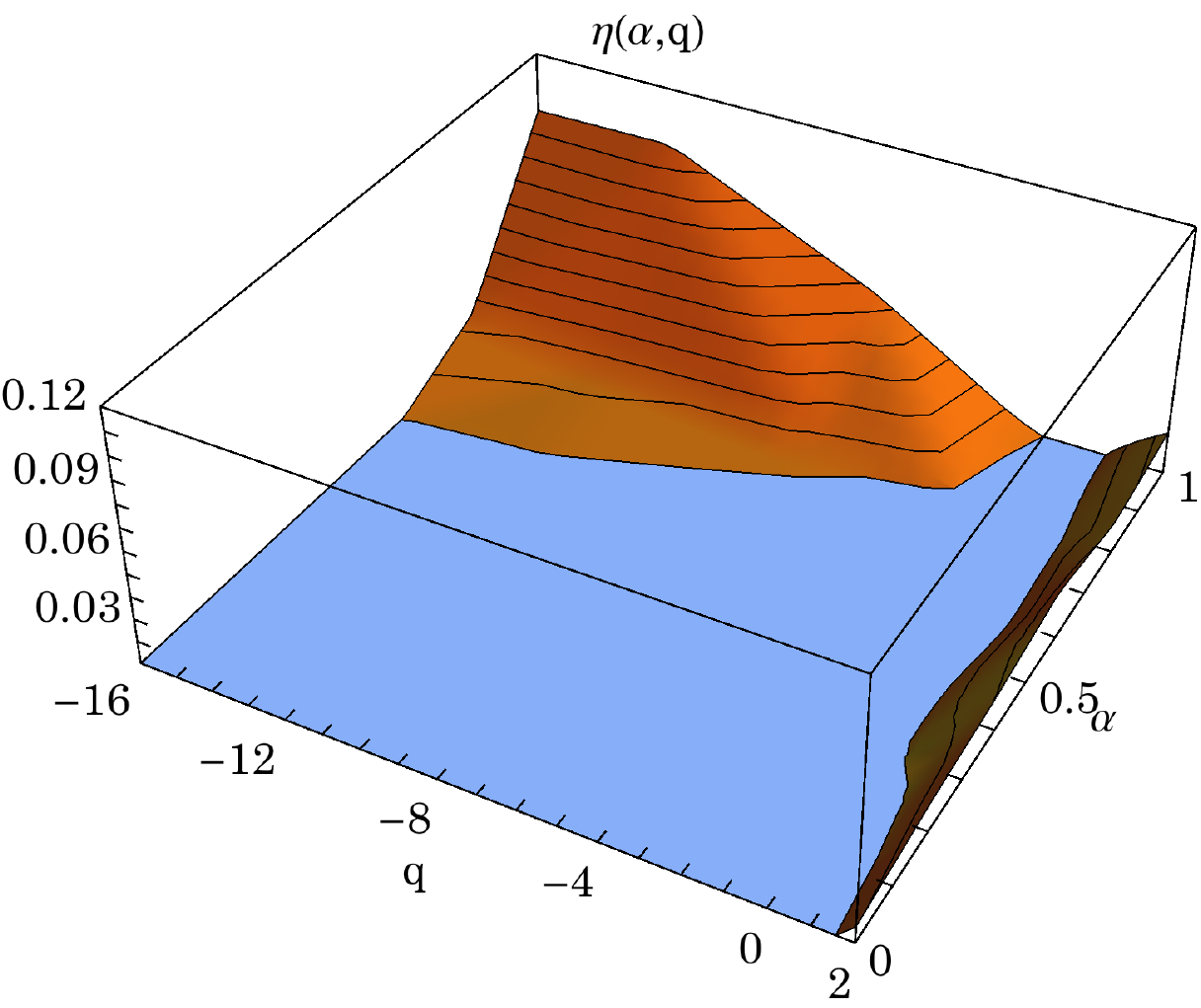} 
              \label{subfig:EtaSurfaceNbody}}
       \vspace{0.2cm}
   \subfloat[Oscillation frequency, matrix method.]{    \centering
        \includegraphics[width=0.44\textwidth]{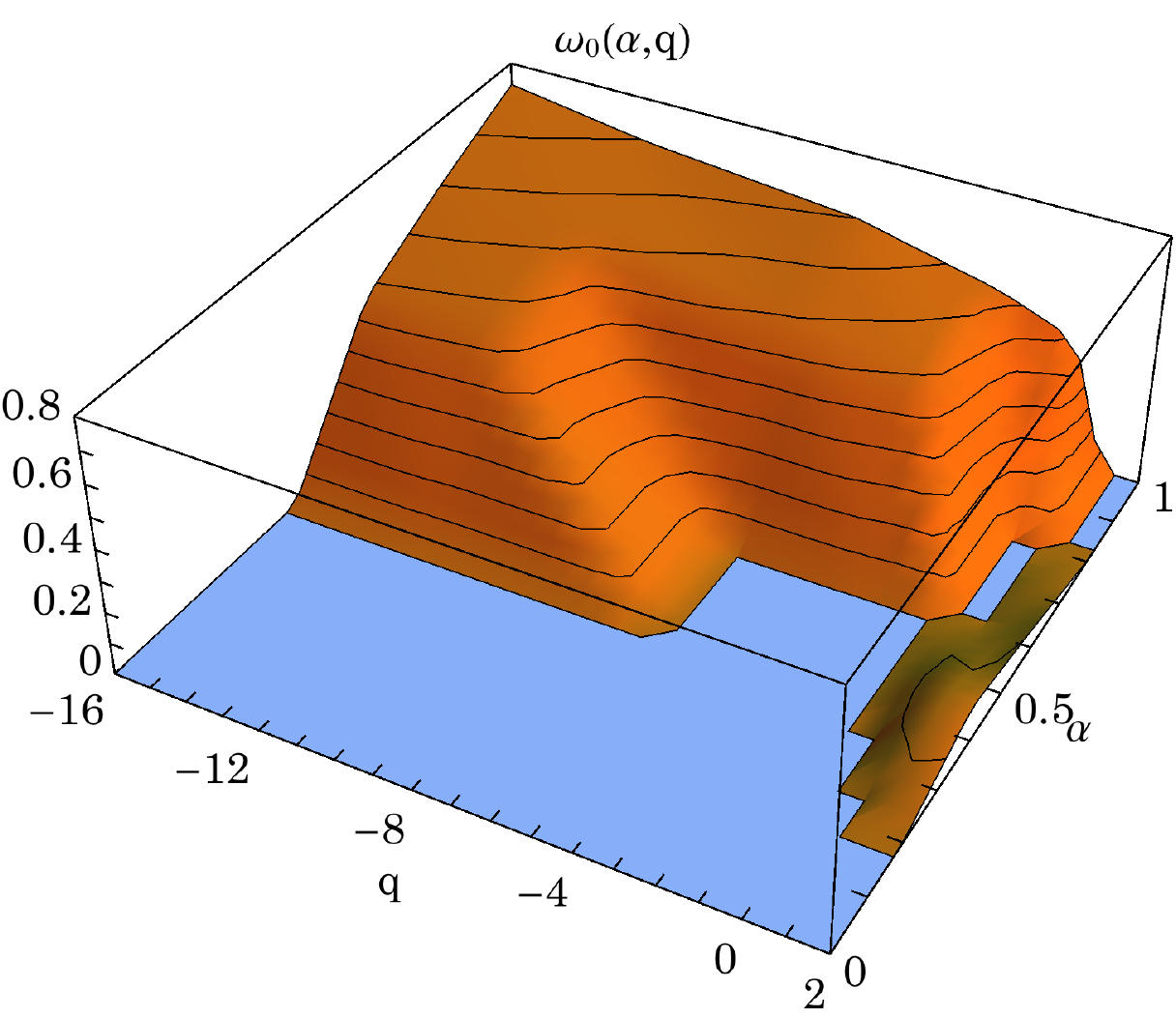} 
               \label{subfig:OmegaSurfaceMatrix}}
        \hspace*{1cm}
   \subfloat[Oscillation frequency, $N$-body measurements.]{        \centering
        \includegraphics[width=0.44\textwidth]{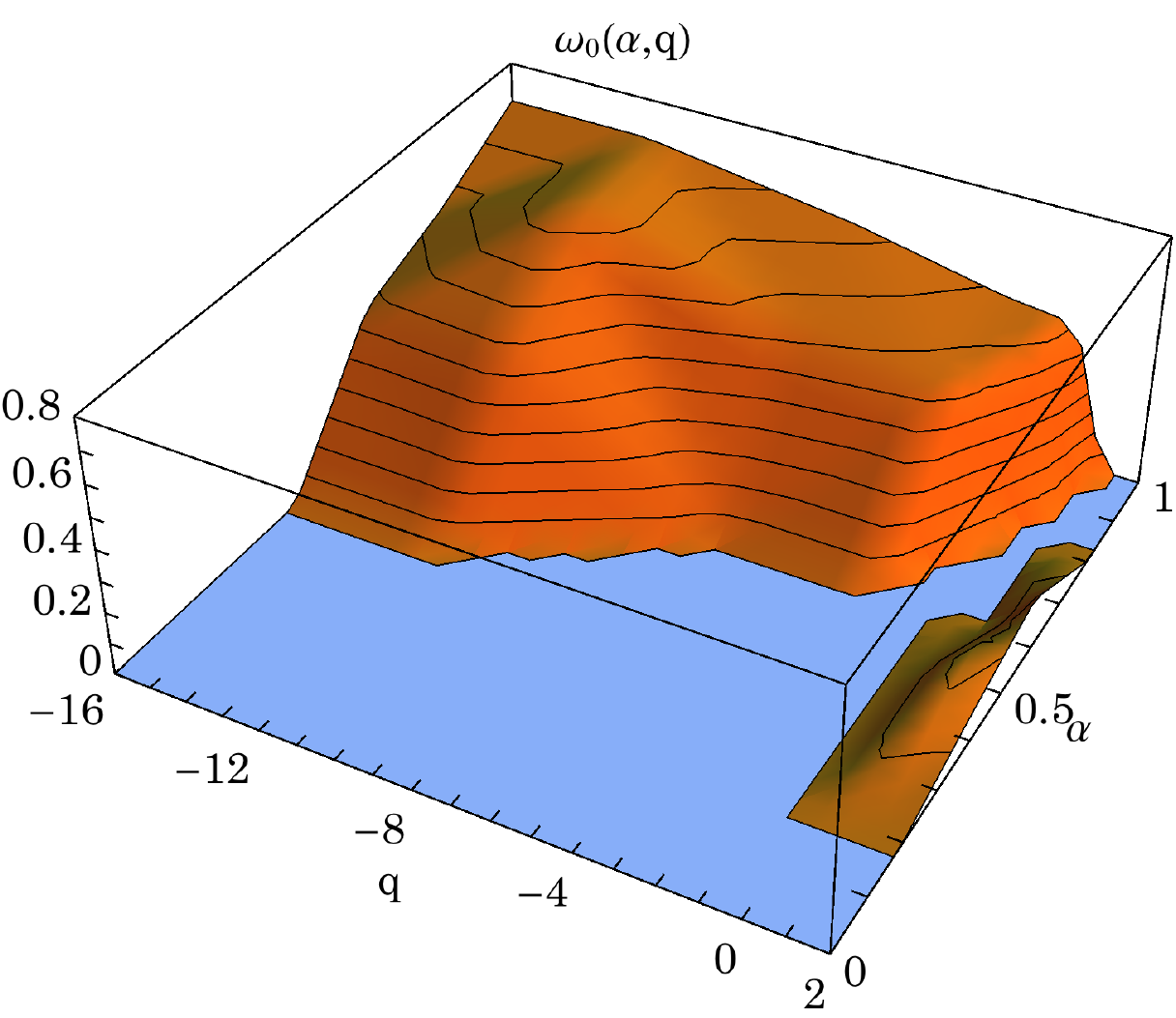} 
              \label{subfig:OmegaSurfaceNbody}}
       \caption{Illustration of the dependence of the growth rates $\eta$ (top panels)
    and oscillation frequencies $\omega_{0}$ (bottom panels),
    as a function of the cluster's parameters ${ (\alpha , q) }$,
     measured using the response matrix method (left panels)
     and $N$-body methods (right panels).
     We searched for unstable modes on a ${ (\alpha,q)}$-grid
     composed of the locations
     ${\alpha = 0, 0.2, 0.4, 0.6, 0.8, 1}$
and ${ q = -16, -12, -6, -2, 0, 1, 2}$.
The blue plane represents our growth rate threshold, ${ \eta = 0.01 }$, for mode selection.
 At this level of accuracy, it appears that rotation introduces a new island of instability 
in the top-left region of each panel (corresponding to ${ q \!<\! 0, \alpha \!>\! \tfrac{1}{2} }$)
and an extension of the known ROI instability island
on the right (near ${ q \!\sim\! 2 }$).
The comparisons of the matrix method's and $N$-body
measurements highlights again the reliability of our implementation
of the linear theory.
}
\label{fig:4panelsSurfaces}
\end{figure*}

We reproduced the methodology of Section~\ref{sec:MatrixMethodology}
to chart the ${(\alpha, q)}$-space using the linear matrix method,
as well as $N$-body simulations.
The resulting frequencies are illustrated in Figure~\ref{fig:4panelsSurfaces}.

In that figure, one can straightforwardly
identify two types of instabilities.
The first one, corresponding to the pyramid on the top of the ${ \eta(\alpha , q) }$
panel (${ q \!<\! 0, \alpha \!\gtrsim\! \tfrac{1}{2}}$) appears to be specific to tangentially-biased and rotating systems.
These instabilities are studied in detail in Section~\ref{sec:q<0space}.
A second type of instability can be also be noted in the bottom right part
of the ${ \eta(\alpha,q) }$ panel (${ q \!>\! 0 }$).
These instabilities appear to be specific to radially-biased systems.
We investigate these instabilities in Section~\ref{sec:q>0space}.
The two right panels of Figure~\ref{fig:4panelsSurfaces} show the $N$-body counterparts
of the measurements using the matrix method.
Again, the quantitative matchings between the left and right panels validate
our generalisation of the matrix method. 
On all these panels, for both $\eta$ and $\omega_{0}$, the lower blue plane
corresponds to a growth rate threshold, ${ \eta = 0.01 }$,
below which we stopped the search for unstable modes,
deeming that the accuracy of the methods
is not sufficient anymore, at least with the choice of accuracy parameters made in this work (see Appendix~\ref{appendix:Convergence})\footnote{Indeed, we found that the matrix method, while potentially
optimal to identify linear growth rates, becomes cumbersome in the small growth rate
regime~\citep[see also][]{Merritt1999}.}.

Having validated the methodology  against $N$-body simulations,
we may now carry out a more thorough exploration of
the ${(\alpha, q)}$-space using only the matrix method,
successively for tangentially-biased (${ q \!<\! 0}$)
and radially-biased systems (${ q \!>\!0 }$).

\subsection{Instability of tangentially-biased systems: ${q<0}$}
\label{sec:q<0space}

\begin{figure}
\includegraphics[width=0.45\textwidth]{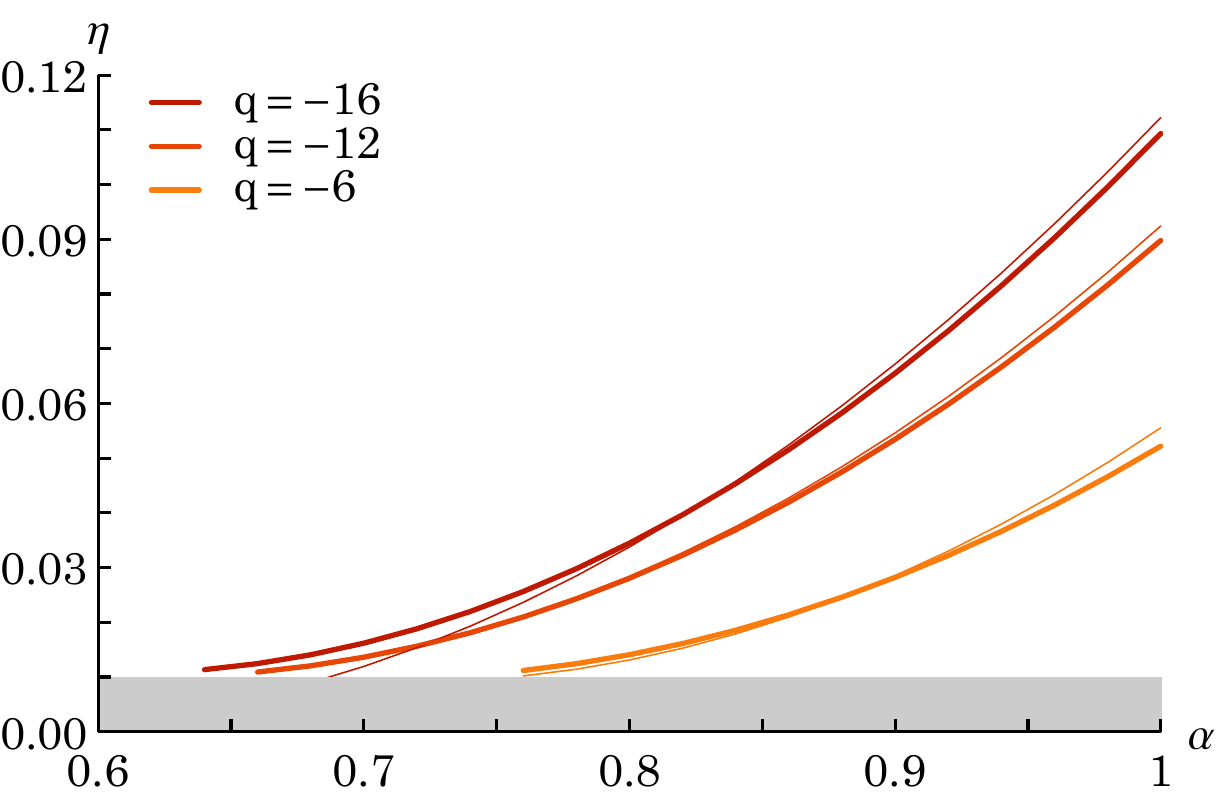}
\includegraphics[width=0.45\textwidth]{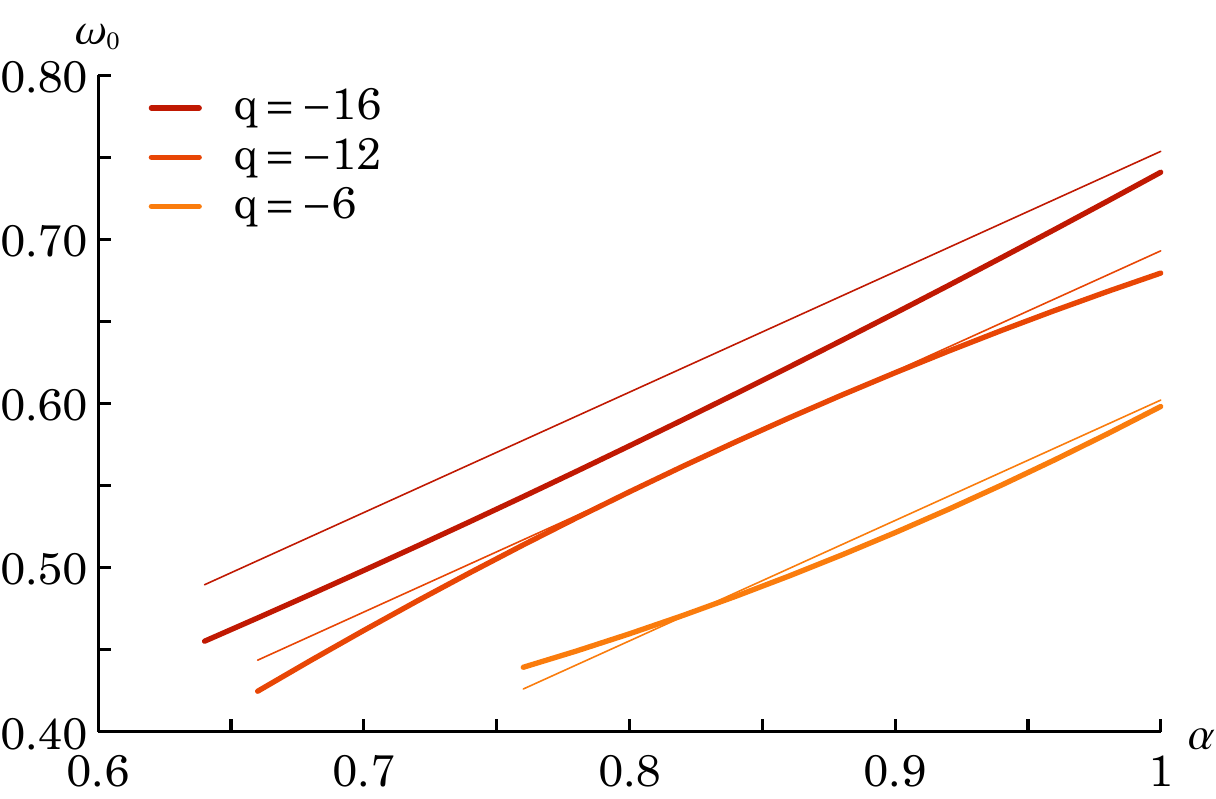}
\caption{Illustration of the growth rate (top panel) and
oscillation frequency (bottom panel) as a function of $\alpha$ for fixed values of $q$,
in the ${ q \!<\! 0 }$ region.
The grey region represents the threshold for the measurement of instabilities.
At this level of accuracy, the $\eta$ and $\omega_{0}$ surfaces can be fitted
globally by quadratic functions of $\alpha$ and $q$, represented with the thin lines.}
\label{fig:etaofalphaq<0}
\end{figure}

In Figure~\ref{fig:etaofalphaq<0}, we illustrate cuts of constant $q$
in the ${\eta(\alpha, q)}$ and ${\omega(\alpha,q)}$ surfaces in the ${q<0}$ region.
In the relevant region, the instability surface ${ (\alpha , q) \mapsto \omega_{0} , \eta }$
can be fitted globally by the simple polynomial expressions
\begin{equation}
\omega_{0} \!=\!  \tfrac{11 \alpha}{15} \!-\! \tfrac{q}{66} \!-\! \tfrac{2}{9} , \;\;
\eta \!= \tfrac{29 \alpha^{2}}{50} \!-\! \tfrac{q^{2}}{8333} \!-\! \tfrac{2  q \alpha }{111} \!+\! \tfrac{q}{103} \!-\! \tfrac{47 \alpha}{50} \!+\! \tfrac{37}{100} .
\label{fitqneg}
\end{equation}
whose cuts are also represented in Figure~\ref{fig:etaofalphaq<0}.

\begin{figure*}
\includegraphics[width=0.95 \textwidth]{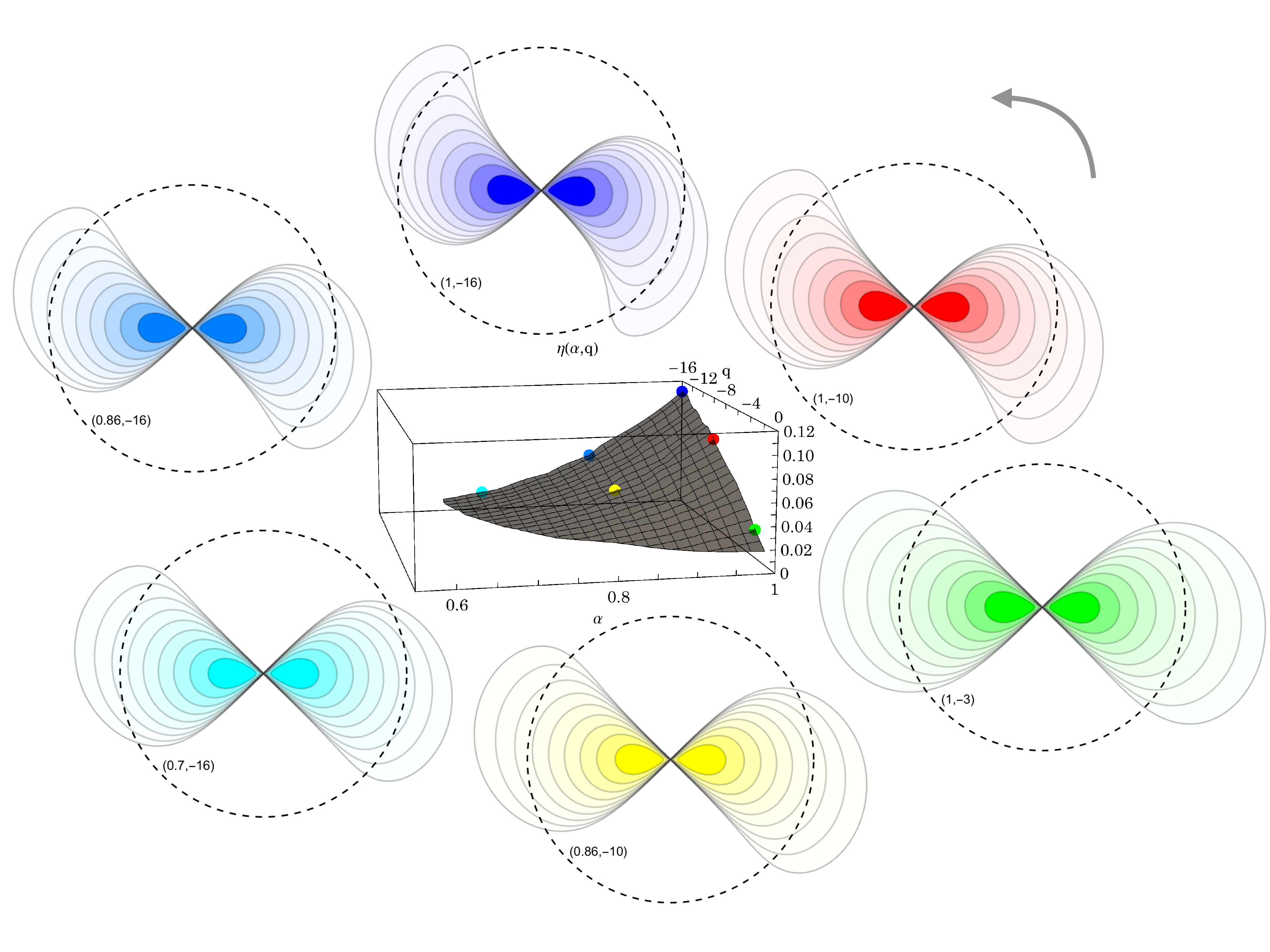}
\caption{Illustration of the various mode shapes as one explores the plane of instability in the ${q\!<\!0}$ region. The black dashed circle represents the scale radius of the Plummer potential. The grey arrow indicates the direction of rotation of both the global rotation and the instability patterns. In addition to matching the colour of the mode with the points in the central figure, the value of ${(\alpha, q)}$ is written inside each circle. With increasing $\alpha$ and decreasing $q$, the spiral patterns get more and more wound.
}
\label{fig:SliceSurfaceq<0}
\end{figure*}

In Figure~\ref{fig:SliceSurfaceq<0}, we represent the ${\eta(\alpha,q)}$ surface
for tangentially-biased systems.
We emphasise that in this regime, both $\eta$ and $\omega_0$ increase with $\alpha$,
and increase as $q$ decreases.
We also note that, so far, these instabilities appear to emerge only in the presence of non-vanishing angular momentum (i.e.\@, they are not present in configuration with ${ \alpha \to 0 }$). We also emphasise that their specific behaviour depend on the adopted choice of the angular momentum distribution (i.e.\@, as resulting from the application of the ``Lynden-Bell trick''), as discussed in Section~\ref{sec:discontinuity}.
In Figure~\ref{fig:SliceSurfaceq<0}, we also show the shape of the modes
at different locations of the instability surface.
The absence of any significant variation in the modes' shape
suggests that the underlying instability mechanism remains the same.

\subsection{Instability of radially-biased systems: ${q>0}$}
\label{sec:q>0space}

\begin{figure}
\includegraphics[width=0.45\textwidth]{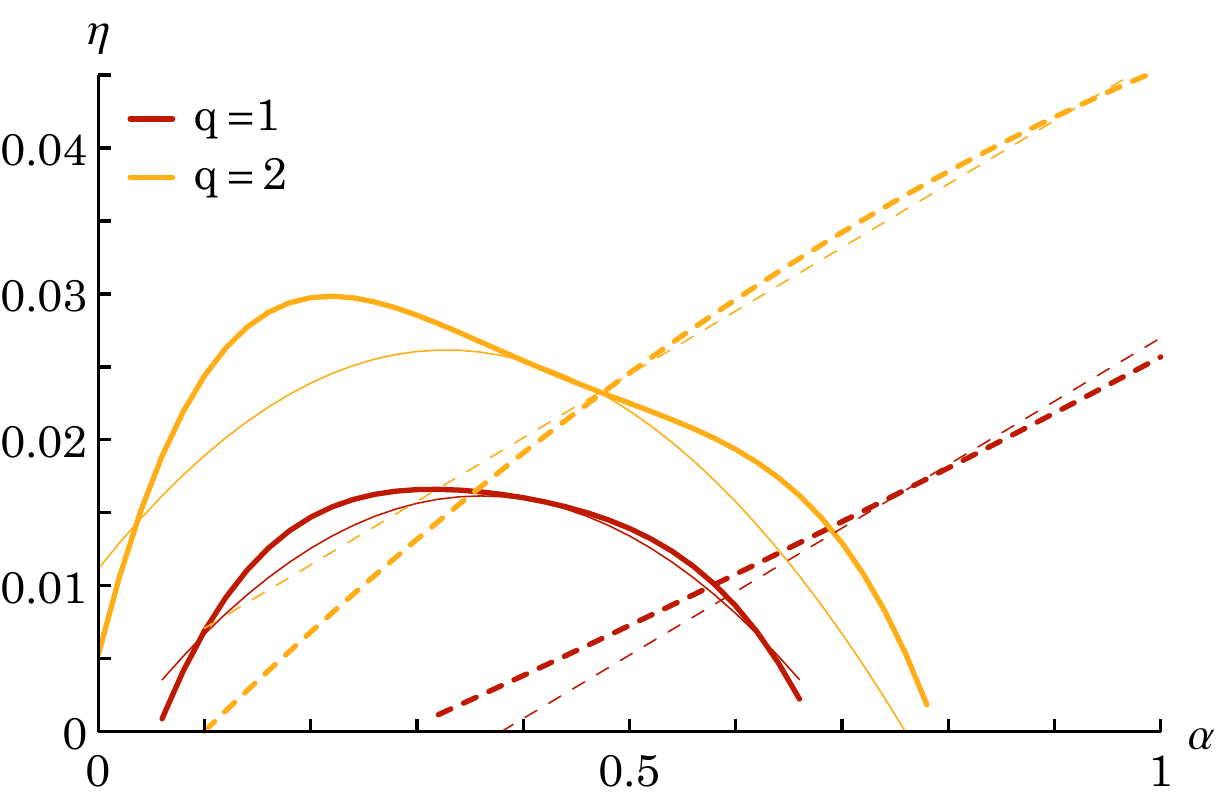}
\includegraphics[width=0.45\textwidth]{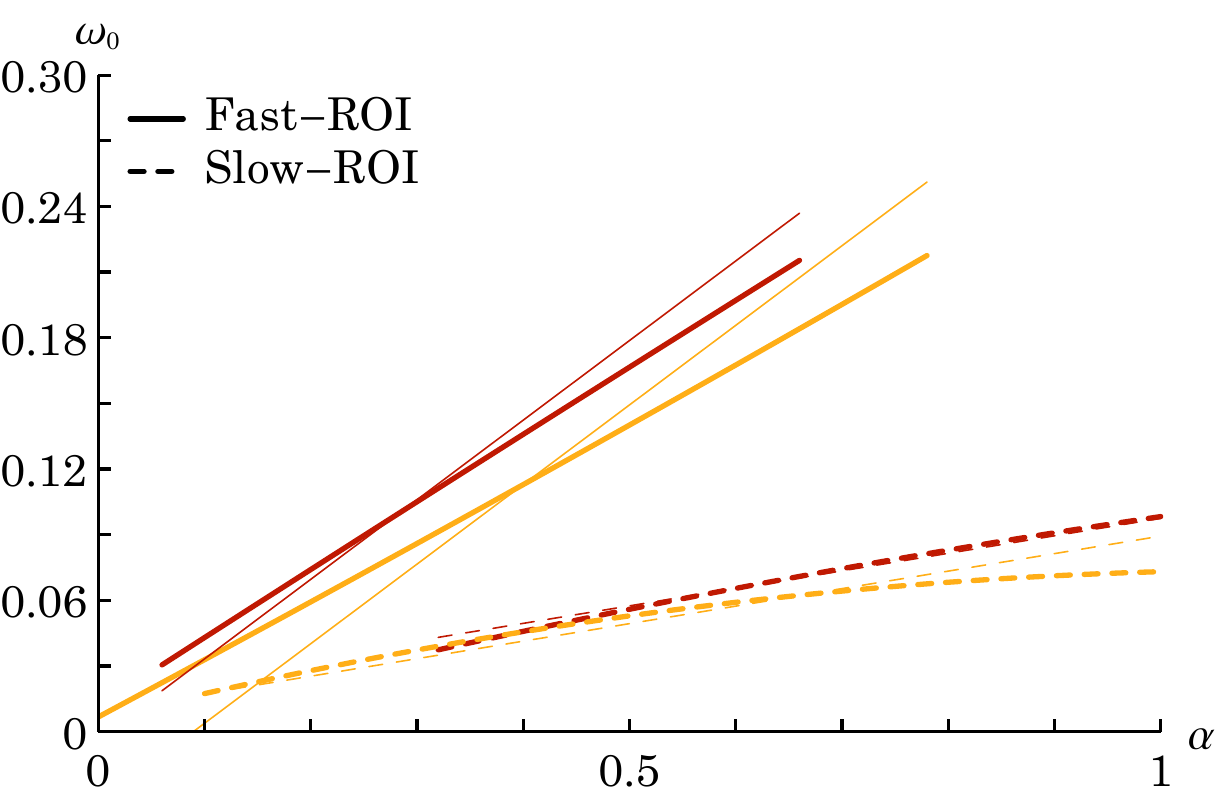}
\caption{Illustration of the growth rate (top panel) and oscillation frequency (bottom panel) as a function of $\alpha$ for fixed values of $q$, in the ${ q \!>\! 0 }$ region.
The thin lines represent the global polynomial fits from equation~\eqref{fitposq}.}
\label{fig:etaofalphaq>0}
\end{figure}

\begin{figure*}
\centering\includegraphics[width=1 \textwidth]{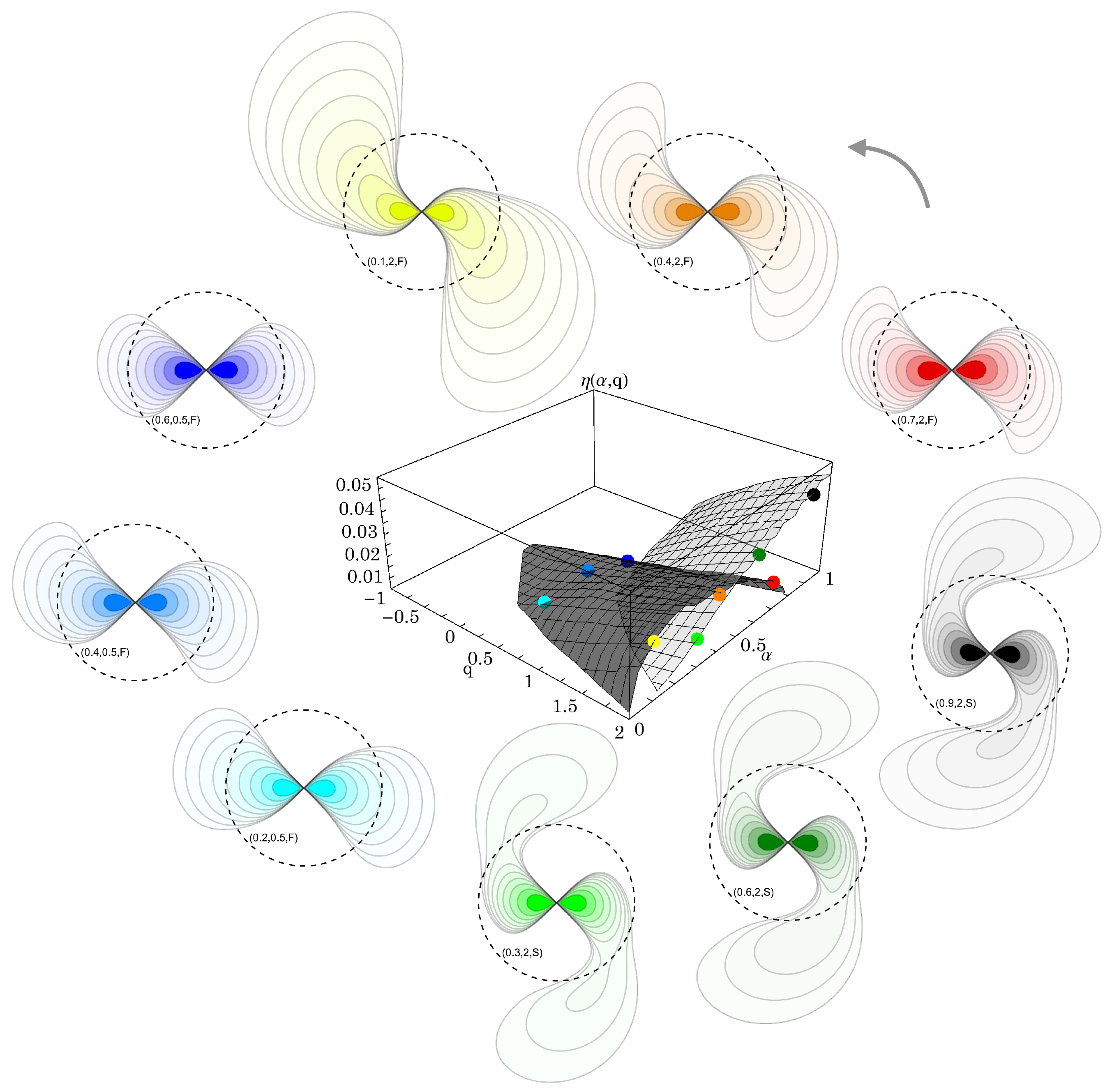}
\caption{Illustration of various modes' shapes as one explores
both surfaces of instability in the ${q \!>\! 0}$ region.
For each coloured point, the mode's density in the equatorial plane is represented.
Two lines of points were chosen with constant $q$ and varying $\alpha$: ${q\!=\! 0.5}$,
and ${ q \!=\! 2 }$.
The grey arrow indicates the direction of rotation of both the global rotation and the instability patterns.
On each mode, the value of ${(\alpha, q)}$ is written inside the circle, as well as a letter indicating whether the mode is on the Fast-ROI (F) or Slow-ROI (S) surface.
In contrast to Figure~\ref{fig:SliceSurfaceq<0}, the shapes of the modes of the ``Fast'' and ``Slow'' families
are strikingly different.
}
\label{fig:SliceSurfaceq>0}
\end{figure*}

In Figure~\ref{fig:etaofalphaq>0}, we illustrate cuts of constant $q$ in the ${\eta(\alpha, q)}$
and ${\omega(\alpha,q)}$ surfaces in the ${ q \!>\! 0 }$ region.
Similarly, in Figure~\ref{fig:SliceSurfaceq>0},
we explore the ${ \eta (\alpha , q) }$ surface,
and represent the shapes of the associated unstable modes.
The shapes of the modes suggest
in this regime the existence of two types of instability,
depending on the region considered.

The first type of instability dominates for slow-rotating systems (${ \alpha \!\lesssim\! \tfrac{1}{2}}$).
It is represented with the solid lines in Figure~\ref{fig:etaofalphaq>0},
and the dark grey surface in Figure~\ref{fig:SliceSurfaceq>0}.
It can be identified as the continuation in the 
rotating regime of the standard radial orbit instability \cite[ROI,][]{Polyachenko1981}.
This type will be referred to hereafter as ``Fast-ROI'', since the oscillation frequency of the instability is large.
For that particular instability, the modes evolve from a bar-like shape around ${ q \!=\! 0 }$ to a loosely-wound
trailing spiral around ${ q \!=\! 2}$, as shown in Figure~\ref{fig:SliceSurfaceq>0}.

The second type of instability dominates for fast-rotating systems (${ \alpha \!\gtrsim\! \tfrac{1}{2}}$).
It is represented with the dashed lines in Figure~\ref{fig:etaofalphaq>0},
and the light grey surface in Figure~\ref{fig:SliceSurfaceq>0}.
We note that the oscillation frequencies of these modes are small,
and we therefore refer to this instability as ``Slow-ROI''.
As illustrated in Figure~\ref{fig:SliceSurfaceq>0},
for that second instability regime, the modes present a barred shape,
with a more wound trailing spiral departing from the bar,
that qualitatively differs from the shape of the Fast-ROI modes.

Both instability surfaces ${ (\alpha , q) \mapsto \omega_{0} , \eta }$
can be fitted globally by simple polynomial functions as
\begin{align}
& \omega_{0} =  \tfrac{4 \alpha}{11}  \!-\! \tfrac{q}{34} \!+\! \tfrac{1}{38} ,\,\,\,
&& \eta =
   \tfrac{q}{76} \! +\! \tfrac{11 \alpha}{100}  \!-\! \tfrac{ q \alpha}{109} \!-\! \tfrac{7  \alpha^{2}}{50} \!-\! \tfrac{1}{66} ,
\nonumber
\\
& \omega_{0} = \tfrac{2 \alpha}{25}  \!-\! \tfrac{q}{123} \!+\! \tfrac{1}{39}  ,\,\,\,\,
 && \eta =
 \, \tfrac{q}{52} \!+\! \tfrac{\alpha}{23} \!-\! \tfrac{1}{28} ,
\label{fitposq}
\end{align}
for the Fast- and the Slow-ROI respectively.

\begin{figure}
\centering\includegraphics[width=0.4\textwidth]{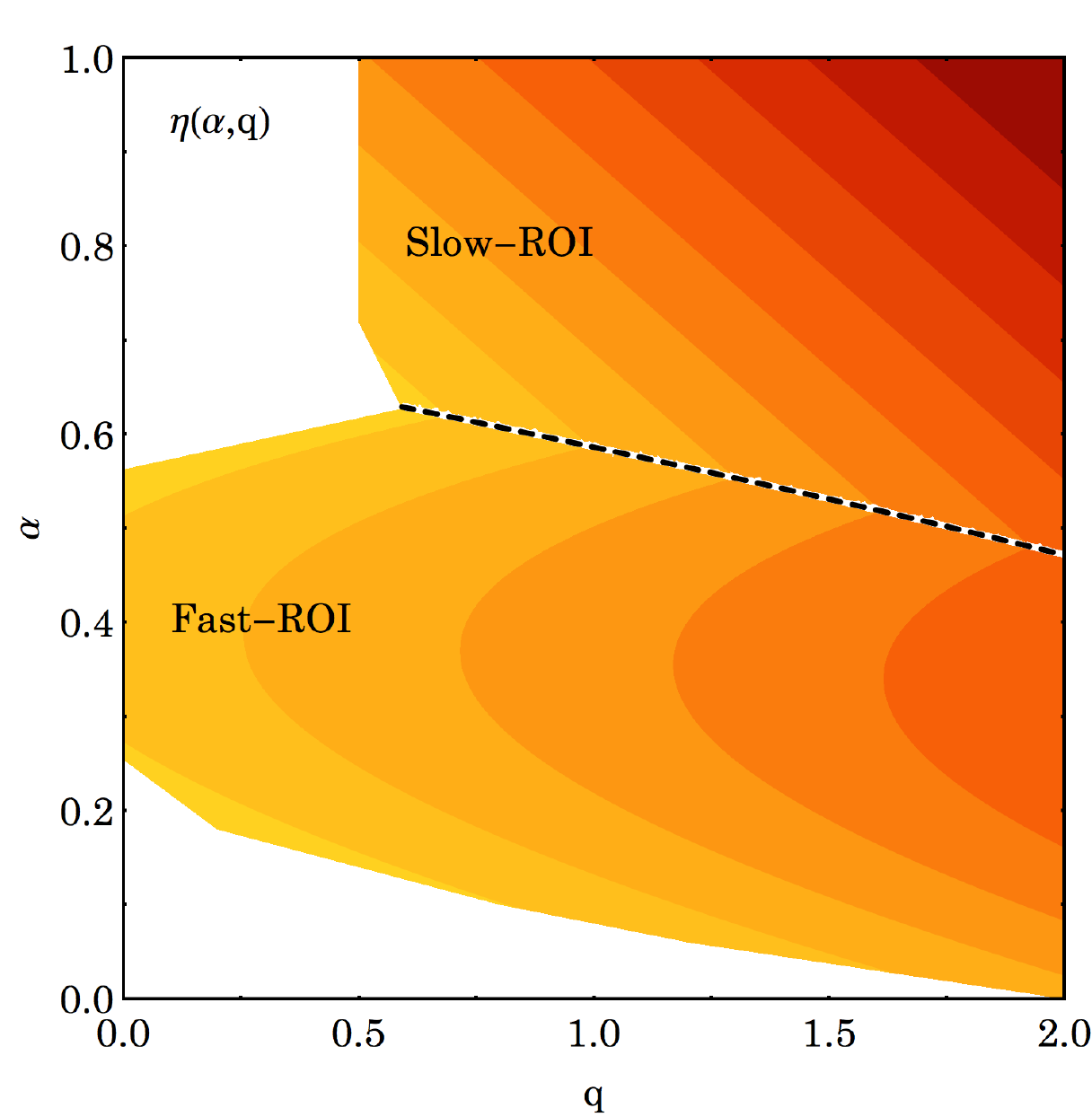}
\caption{Isocontours of the growth rate in the ${(\alpha, q)}$-space for ${ q \!>\! 0 }$,
as given by the fits from equation~\eqref{fitposq}.
Contours are spaced linearly between the maximum growth rate and zero.
The dashed line highlights the transition
between the Fast-ROI and Slow-ROI instabilities,
as already visible in Figure~\ref{fig:SliceSurfaceq>0}.}
\label{fig:isocontoursq>0}
\end{figure}

In Figure~\ref{fig:isocontoursq>0}, we finally illustrate isocontours of the growth rates of the dominant instabilities in the ${(\alpha, q>0)}$-space.
This figure clearly highlights the transition between the
two regions of instability, Fast-ROI vs. Slow-ROI.

\section{Discussion}
\label{sec:Discussion}

The previous sections identified various islands of instability
in the two parameters ${ (\alpha , q) }$ for the model from Section~\ref{sec:DF}.
Such a charting is costly for rotating spheres,
because the broken symmetry makes the matrix problem larger,
as azimuthal harmonics couple via gravity (see Appendix~\ref{appendix:MatrixMethod}).
Moreover, the previous charting is clearly particular to our choice of potential,
as well as our parametrisation of the \DF\ and the angular momentum distribution.
However, it nonetheless displays some diversity, which hopefully should be of general character.
Let us therefore remap the stability chart in terms of physically-motivated properties of rotating spheres,
before commenting on the origin and the topography of the instability islands.

\subsection{Remapping the instability chart}
\label{sec:remapping}

\begin{figure}
\includegraphics[width=0.45 \textwidth]{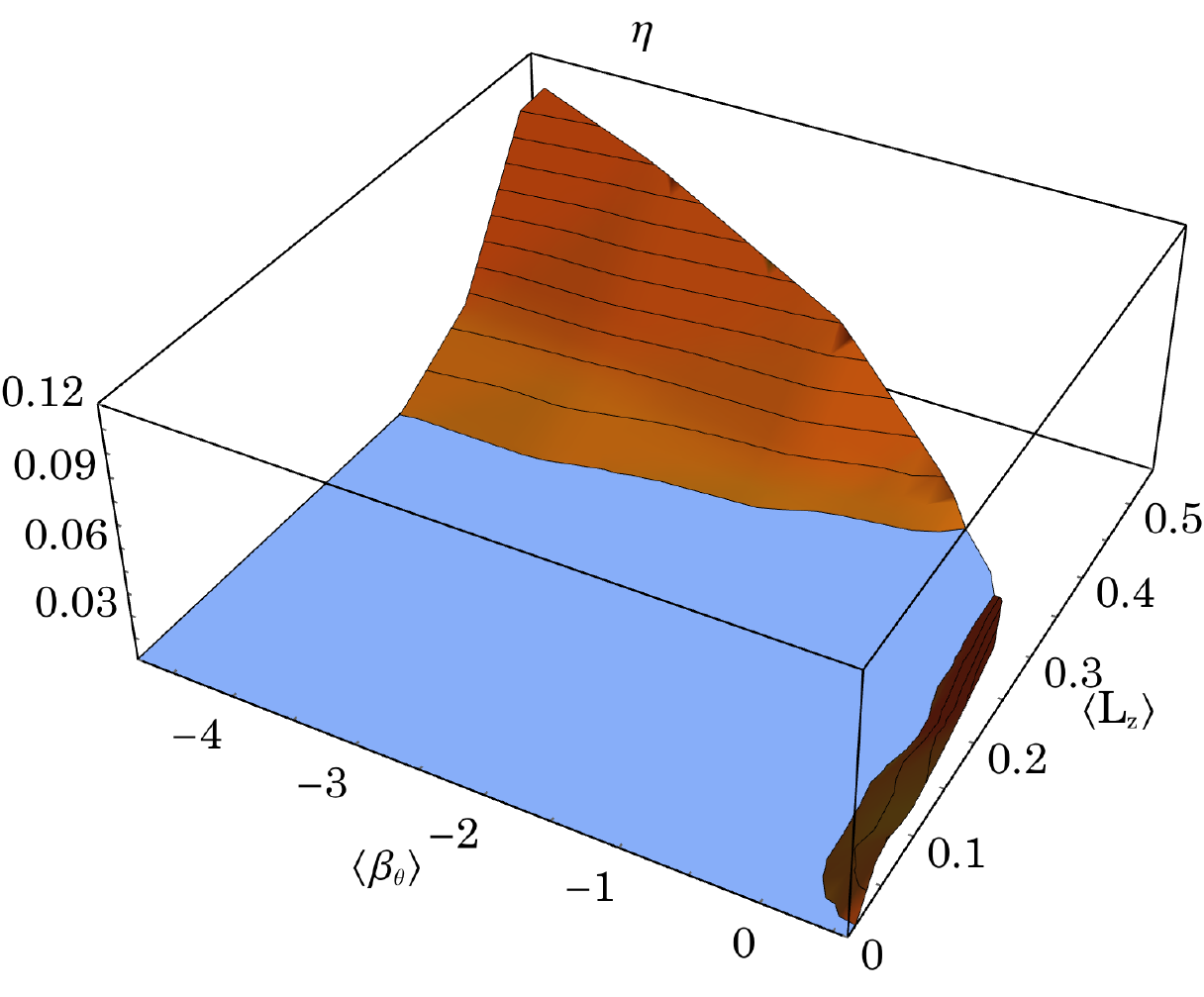}
\includegraphics[width=0.45 \textwidth]{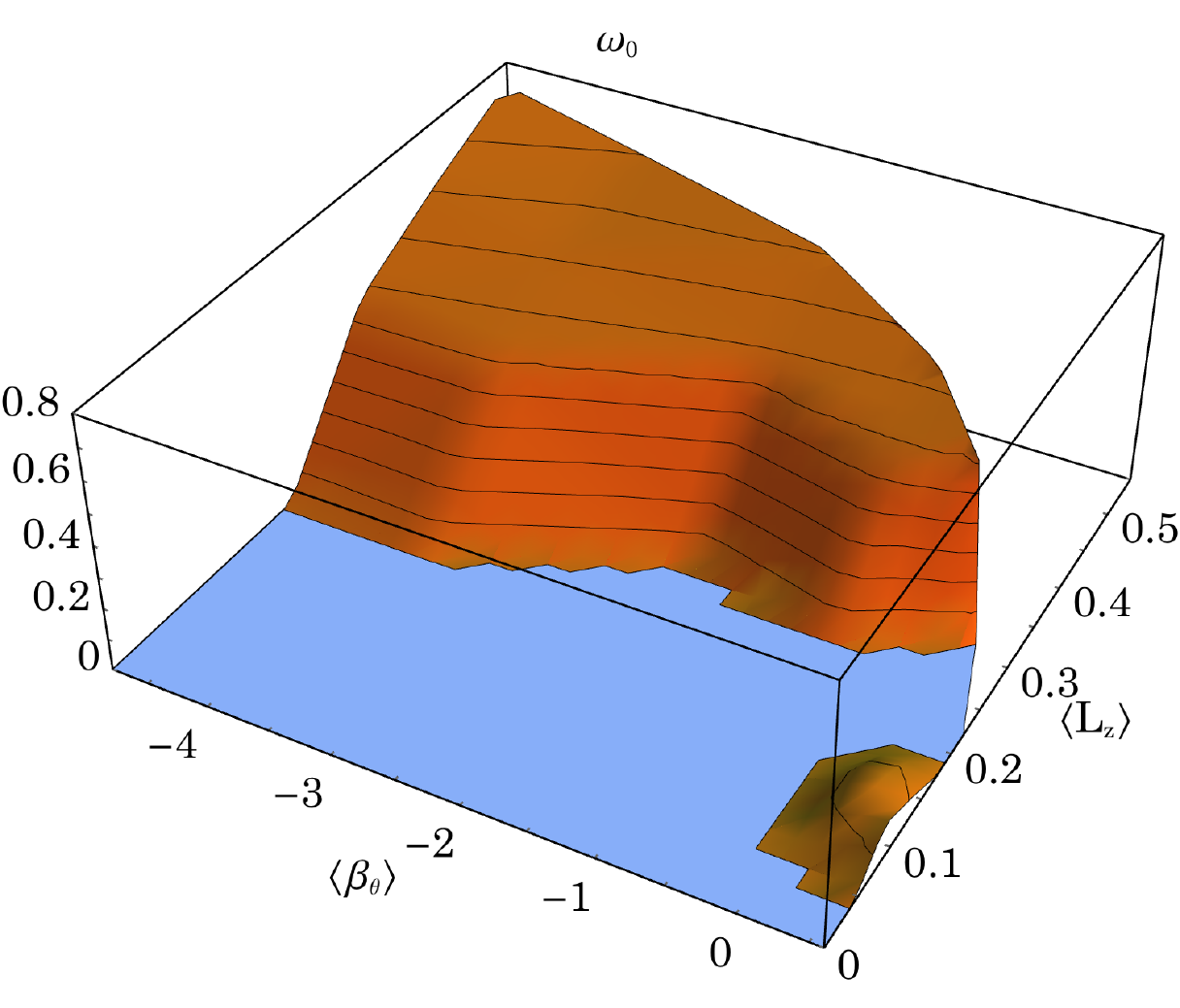}
\caption{Illustration of the surfaces of instability in the ${ (\langle \beta_{\theta} \rangle, \langle L_z \rangle) }$-space.
This figure is to be compared to Figure~\ref{fig:4panelsSurfaces}.
}
\label{fig:stability-in-beta-Lz}
\end{figure}

In order to rephrase our results in a more general framework,
we shall define macroscopic kinematic parameters of the sphere 
via the mass-weighted radial average of $\beta_\theta$ and $L_z$ respectively,
using equations~\eqref{eq:defbeta} and~\eqref{solid_DF}.
They read
\begin{align}
\langle \beta_\theta \rangle & \, \equiv \frac{1}{M_{\rm tot}} \!\! \int \!\! \rd r \,  4\pi r^2 \rho(r) \beta_\theta(r)
 = \tfrac{3}{10}\,q \, ,
\label{eq:defavgbeta}
 \\
\langle L_{z}\rangle & \, = \alpha \sqrt{G M_{\rm tot} R_s} {l}(q),\,
\mbox{with} \,\,\, l (q) \simeq \tfrac{3}{5} - \tfrac{10}{9} \left(\tfrac{69}{14} - q\right)^{-1} ,
\nonumber
 \end{align}
 where ${ l(q) }$ is a good fit decreasing monotonically from ${0.6}$ to ${0.22}$ as $q$ increases.
From equation~\eqref{eq:defavgbeta}, we note that to zeroth order, $q$ and $\alpha$ are direct proxy
for ${ \langle \beta_\theta \rangle }$ and ${ \langle L_z \rangle }$.
Using this reparametrisation,
we present in Figure~\ref{fig:stability-in-beta-Lz} a remapping of the stability manifolds
in the ${ ( \langle \beta_\theta \rangle , \langle L_{z} \rangle ) }$-space.

As expected the qualitative differences
between Figures~\ref{fig:4panelsSurfaces}
and~\ref{fig:stability-in-beta-Lz} are minor.
The following simple fits represent the three manifolds 
\begin{align}
& \omega_{0}= \tfrac{10 L}{7}  \!-\! \tfrac{\beta}{43}  \!-\! \tfrac{1}{7} ,
&&\eta =  \tfrac{117 L^{2}}{55}  \!-\! \tfrac{\beta^{2}}{1623}  \!-\! \tfrac{ L \beta }{26}   \!-\! \tfrac{141 L}{89}  \!+\! \tfrac{\beta}{107}
\!+\! \tfrac{14}{47},\nonumber
\\
& \omega_{0} = L ,
&& \eta = \tfrac{11 L}{21} \!-\! \tfrac{63 L^{2}}{38} \!+\! \tfrac{2 \beta}{17} \!-\! \tfrac{18 L \beta}{37} \!-\! \tfrac{\beta^{2}}{20} \!-\! \tfrac{1}{28},
\nonumber
\\
& \omega_{0} = \tfrac{L}{3} ,
&& \eta  =
 \,  \tfrac{L}{7} \!+\! \tfrac{\beta}{10} \!-\! \tfrac{1}{21},
\label{fits_betaLz}
\end{align}
respectively for negative $q$, Fast-ROI, Slow-ROI,
using the shortening notations ${ \beta\!=\! \langle \beta_{\theta} \rangle }$
and ${ L \!=\! \langle L_{z} \rangle }$.
From equation~\eqref{fits_betaLz}, the Ostriker-Peebles criterion~\citep{OstrikerPeebles1973} can be generalised 
by requesting that $\eta$ is above a given threshold $\eta_{\mathrm{crit}}$.
For instance, for the Slow-ROI instability, it reads
${ \langle L_{z} \rangle/7 + \langle \beta_{\theta} \rangle/10\ge \eta_{\mathrm{crit}}  + \tfrac{1}{21} }$.

\subsection{The role of the DF's discontinuity}
\label{sec:discontinuity}

It is generally now accepted that \DFs\ produced by violent or secular relaxation
need not be very smooth in action space~\citep[e.g.][]{Fouvry2015}. Yet, typically, these discontinuities are 
not of the form determined by the application of the Lynden-Bell trick from equation~\eqref{solid_DF} which may  partially impact 
artificially the distribution of orbits near ${ L_z = 0 }$,
as can be seen in Figure~\ref{fig:PDFvrvphi}.
This may prevent other 
instabilities from playing a role within the top left part of the chart from Figure~\ref{fig:4panelsSurfaces},
though it should be noted that clusters with ${ q \!\sim\! 0 }$
and ${ \alpha \!\sim\! 1 }$ are not  unstable, at least according to the criterion adopted here.

In order to check if the  specific discontinuity in equation~\eqref{solid_DF} is  responsible for the existence of the modes, $Q$ was set to zero in equation~\eqref{expression_M1}, effectively eliminating the contribution of the discontinuity of $\Sign (L_z) $
in equation~\eqref{solid_DF}. All three island are still present, though the growth rates are lower.
As  a further check, a distribution whose odd part in $L_z$ scales like ${ L_z/L }$  was also investigated.
Once again, qualitatively, the three instability islands remain.

\subsection{Origin of the instability islands}
\label{sec:origin}

It is beyond the scope of this paper to characterise in detail the origin
of the various manifolds. Let us nonetheless speculate on the nature
of the three types of instability.

From Figure~\ref{fig:KinematicMoments}, the instability island for ${ q \!<\! 0 }$
corresponds to spheres where the centrifugal support is significant
and where the azimuthal pressure is low.
Such systems may redistribute their orbits via an instability,
in order to increase their entropy towards ``hotter'' configurations. 
More generally, even for $q>0$,  mean rotation provides a reservoir of free energy 
for the sphere. 
Beyond this entropic argument, the origin of the  $q<0$ manifold is not understood in detail at this stage.

As for the $q>0$ manifolds, the matrix method and angle-action variables
allow us to emphasise the role played by orbits around 
specific resonances (ILR, COR and OLR for bi-symmetric instabilities)
in driving the instability. 
It provides a mathematical framework to study the tumbling instability scenario
brought forward by~\cite{DLB79} for discs,
but  which also applies to spheres dominated by quasi-radial orbits.
Under the assumption of the toy model presented in~\cite{Pichon:1993tx}, 
which only focuses on the ILR, and assumes all orbits are identical with positive effective 
orbital moment of inertia\footnote{We note that the role played by the moment of inertia is implicit in 
equation~\eqref{Fourier_M} via the integral over $\bJ$ which can be 
rewritten as an integral over, say, $\Omega_{\rm ILR}$ via the corresponding moment of inertia.}
\citep{Earn1996}, the instability criterion becomes 
the azimuthal analogue of Jeans's~\citep{Penrose:1960ip}.
The distribution of tumbling orbits has to be 
narrow enough around a given maximum to allow their relative torque to overcome
 their spread in azimuthal tumbling rates (azimuthal pressure),
 which otherwise phase mixes any disturbance.
The standard (non-rotating) ROI seems in most cases
to qualitatively correspond to such a scenario~\citep{Polyachenko:2015ii},
so long as the adiabaticity of the fast action ${ J_{\rf} = J_r \!+\! \tfrac{1}{2} L }$
allows us to pin down the ILR as the unique relevant
resonance~\citep[but see][for a discussion of counter examples with fast growing modes]{Polyachenko:2015ii}.

\begin{figure}
\includegraphics[width=0.4 \textwidth]{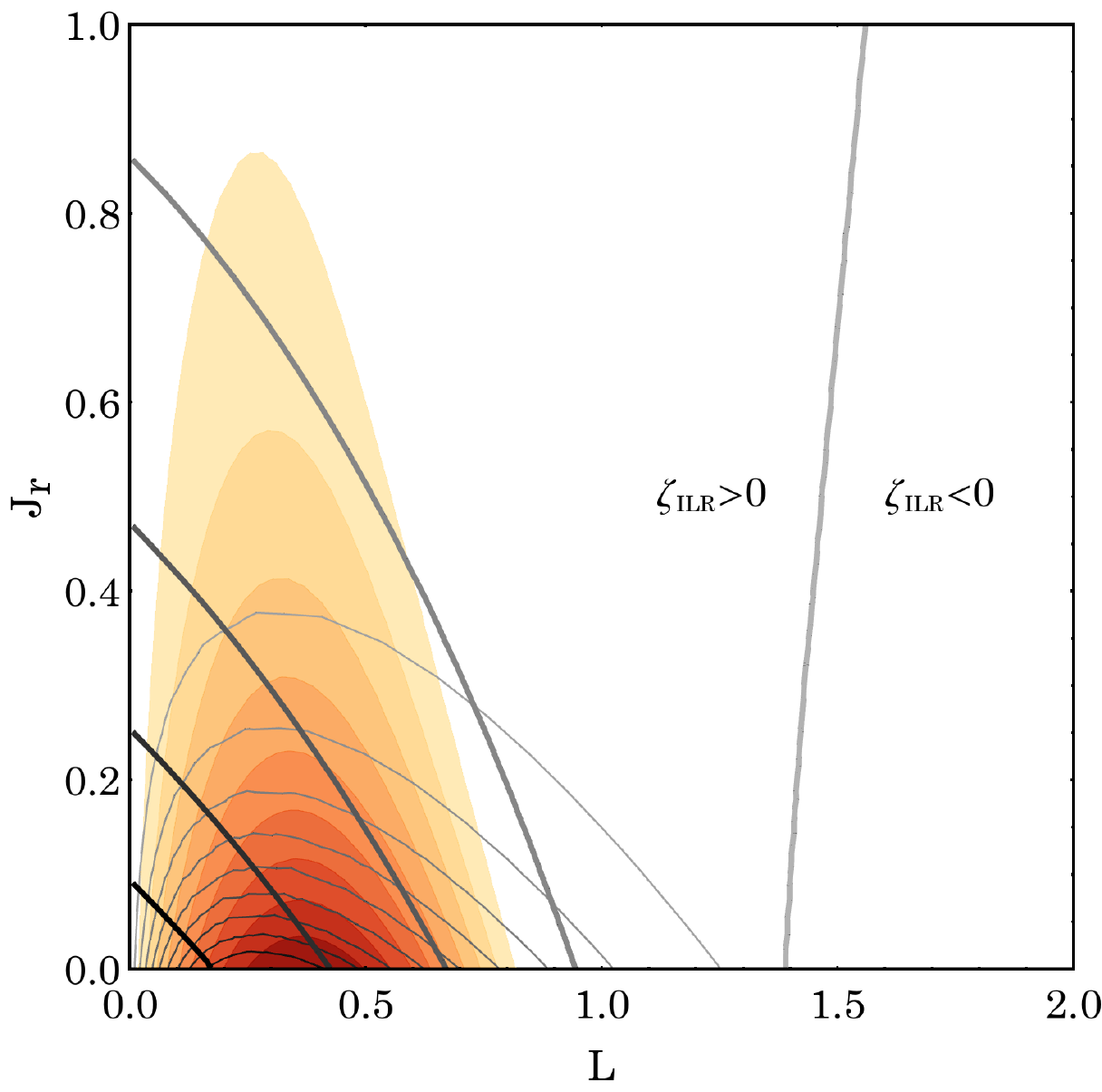}
\caption{Superimposed isocontours of phase space DFs (equation~\eqref{def_F0}) and cooperation parameter ${\zeta_{\text{ILR}}}$ in ${(J_r,L)}$ space. \textit{Filled contours:} phase space DF of the ${q=2}$ system. The coloured regions are linearly spaced from 10\% (yellow) to 90\% (red) of the maximum value. \textit{Thin contours:} phase space DF of the ${q=0}$ system. The contours are linearly spaced from 10\% (grey) to 90\% (black) of the maximum value. \textit{Thick contours:} Cooperative parameter ${\zeta_{\text{ILR}}}$. The contours are linearly spaced from 0\% (grey) to 80\% (black) of the maximum value. The large amount of radial cooperative orbits seems responsible for the instability of the ${ q = 2 }$ systems.
}
\label{fig:DFzeta}
\end{figure}

\begin{figure}
\includegraphics[width=0.45 \textwidth]{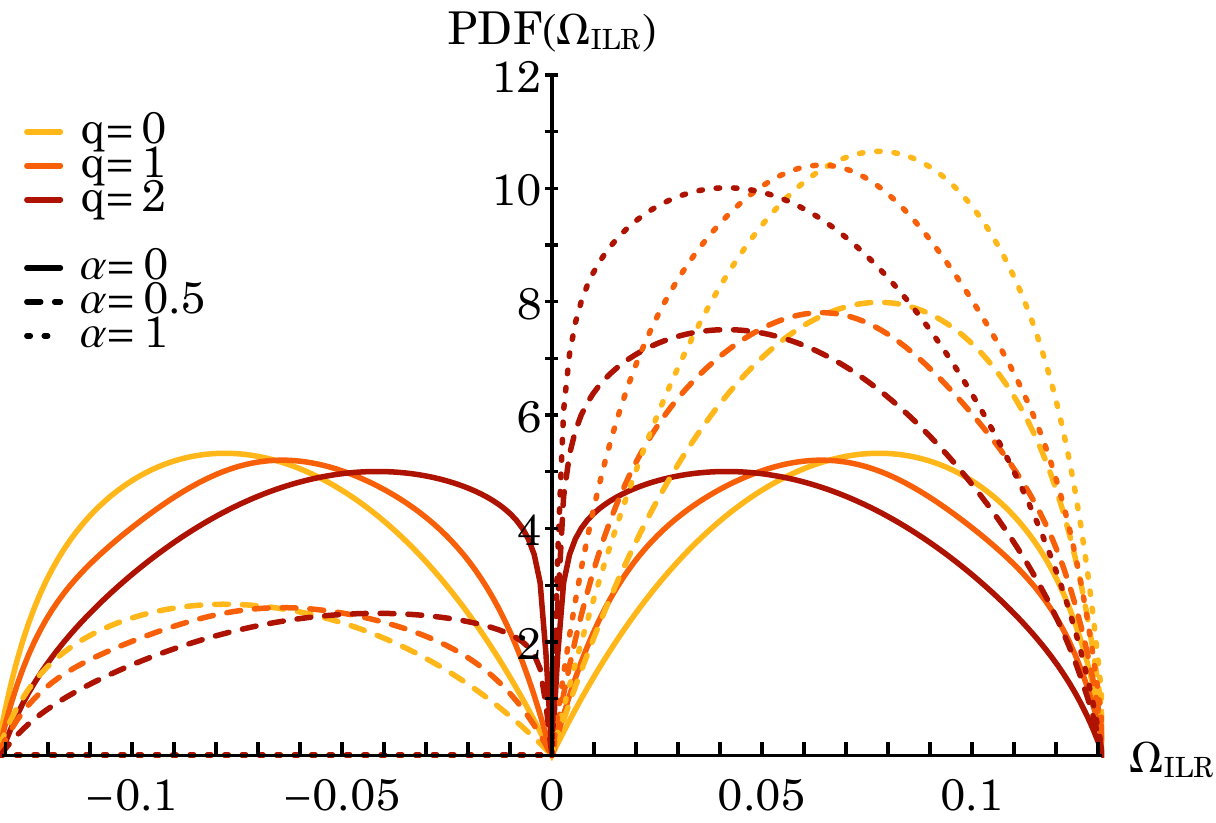}
\caption{PDF of the tumbling frequency ${\Omega_{\mathrm{ILR}}}$ for various values of ${q\geq 0}$ and ${\alpha}$. When $q$ increases, the PDF symmetrically concentrates around ${\Omega_{\mathrm{ILR}}=0}$. When $\alpha$ increases, some of the left hill is transferred into the right hill, concentrating the PDF around positive values of ${\Omega_{\mathrm{ILR}}}$.
As $\alpha$ grows, the effective frequency PDF can become narrow enough to drive a tumbling instability.
}
\label{fig:DFomegaILR}
\end{figure}

Indeed, Figure~\ref{fig:DFzeta} shows that most orbits have ${\zeta_{\mathrm{ILR}} \!>\! 0}$,
with ${\zeta_{\mathrm{ILR}} \!=\! ( {\partial \Omega_{\mathrm{ILR}}}/{\partial L} )_{\!J_\rf}}$
the cooperation parameter at the ILR,
and the tumbling frequency ${ \Omega_{\mathrm{ILR}} \!=\! \Omega_{\phi} \!-\! \tfrac{1}{2} \Omega_{r} }$.
Hence, 
the system is likely to support a tumbling instability,
provided the resonance frequency distribution is sufficiently narrow~\citep{DLB79}.
This is illustrated in Figure~\ref{fig:DFomegaILR},
where the PDF of the tumbling frequency $\Omega_{\mathrm{ILR}}$
is represented for various values of ${ q \geq 0 }$ and $\alpha$.
As highlighted in these figures, the effect of increasing ${q}$ is both to concentrate the orbits in regions of high cooperation parameter and to lower the spread in azimuthal precession rates. Rotation also narrows the spread in azimuthal precession rates, which favours the development of instabilities.
In order to take into account the direction of rotation of the orbits and make this reasoning slightly more quantitative, we classified orbits by the sign of ${L_z}$\footnote{In such a system, all orbits precess in the same direction as they revolve.}, mapping them into DF$({\Omega_{\mathrm{ILR}} \Sign(L_z)})$. Without rotation (${\alpha=0}$), the DF is symmetric, since the phase space DF is independent of ${L_z}$. In this case, increasing $q$ concentrates the DF in the regions of low (absolute) frequency, consequently decreasing the overall spread. When ${\alpha>0}$, the left bump of ${\mathrm{PDF}(\Omega_{\mathrm{ILR}})}$ will be multiplied by ${ 1 \!-\! \alpha}$ and the right bump by ${1 \!+\!\alpha}$. This  rotation-induced tilting decreases the spread in ${\Omega_{\mathrm{ILR}}}$ and is likely to be responsible for the seemingly linear growth of the instability pattern speed with $\alpha$ in the Fast-ROI regime,
as plotted in Fig.~\ref{fig:etaofalphaq>0}.

While increasing  $q$  increases the amount of radial and cooperating orbits,  decreases the spread in precession rates, and makes the system more unstable, the evolution of the growth rate with ${\alpha}$ is more complicated. 
For ${ \alpha \!>\! 0 }$, rotation provides a reservoir of rotational kinetic energy
to feed possible unstable modes~\citep{BinneyTremaine2008}.
A naive expectation would be that, as one introduces positive $\alpha$
the instability gets stronger and rotates more (while preserving adiabaticity),
which it does on the Slow-ROI manifold.
 However, for ${\alpha > 0.2}$, the pattern is spinning too fast to carry a large amount of orbits, which in turn lowers the destabilising power of this process. When the pattern speed reaches the edge of the ${\Omega_{\mathrm{ILR}}}$ distribution (${\alpha \simeq 0.8}$ in the ${q = 2}$ case, where ${ {\omega_{0}}/{m} \simeq 0.11}$), no more orbits can contribute to the instability and it vanishes.

Orbital torquing between the COR and OLR resonances,
where orbits typically have negative 
moment of inertia, may also make the sphere unstable,
if the spread in azimuthal tumbling rates is centred on a given narrow
minimum~\citep[the galactic analogue of the two-stream instability in plasma physics,][]{Penrose:1960ip}.
In practice, the exact condition from equation~\eqref{eq:det0}
is more complex than Lynden-Bell's criterion since in equation~\eqref{Fourier_M},
${ \wbM (\omega) }$ involves all couplings simultaneously.

\section{Conclusion}
\label{sec:Conclusion}

We investigated the stability of a two-parameter sequence of modified Plummer spheres
in which the velocity anisotropy (parameter $q$) and the net amount of total angular momentum (parameter $\alpha$) can be varied independently.
For the first time, we implemented the matrix method to compute the growth rate,
oscillation frequency and shape of the unstable modes of a rotating stellar sphere,
and compared those favourably to $N$-body simulations.
We then charted the marginal stability boundary in the parameter
space of velocity anisotropy and total angular momentum,
using not only ${ \alpha , q }$, but also
${ \langle \beta_{\theta} \rangle , \langle L_{z} \rangle }$,
which are intended to help in applications to wider classes of model
than the modified Plummer sphere.

When rotation was introduced, two apparently separate sets of modes were identified,
corresponding to models with predominantly radial and tangential velocity anisotropy respectively.
For radially dominated spheres, the growing modes themselves separate into two families,
which appear as two intersecting surfaces in a plot of growth rate against the two parameters,
corresponding to ``Fast'' and ``Slow'' modes,
depending on the net rotation rate of the sphere.
The sequence of unstable modes that dominates much of the parameter space
of tangentially-biased models relies essentially on the rotational kinetic energy of the models, and may be novel.
On the other hand the ${ q \!>\! 0 }$ modes appear to extend the known ROI to rotating configurations,
with a slowly rotating mode dominating at higher system rotation (${ \alpha \!\sim\! 1 }$),
perhaps surprisingly.
We conjecture that the ${ q \!<\! 0 }$ modes (tangential anisotropy) correspond to the conversion of rotational support
into heat, which is made possible by the low azimuthal and radial pressure.
Equation~\eqref{fits_betaLz} offers a first generalisation of the classical results of~\cite{OstrikerPeebles1973} to rotating spheres.

One possible limitation of the present mapping of instabilities is the procedure in  equation~\eqref{solid_DF}
for the introduction of rotation, as it produces a sharp discontinuity in orbital space at ${ L_z \!=\! 0 }$.
It would be of interest to study a more generic class of rotating distributions,
without such a discontinuity (see, for example, Section~\ref{sec:discontinuity}).
The lack of control of differential rotation at fixed mean rotation is also a limitation.
An extension of the present work to a more general class of models parametrising the shear and smoothing the discontinuity is  the topic of ongoing work, exploiting the flexibility of the $N$-body method~\citep{Breen+2019b}.

An additional dimension of this problem which deserves further attention concerns the detailed investigation of the role of corotation and other resonances, which, at this stage, has been only partly addressed in Section~\ref{sec:origin}. Indeed, the behaviour of differentially rotating polytropes (e.g., see, among others, \citealt{Pickett1996, New2001, Centrella2001, Shibata2002, Ou2006, Galeazzi2012}) and the suggested connection between the emergence of an unstable mode and the existence of the associated corotation point (see, e.g, \citealt{Watts2005, Passamonti2015,Yoshida2017}) may offer a meaningful physical intuition for the development of a comparative analysis of the stability of stellar and fluid dynamical systems.

In the present study the focus has been on the Plummer sphere, which is appropriate to describe stellar systems with a central core, but it does not capture properties of cuspy ellipticals.
In that respect, the current stability charting could be extended to, e.g.\@, rotating Hernquist models.
A subsequent major step in this line of investigation could be the exploration of the properties of non-spherical equilibria (e.g., for a study of selected axisymmetric models, see~\citealt{Varri2012}).
One difficulty in this context is the lack of angle-action variables,
but St\"{a}ckel families~\citep{Stackel1890}, with more general classes of \DF\ 
than the thin shell configurations investigated in~\cite{Robijn1995}, could be a possible line of attack.

In a similar vein, it would also be of interest to investigate the impact of rotation on secular evolution.
The effect of such an additional physical ingredient could be substantial, as rotation directly impacts the set of available frequencies, allowing resonant relaxation to reshuffle the system's orbital structure. The investigation presented by~\cite{Hamilton2018} could be naturally extended in that respect.
Certainly, $N$-body studies show that the effect on the time to core collapse can exceed 20\%~\citep{Breen+2019a}.

Finally, ongoing~\citep{GaiaDR2} and upcoming proper motion and spectroscopic surveys will soon provide observational constraints  on the level of rotation within the several nearby globular clusters of our Milky Way. In addition, ongoing (SAMI,~\citealt{sami2012}, MANGA,~\citealt{Manga2015}) and upcoming~\citep[HECTOR,][]{Hector2015} integral field spectroscopy surveys
will provide large statistical samples of the resolved kinematics of elliptical galaxies. It will, therefore, be of great interest to characterise the stability boundaries identified in the present study in terms of relevant kinematic observables, with an attention to possible degeneracies associated with projection effects. Such a development could provide a physically-based framework to interpret the rich kinematics of several classes of spheroidal stellar systems which will result from upcoming empirical investigations.

\subsection*{Acknowledgements}
This work was motivated in part by D. Lynden-Bell's interest in the stability of rotating spheres. 
We dedicate this paper to his memory. 
JBF acknowledges support as a Hubble Fellow from Program number HST-HF2-51374 was provided by NASA through
a grant from the Space Telescope Science Institute, which is operated by the
Association of Universities for Research in Astronomy, Incorporated,
under NASA contract NAS5-26555. This research is part of ANR grant Spin(e)
(ANR-13-BS05-0005, \href{http://cosmicorigin.org}{http://cosmicorigin.org})
and ERC grant 70193 (COSFORM).
PGB, DCH and ALV acknowledge support from the Leverhulme Trust
(Research Project Grant, RPG-2015-408), ALV also from
a Marie Sklodowska-Curie Fellowship (MSCA-IF-EF-RI NESSY 658088). This work was supported by the Programme National Cosmology et Galaxies (PNCG) of CNRS/INSU with INP and IN2P3, co-funded by CEA and CNES.
SR and ALV are grateful to the School of Mathematics and Institute for Astronomy in Edinburgh as well as the Institute d'Astrophysique de Paris for hospitality during mutual visits in Fall 2018.
This work has made use of the Horizon Cluster hosted by Institut d'Astrophysique de Paris.

\appendix

\section{The Matrix method}
\label{appendix:MatrixMethod}

The starting point of this Appendix is equation~\eqref{Fourier_M},
the generic expression of the response matrix of a spherically symmetric system.
It involves a biorthogonal basis of potentials and densities, ${ \psi^{(p)} (x) }$ and ${ \rho^{(p)} (x) }$, defined as
\begin{align}
& \, \psi^{(p)} (\bx) = \!\! \int \!\! \rd \bxp \, U (\bx , \bxp) \, \rho^{(p)} (\bxp) ,
\nonumber
\\
& \, \!\! \int \!\! \rd \bx \, \big[ \psi^{(p)} (\bx) \big]^{*} \, \rho^{(q)} (\bx) = - \delta_{p}^{q} ,
\label{def_basis}
\end{align}
with ${ U (\bx , \bxp) = - G / |\bx - \bxp| }$ the pairwise Newtonian interaction. For a spherically symmetric system, it is natural to separate the azimuthal dependence of the basis elements with spherical harmonics and write
\begin{align}
\psi^{(p)} (\bx) = \psi_{\ell m n} (r , \vtheta , \phi) = Y_{\ell}^{m} (\vtheta , \phi) \, U_{n}^{\ell} (r) ,
\nonumber
\\
\rho^{(p)} (\bx) = \rho_{\ell m n} (r , \vtheta , \phi) = Y_{\ell}^{m} (\vtheta , \phi) \, D_{n}^{\ell} (r) ,
\label{shape_basis}
\end{align}
where a given basis element is characterised by three indices,
${ \ell \geq 0 }$, ${ |m| \leq \ell }$, and ${ n \geq 0 }$,
and we follow the convention ${ \! \int \! \rd \vtheta \rd \phi \sin ( \vtheta) |Y_{\ell}^{m}|^{2} = 1 }$.
We also introduced the (real) radial functions ${ U_{n}^{\ell} (r) }$ and ${ D_{n}^{\ell} (r) }$,
and many examples can be found
in the literature~\citep{CluttonBrock1973,Weinberg1989,HernquistOstriker1992,RahmatiJalali2009}.
Finally, in equation~\eqref{Fourier_M}, for any resonance vector ${ \bn \in \mathbb{Z}^{3} }$,
we introduced the Fourier transform w.r.t.\ the angle $\bT$ as
\begin{equation}
f (\bT , \bJ ) = \sum_{\bn} f_{\bn} (\bJ ) \, \re^{\ri \bn \cdot \bT} , \; f_{\bn} (\bJ) = \!\! \int \!\! \frac{\rd \bT}{(2 \pi)^{3}} \, f (\bJ) \, \re^{- \ri \bn \cdot \bT} .
\label{def_Fourier_bT}
\end{equation}
Following~\cite{TremaineWeinberg1984}, the Fourier transforms of the basis elements
from equation~\eqref{shape_basis} read
\begin{equation}
\psi^{(p)}_{\bn} (\bJ) = \delta_{m^{p}}^{n_{3}} \, \ri^{m^{p} - n_{2}} \, Y_{\ell^{p}}^{n_{2}} (\pi / 2 , 0) \, R^{\ell^{p}}_{n_{2} m^{p}} (i) \, W^{\tbn}_{\ell^{p} n^{p}} (\tbJ) ,
\label{FT_basis}
\end{equation}
with the reduced resonance vector ${ \tbn \!=\! (n_{1} , n_{2}) }$.
Here, we introduced the rotation matrix ${R_{n m}^{\ell} (i) }$ as
\begin{align}
R_{n m}^{\ell} (i) = & \, \sum_{t} (-1)^{t} \, \frac{\sqrt{(\ell + n)! \, (\ell - n)! \, (\ell + m)! \, (\ell - m)!}}{(\ell - m - t)! \, (\ell + n - t)! \, t! \, (t + m - n)!} \,
\nonumber
\\
& \;\; \times \big[ \cos (i / 2) \big]^{2 \ell + n - m - 2 t} \, \big[ \sin (i / 2) \big]^{2 t + m - n} ,
\label{RotMat}
\end{align}
where the sum over $t$ is restricted to the values such that the arguments
of the factorials are positive, i.e. ${ t_{\rm min} \!\leq\! t \!\leq\! t_{\rm max} }$,
with ${ t_{\rm min} \!=\! \text{Max} [0 , n - m] }$ and
${ t_{\rm max} \!=\! \text{Min} [\ell - m , \ell + n] }$.
In equation~\eqref{FT_basis}, we also introduced the Fourier-transformed
``in-plane'' radial coefficients ${ W^{\tbn}_{\ell n} (\tbJ) }$ defined as
\begin{equation}
W^{\tbn}_{\ell n} (\tbJ) = \frac{1}{\pi} \!\! \int \!\! \rd \theta_{1} \, U_{n}^{\ell} (r (\theta_{1})) \, \cos [n_{1} \theta_{1} + n_{2} (\theta_{2} - \psi)] ,
\label{def_W}
\end{equation}
which are real for real radial basis functions.
This integral asks for the radial dependence of the angles $\theta_{1}$
and ${ (\theta_{2} - \psi ) }$ for a given action $\tbJ$.
They are given by
\begin{align}
\theta_{1} & \, = \Omega_{1} \!\! \int_{\mC} \!\! \rd r \, \frac{1}{\sqrt{2 (E - \psi (r)) - J_{2}^{2} / r^{2}}} ,
\nonumber
\\
\theta_{2} - \psi & \, = \!\! \int_{\mC} \!\! \rd r \, \frac{\Omega_{2} - J_{2} / r^{2}}{\sqrt{2 (E - \psi (r)) - J_{2}^{2} / r^{2}}} ,
\label{angles_radial}
\end{align}
where $\mC$ is the integration contour going from the pericentre $\rperi$ up to the current position ${ r = r (\theta_{1}) }$ along the radial oscillation. Computing the radial coefficients from equation~\eqref{def_W} is numerically challenging. In Appendix~\ref{appendix:FastWMat}, we present an efficient implementation of this computation, that improves upon~\cite{Fouvry2015}. Finally, we note that these basis elements satisfy the symmetry relation ${ \psi^{[\ell^{p} , - m^{p} , n^{p}]}_{\bn} = (-1)^{m^{p}} (\psi^{[\ell^{p} , m^{p} , n^{p}]}_{- \bn})^{*} }$.

\begin{figure}
\includegraphics[width=0.45\textwidth]{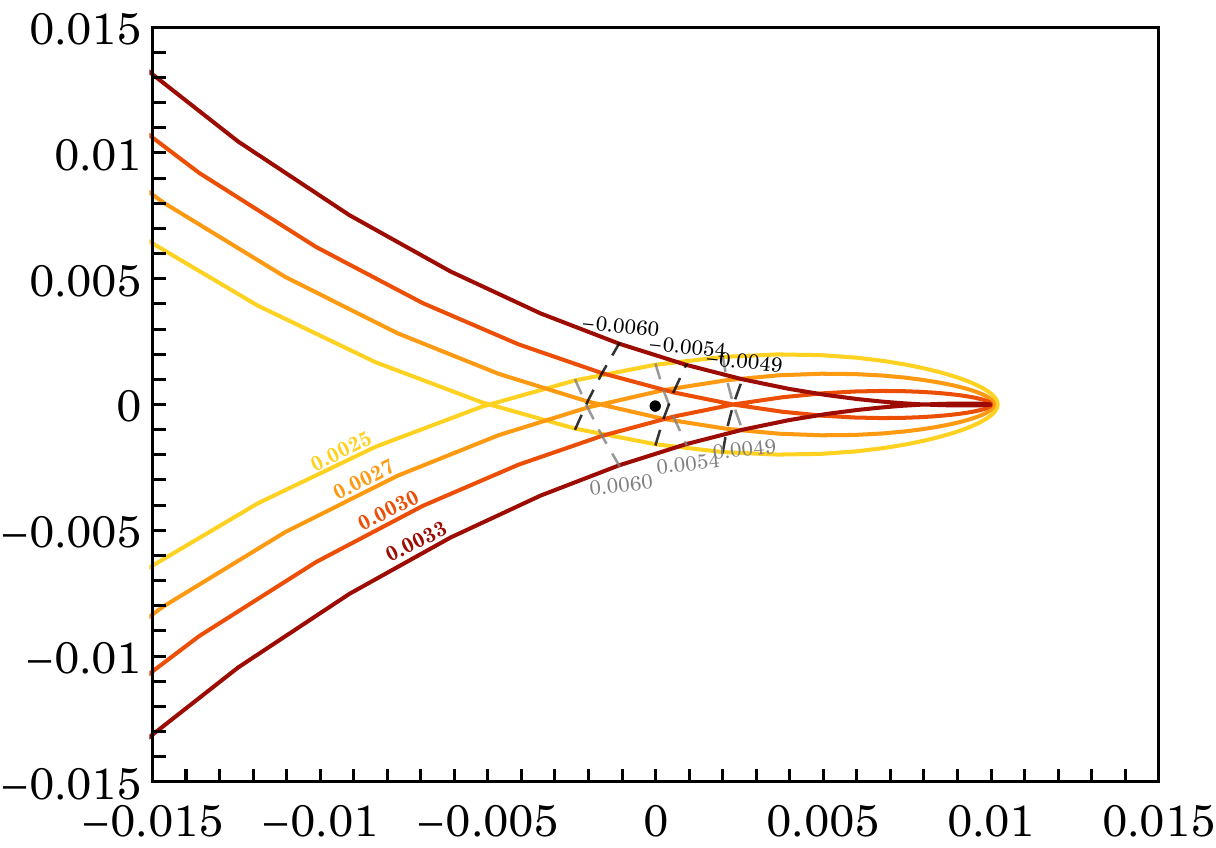}
\includegraphics[width=0.45\textwidth]{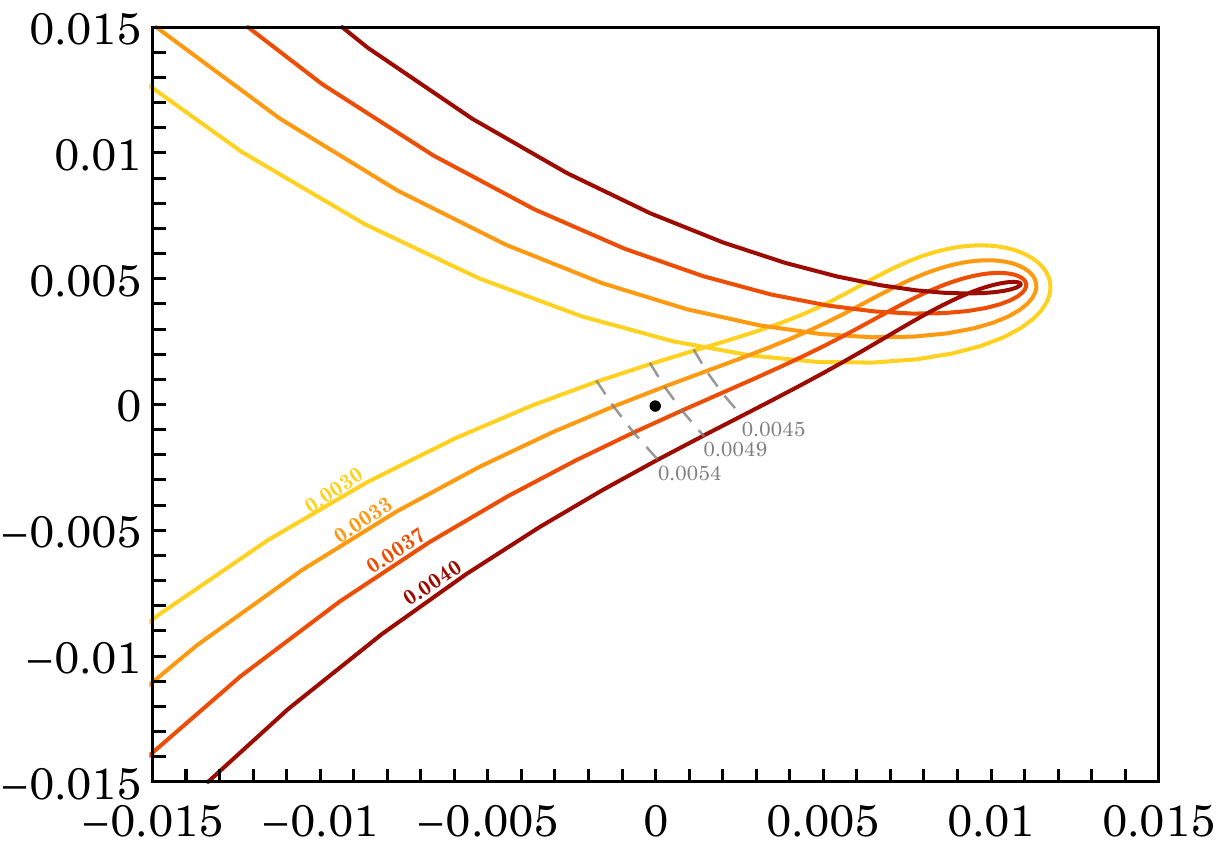}
\caption{Zoom of the Nyquist contours around the instability for ${ q \!=\! 2 }$,
${ \alpha \!=\! 0 }$ (top panel) and ${ \alpha \!=\! 4 \!\times\! 10^{-4} }$ (bottom panel). 
On the top panel, we note that at ${ \eta \!=\! 0.030 }$ a loop was created (red curve).
When $\eta$ reaches ${0.0027}$, the corresponding neck of the loop 
has crossed the origin.
Even without rotation, this simultaneous crossing of the two branches reflect the existence
of two growing modes with opposite (non-zero) oscillation frequencies.
On the bottom panel, rotation breaks the horizontal symmetry of the diagram.}
\label{fig:nyquistq2}
\end{figure}

Let us now follow equation~\eqref{Fourier_M} to compute the response matrix associated with the \DF\ from~\eqref{solid_DF}.
Owing to its linear dependence w.r.t.\ $\alpha$,  the matrices ${ \wbM_{0} (\omega) }$ and ${ \wbM_{1} (\omega) }$ are defined as
\begin{align}
\wM_{pq}^{0} (\omega) & \!=\! (2 \pi)^{3} \sum_{\bn} \!\! \int \!\! \rd \bJ \frac{\bn \cdot \partial F_{0} / \partial \bJ}{\omega - \bn \cdot \bO (\bJ)} \big[ \psi^{(p)}_{\bn} (\bJ) \big]^{*} \psi_{\bn}^{(q)} (\bJ) ,
\label{def_M0_M1}
\\
\wM_{pq}^{1} (\omega) & \!=\! (2 \pi)^{3} \sum_{\bn} \!\! \int \!\! \rd \bJ \frac{\bn \cdot \partial ( F_{0} \Sign (L_{z})) / \partial \bJ}{\omega - \bn \cdot \bO (\bJ)} \big[ \psi^{(p)}_{\bn} (\bJ) \big]^{*} \psi_{\bn}^{(q)} (\bJ) .
\nonumber
\end{align}
These coefficients satisfy the symmetry relations
\begin{align}
\wM_{- m^{p} , - m^{p}} (\alpha , \omega_{0} + \ri \eta) & \, = \big[ \wM_{m^{p} , m^{p}} (\alpha , - \omega_{0} + \ri \eta) \big]^{*} ,
\nonumber
\\
\wM_{- m^{p} , - m^{p}} (\alpha , \omega) & \, = \wM_{m^{p} , m^{p}} (- \alpha , \omega) .
\label{symmetry_M}
\end{align}
In particular, for ${ \alpha = 0 }$, we recover the invariance of the Nyquist contours
w.r.t.\ complex conjugation. Indeed, an immediate consequence of equation~\eqref{symmetry_M} is
\begin{equation}
\wM_{m^{p} , m^{p}} (0 , \omega) \, = \big[ \wM_{m^{p} , m^{p}} (0 , - \omega_{0} + \ri \eta) \big]^{*} \!.
\end{equation}
When a small global rotation is introduced in an originally unstable system,
the Nyquist contours depart from this symmetric case,
consequently shifting the values of the growth rate and
oscillation frequency of instability.
This behaviour is represented in Figure~\ref{fig:nyquistq2},
in the mildly unstable ${ q \!=\! 2, \alpha \!=\! 0 }$ case.

Owing to the Kronecker symbol, $\delta_{m^{p}}^{n_{3}}$, in equation~\eqref{FT_basis},
the sums over $n_{3}$ in equation~\eqref{def_M0_M1} can be straightforwardly carried out.
In equation~\eqref{def_M0_M1}, all the dependences w.r.t.\ $L_{z}$
are in the Fourier transformed basis elements,
via the rotation matrices from equation~\eqref{RotMat}.
We recall that they satisfy in particular the orthogonality relation
\begin{equation}
\int_{0}^{\pi} \!\!\!\! \rd i \, \sin (i) \, R^{\ell^{p}}_{n_{2} m^{p}} (i) \, R^{\ell^{q}}_{n_{2} m^{p}} (i) = \delta_{\ell^{p}}^{\ell^{q}} \, \frac{2}{2 \ell^{p} + 1} .
\label{orthogonality_RotMat}
\end{equation}
Following the same calculation as in~\cite{Hamilton2018},
this allows us to immediately rewrite ${ \wbM_{0} (\omega) }$ as
\begin{equation}
\wM_{pq}^{0} (\omega) = \delta_{\ell^{p}}^{\ell^{q}} \, \delta_{m^{p}}^{m^{q}} \, \sum_{\tbn} C_{\ell^{p}}^{n_{2}} \, P^{\tbn}_{\ell^{p} n^{p} \ell^{p} n^{q}} (\omega) ,
\label{rewrite_M0}
\end{equation}
with ${ C_{\ell}^{n} \!=\! 2 \,(2\pi)^{3} |Y_{\ell}^{n} (\pi /2 , 0)|^{2}/(2 \ell+1) }$. In equation~\eqref{rewrite_M0},    the coefficient ${ P (\omega) }$ is also introduced as
\begin{equation}
P^{\tbn}_{\ell^{p} n^{p} \ell^{q} n^{q}} (\omega) \!=\! \!\! \int \!\! \rd \tbJ L \frac{\tbn \cdot \partial F_{0} / \partial \tbJ}{\omega - \tbn \cdot \tbO (\tbJ)} W^{\tbn}_{\ell^{p} n^{p}} (\tbJ) W^{\tbn}_{\ell^{q} n^{q}} (\tbJ) ,
\label{def_P}
\end{equation}
where the prefactor $L$ originates from the volume element
${ \rd \bJ = \rd \tbJ L \rd i \sin (i) }$.
Equation~\eqref{rewrite_M0} corresponds to the generic response
matrix for a spherically symmetric and non-rotating system.
One can proceed similarly to compute the matrix ${ \wbM_{1} (\omega) }$
from equation~\eqref{def_M0_M1}. To do so, the needed gradient can be computed as
\begin{equation}
\bn \cdot \frac{\partial (F_{0} (\tbJ)  \Sign (L_{z}))}{\partial \bJ} \!=\! \tbn \cdot\! \frac{\partial F_{0}}{\partial \tbJ}  \Sign (L_{z}) \! +\! 2 n_{3} \deltaD [L_{z}]  F_{0} (\tbJ)
\label{gradient_M1}
\end{equation}
using ${ \partial \Sign (x) / \partial x = 2 \deltaD (x) }$, and ${ \bn \cdot \partial E / \partial \bJ = \tbn \cdot \bO (\tbJ) }$.
It is then straighforward to perform the integration of equation~\eqref{gradient_M1}. As such, we define the coefficients 
\begin{align}
I_{n_{2} m^{p}}^{\ell^{p} \ell^{q}}& \,  =  (2 \pi)^{3} \, Y_{\ell^{p}}^{n_{2}} (\pi / 2 , 0) \, Y_{\ell^{q}}^{n_{2}} (\pi / 2 , 0)
\nonumber
\\
\times & \!\! \int_{0}^{\pi} \!\! \rd i \, \sin (i) \, \Sign (\cos (i)) \, R_{n_{2} m^{p}}^{\ell^{p}} (i) \, R_{n_{2} m^{p}}^{\ell^{q}} (i) .
\label{def_I}
\end{align}
and
\begin{align}
J_{n_{2} m^{p}}^{\ell^{p} \ell^{q}} = & \, 2 \, m^{p} \, (2 \pi)^{3} \, Y_{\ell^{p}}^{n_{2}} (\pi / 2 , 0) \, Y_{\ell^{q}}^{n_{2}} (\pi / 2 , 0)
\nonumber
\\
\times & \, R_{n_{2} m^{p}}^{\ell^{p}} (\pi/2) \, R_{n_{2} m^{p}}^{\ell^{q}} (\pi/2) .
\label{def_J}
\end{align}
These should be seen as geometrical prefactors that can be pre-computed once and for all, and do not depend on the details of the considered system.
In particular, they satisfy the symmetry relations
\begin{equation}
\begin{cases}
I_{- m^{p}} = - I_{m^{p}} ,
\\
J_{- m^{p}} = - J_{m^{p}} ,
\end{cases}
; \;
\begin{cases}
I_{- n_{2}} = - I_{n_{2}} ,
\\
J_{- n_{2}} = J_{n_{2}} .
\end{cases}
\label{symmetry_I_J}
\end{equation}
Similarly to equation~\eqref{def_P},  let us introduce
\begin{equation}
Q^{\tbn}_{\ell^{p} n^{p} \ell^{q} n^{q}} (\omega) \!=\! \!\! \int \!\! \rd  \tbJ \frac{F_{0} (\tbJ)}{\omega - \tbn \cdot \tbO (\tbJ)} W^{\tbn}_{\ell^{p} n^{p}} (\tbJ) W^{\tbn}_{\ell^{q} n^{q}} (\tbJ) ,
\label{def_Q_gamma}
\end{equation}
Gathering all these notations, the matrix ${ \wbM_{1} (\omega) }$ reads
\begin{align}
\wM^{1}_{pq} (\omega) \!=\! \delta_{m^{p}}^{m^{q}} \sum_{\tbn} \!  \bigg\{ & \, P^{\tbn}_{\ell^{p} n^{p} \ell^{q} n^{q}} (\omega) \, I_{n_{2} m^{p}}^{\ell^{p} \ell^{q}}
\nonumber
\\
+ & \, Q^{\tbn}_{\ell^{p} n^{p} \ell^{q} n^{q}} (\omega) \, J_{n_{2} m^{p}}^{\ell^{p} \ell^{q}} \bigg\} .
\label{expression_M1}
\end{align}

\begin{figure}
\includegraphics[width=0.22\textwidth]{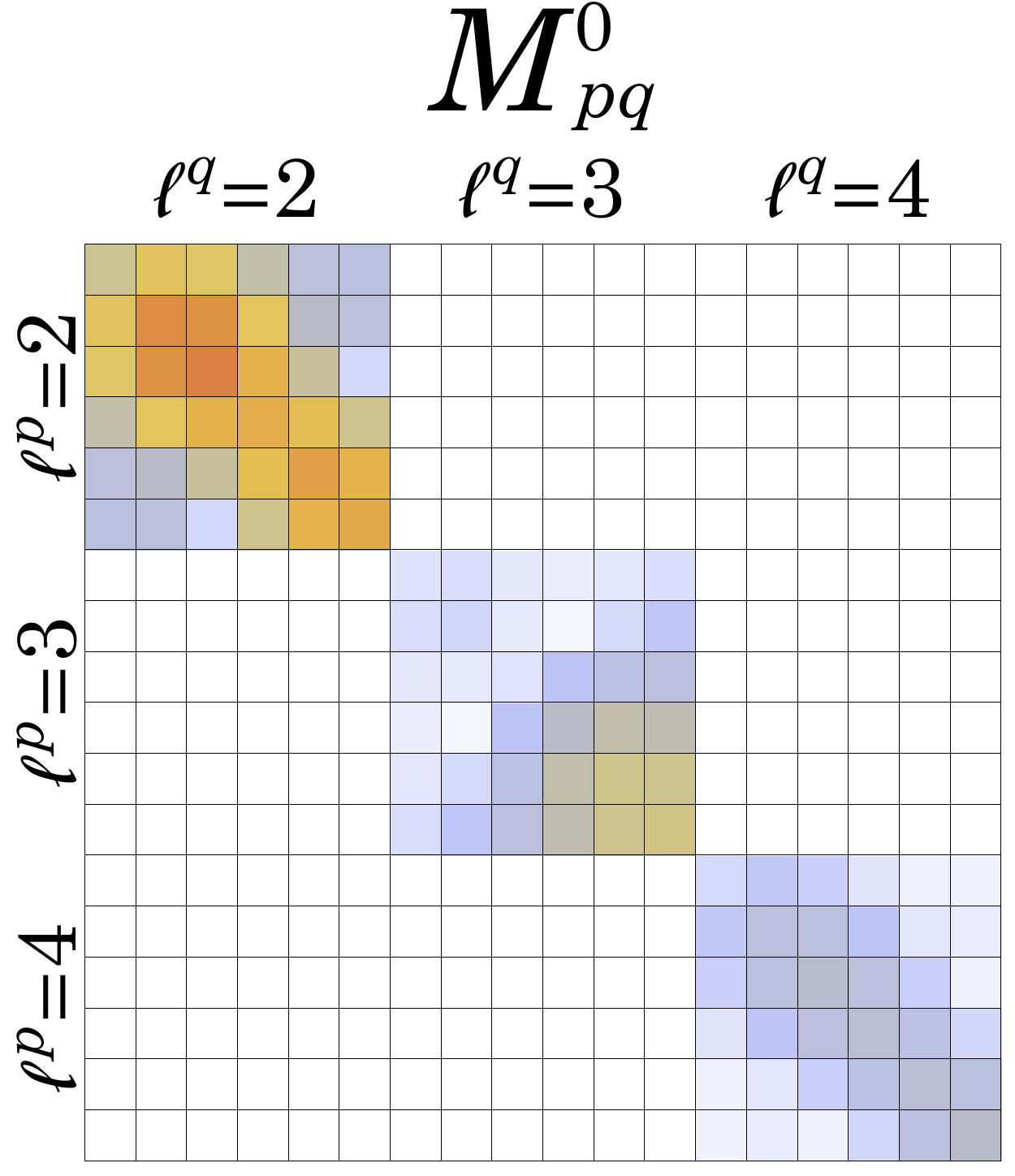}
\includegraphics[width=0.22\textwidth]{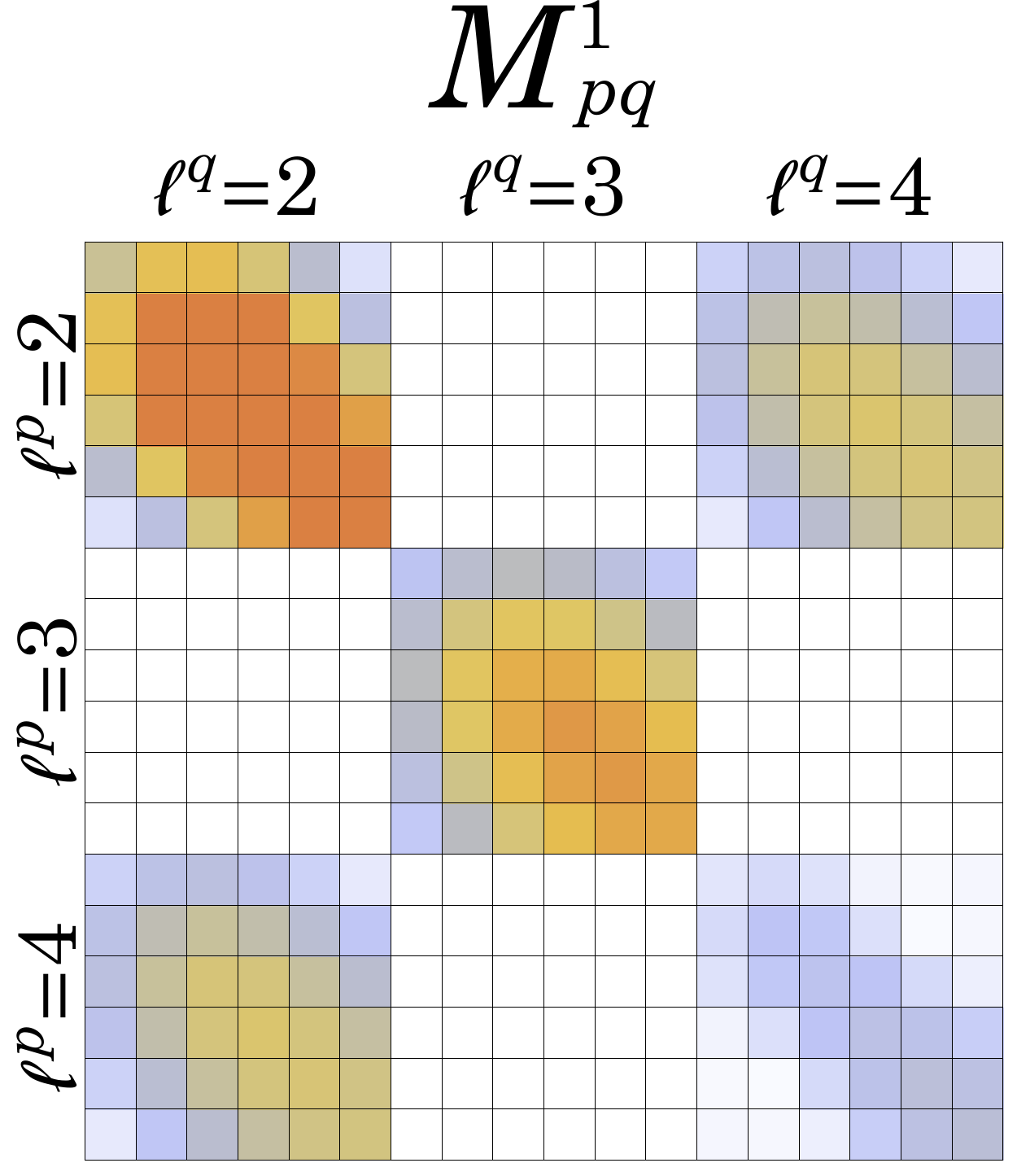}
\caption{Sketch of the $\wM_{pq}^{0}$ and $\wM_{pq}^{1}$ matrices in
the ${ q \!=\! -6 }$, ${ \alpha \!=\! 1}$, ${ \nmax \!=\! 5 }$, ${ \ell_{\max} \!=\! 4 }$ case,
with ${ (\eta, \omega_0) \!=\! (0.074,0.60) }$, corresponding to an ${ m \!=\! 2 }$ unstable mode.
The colours scale with the norm of each cell from white, light blue up to orange.
We note that  $\wM_{pq}^{1}$ is not block-diagonal anymore.}
\label{fig:Msketch}
\end{figure}

Figure~\ref{fig:Msketch} sketches $\wM_{pq}^{0}$ and $\wM_{pq}^{1}$
for the academic case of ${\nmax=5}$
and ${\ell_{\max} = 4}$ (so the corresponding block is
${ 6 \!\times\! 6 }$ because indexation starts at $0$).
Both matrices are made of blocks labeled by the pair ${(\ell^p,\ell^q)}$,
while each element of a block is labeled by ${(n^p,n^q)}$. Both matrices are complex
(the norm of each element is shown in Figure~\ref{fig:Msketch}) and symmetric,
as can be seen by swapping $p$ and $q$ in equations~\eqref{rewrite_M0} and~\eqref{expression_M1}.
The off-diagonal blocks in $\wM_{pq}^{1}$ reflects interaction between different harmonics $\ell$.
The  ${\delta_{\ell^{p}}^{\ell^{q}}}$ in equation~\eqref{rewrite_M0} makes
$\wM_{pq}^{0}$ diagonal by blocks, while the chessboard-like structure of $\wM_{pq}^{1}$ is due to
the ${ Y_{\ell^{p}}^{n_{2}} (\pi / 2 , 0) \, Y_{\ell^{q}}^{n_{2}} (\pi / 2 , 0) }$ term in both
${I_{n_{2} m^{p}}^{\ell^{p} \ell^{q}}}$ and ${J_{n_{2} m^{p}}^{\ell^{p} \ell^{q}}}$
(equations~\eqref{def_I} and~\eqref{def_J}),
which is non-zero only when ${\ell^p}$ and ${\ell^q}$ have the same parity.

To numerically compute the integrals appearing in equation~\eqref{expression_M1},
the approach of~\cite{Fouvry2015} is followed, modulo two main changes:
in the sampling of the orbital space $\tbJ$,
and in the computation of the elements ${ W^{\tbn}_{\ell \MakeLowercase{n}} (\tbJ) }$,
as presented in Appendix~\ref{appendix:FastWMat}.

In order to improve the quality of the sampling of the orbital space $\tbJ$,
we use a log-linear sampling in the ${ (\rperi , \rapo)-}$space.
This is done via an additional change of coordinates ${ (u,v) \mapsto (\rperi , \rapo) }$,
with ${(u , v)}$ sampled linearly and
\begin{align}
\rperi(u) & \, = \frac{\Rb}{2 (u_0 + 1)}  \left[ \re^{u_0-u} \left( \! \rm{erf}\!\left(\!\frac{u_0}{\sigma}\!\right) + erf\!\left(\!\frac{u-u_0}{\sigma}\!\right)\!\right) \right.
\nonumber
\\
& \left. + (u_0 - u + 1) \left( \!\rm{erfc}\!\left(\!\frac{u_0}{\sigma}\!\right) + erfc\!\left(\!\frac{u-u_0}{\sigma}\!\right)\!\right) \right] \!\! ,
\nonumber
\\
\rapo (u , v)  &\, = \rperi(u) + \rperi(v),
\label{mapping_uv}
\end{align}
where $\rm{erf}$ (resp. $\rm{erfc}$) are the (resp. complementary) error functions,
$\sigma$ is an ad hoc width, $\Rb$ is the radial scale of the basis functions.
We also introduced ${u_0 \!=\! \pb \, \nmax \, \Delta u - 1}$, with ${\Delta u}$
the step size in the ${ (u , v) }$-grid, $\nmax$ the maximum radial number for
the basis functions and $\pb$ a control parameter for the step size of the ${(\rperi , \rapo)}$-grid
where it is linearly sampled.
This transformation allows for $\rperi$ to be sampled logarithmically near $0$ and linearly further out,
while $\rapo$ is sampled logarithmically near $\rperi$ (close to circular orbits) and linearly further out.
On top of having an explicit Jacobian, equation~\eqref{mapping_uv} offers a smooth transition between the
logarithmic and linear samplings, with ${ \sigma \!=\! 10^{-3} }$ the typical width of the transition region.
The sampling in ${ (\rperi , \rapo ) }$ becomes linear when the radial step size reaches ${\Rb / (\pb \, \nmax)}$,
hence the parameter $\pb$ represents the minimal number of points in the $\rperi$-grid per radial oscillation
for the highest order radial basis function.

\section{Computing \texorpdfstring{${ W^{\tbn}_{\ell \MakeLowercase{n}} (\tbJ) }$}{WnlnJ}}
\label{appendix:FastWMat}

In this Appendix, we detail how the radial coefficients, ${ W^{\tbn}_{\ell n} (\tbJ) }$,
from equation~\eqref{def_W} may be efficiently computed,
improving upon the method presented in~\cite{Fouvry2015}.
We note that these coefficients only depend on the mean potential, ${ \psi (r) }$,
and our choice of basis elements, ${ U_{n}^{\ell} (r) }$.
As such, they are independent of the system's \DF\@.
They also satisfy the symmetry relation ${ W_{\ell n}^{- \tbn} (\tbJ) = W^{\tbn}_{\ell n} (\bJ) }$,
which essentially halves the number of coefficients to evaluate.

In equation~\eqref{angles_radial}, we have at our disposal the functions
${ r \mapsto \theta_{1} [r] , (\theta_{2} - \psi) [r] }$, so that it is more appropriate to
perform the integration from equation~\eqref{def_W} w.r.t.\ $r$ rather than $\theta_{1}$.
As a result, we write
\begin{equation}
W_{\ell n}^{\tbn} (\tbJ) = \frac{1}{\pi} \!\! \int_{\rperi}^{\rapo} \!\!\!\! \rd r \, \frac{\rd \theta_{1}}{\rd r} \, U_{n}^{\ell} [r] \! \cos \big[ n_{1} \theta_{1} [r]\! + n_{2} (\theta_{2} - \psi) [r] \big] ,
\label{rewrite_W}
\end{equation}
with the Jacobian ${ \rd \theta_{1} / \rd r \!=\! \Omega_{1} / \sqrt{2 (E \!-\! \psi (r)) \!-\! J_{2}^{2} / r^{2}} }$.
To compute ${ \theta_{1} [r] }$ and ${ (\theta_{2} \!-\! \psi) [r] }$ in equation~\eqref{rewrite_W},
one has to compute the integrals from equation~\eqref{angles_radial},
with the boundaries ${ \!\! \int_{\mathcal{C}} \! \rd \rpp = \!\! \int_{\rperi}^{r} \!\! \rd \rpp }$.

While all these integrals are well-defined, we note that their integrands have a
diverging behaviour at the boundaries ${ r = \rperi , \rapo }$,
originating from the vanishing of the radial velocity at the pericentre and apocentre.
In order to cure these divergences and ease the integration,
one can perform an additional change of variables.
Rather than integrating w.r.t.\ $r$, let us therefore integrate w.r.t.\ an
(explicit) angular anomaly, that we call $u$.
First, we define ${ \Sigma \!=\! (\rperi \!+\! \rapo)/2 }$ and ${ \Delta \!=\! (\rapo \!-\! \rperi)/2 }$.
A natural choice would be to consider the change of coordinates ${ r (u) = \Sigma + \Delta \sin (\tfrac{\pi}{2} u) }$,
with ${ u \in [-1 ; 1]  }$, that is explicit, monotonic and satisfies ${ \rd r / \rd u = 0 }$ for ${ u = \pm 1 }$,
therefore curing the boundary divergences.
In practice, it can be numerically more efficient not to use a mapping based on a trigonometric function,
but rather on a polynomial function of low degree.
As a result, we consider the mapping
\begin{equation}
r (u) = \Sigma + \Delta \, f (u) ; \;\;\; f (u) = u \big( {3}/{2} - {u^{2}}/{2} \big) ,
\label{mapping_r_u}
\end{equation}
which is the appropriate third degree polynomial satisfying the four constraints
${ f (\pm 1) = \pm 1 }$ and ${ f' (\pm 1) = 0 }$~\citep[see equation~{(52)} in][]{Henon1971}.
This mapping has an explicit (and simple) Jacobian and can be evaluated very efficiently.
Following this change of coordinate, we rewrite equation~\eqref{rewrite_W} as
\begin{equation}
W^{\tbn}_{\ell n} (\tbJ) \!=\! \frac{1}{\pi} \!\! \int_{-1}^{1} \!\!\!\! \rd u  \frac{\rd r}{\rd u}  \frac{\rd \theta_{1}}{\rd r}  U_{n}^{\ell} [u] \!\cos \big[ n_{1} \theta_{1} [u] \!+\! n_{2} (\theta_{2} \!-\! \psi) [u] \big]\,,
\label{rewrite_W_II}
\end{equation}
with
\begin{align}
\theta_{1} [u] & \, = \!\! \int_{-1}^{u} \!\!\!\! \rd \upp \, \frac{\rd r}{\rd \upp} \, \frac{\Omega_{1}}{\sqrt{2 (E - \psi [r]) - J_{2}^{2} / r^{2}}} ,
\nonumber
\\
(\theta_{2} - \psi) [u] & \, = \!\! \int_{-1}^{u} \!\!\!\! \rd \upp \, \frac{\rd r}{\rd \upp} \, \frac{\Omega_{2} - J_{2} / r^{2}}{\sqrt{2 (E - \psi [r]) - J_{2}^{2} / r^{2}}} .
\label{rewrite_theta1_theta2_II}
\end{align}

Naively, one could interpret the expression from equation~\eqref{rewrite_W_II}
as involving nested integrals, making the numerical evaluation cumbersome.
Yet, the writing of equation~\eqref{rewrite_W_II} can be improved to make the
calculation even simpler. Forgetting about indices to simplify the notations,
equation~\eqref{rewrite_W_II} asks us to compute ${ W (u = 1) }$, for ${ W (u) }$ of the generic form
\begin{equation}
W (u) = \!\! \int_{-1}^{u} \!\!\!\! \rd \upp \, w \big[ \upp , T_{1} (\upp) , T_{2} (\upp) \big] ,
\label{rewrite_W_III}
\end{equation}
where $T_{1}$ (resp.\ $T_{2}$) stands for ${ \theta_{1} }$ (resp.\ ${ (\theta_{2} - \psi) }$), and are given by expressions of the form
\begin{equation}
T_{1} (u) = \!\! \int_{-1}^{u} \!\!\!\! \rd \upp \, t_{1} [\upp] ; \;\;\; T_{2} (u) = \!\! \int_{-1}^{u} \!\!\!\! \rd \upp \, t_{2} [\upp] . 
\label{rewrite_T_III}
\end{equation}
Because the integration boundaries of ${ T_{1} (u) }$ and ${ T_{2} (u) }$ match the ones for ${ W (u) }$, one can reinterpret equation~\eqref{rewrite_W_III} not as a set of nested integrals, but rather as only one integral. To do so, we note that ${ W (u) }$, ${ T_{1} (u) }$, and ${ T_{2} (u) }$ satisfy the ordinary differential equations
\begin{align}
\frac{\rd W}{\rd u} & \, = w \big[ u , T_{1} (u) , T_{2} (u) \big] ,
\nonumber
\\
\frac{\rd T_{1}}{\rd u} & \, = t_{1} \big[ u \big] ,
\nonumber
\\
\frac{\rd T_{2}}{\rd u} & \, = t_{2} \big[ u \big] ,
\label{ODE_rewrite}
\end{align}
with the initial conditions ${ (W (-1) , T_{1} (-1) , T_{2} (-1)) = (0,0,0) }$. Introducing the state vector ${ \bS = (W , T_{1} , T_{2}) }$, and the velocity vector ${ \bs = (w , t_{1} , t_{2}) }$, one can finally rewrite equation~\eqref{ODE_rewrite} under the compact form
\begin{equation}
\frac{\rd \bS}{\rd u} = \bs \big[ u , \bS (u) \big] ; \;\;\; \bS (-1) = 0 .
\label{ODE_compact}
\end{equation}
As a conclusion, under this form, computing the coefficients ${ W^{\tbn}_{\ell n} (\tbJ) }$
amounts then to computing the value of ${ \bS (1) }$, by integrating forward the
differential equation~\eqref{ODE_compact}. Following such an approach,
one avoids the issue of having to compute nested integrals, and all the relevant quantities
are moved forward simultaneously, which offers a significant speed-up of the numerical
computation. In practice, for the applications presented in the main text,
equation~\eqref{ODE_compact} is integrated with a fourth-order Runge-Kutta
integrator~\cite[see equation~{(17.1.3)} in][]{NumericalRecipes2007}
with ${100}$ fixed-size steps for ${ u \in [-1 + \veps ; 1 - \veps] }$ with ${ \veps = 10^{-4} }$
to avoid possible boundary overflows.

\section{Convergence studies}
\label{appendix:Convergence}

Let us briefly investigate the convergence of our numerical methods
used to characterise unstable modes.
Section~\ref{appendix:ConvergenceMatrix} focuses on the matrix method,
while Section~\ref{appendix:ConvergenceNbody} considers the convergence of the $N$-body runs.
In these two sections, we consider the same fiducial model as in Section~\ref{sec:OneMode}, namely
${ (\alpha , q) \!=\! (1 , -6) }$ .

\subsection{Convergence study of the Matrix Method implementation}
\label{appendix:ConvergenceMatrix}

The different approximations in the computation of the response matrix introduce
various control parameters that can be tuned to increase the precision
(and the computational complexity) of the resulting growth rate and oscillation
frequency measurements. These parameters involve on the one hand the sampling
of action space in the integral of equation~\eqref{def_M0_M1}, which is turned
into a ${(\rperi , \rapo)}$-grid by a change of variables,
and on the other hand the number of harmonics both in the Fourier decomposition in angles
and in the projection onto the biorthogonal set of basis elements.
The parameters controlling the ${(\rperi , \rapo)}$-grid
are $r_{\min}$, $r_{\max}$, $\Delta u$, $\pb$, $\Rb$ and were introduced in Appendix~\ref{appendix:MatrixMethod}.
The parameters controlling the level of projection are
$n_{1}^{\mathrm{max}}$, the maximum wave number in the Fourier decomposition over the radial angle,
$n_{\mathrm{max}}$ the maximum order of radial basis functions,
which also sets the size of a block in the response matrix,
and $\ell_{\mathrm{max}}$ the maximum harmonic number $\ell$,
which also defines the number of blocks in the response matrix (see Figure~\ref{fig:Msketch}).
We note that these two sets of parameters are not exclusive.
Here, $\Rb$ also controls the quality of the projection onto the radial basis set,
and should be chosen so that the important information is contained inside this radius.
Similarly, $n_{\mathrm{max}}$ also controls the grid refinement, as can be seen in equation~\eqref{mapping_uv}.
In the following study, we made sure that varying $n_{\mathrm{max}}$ would not change the grid sampling
but only the level of projection onto the radial basis.

Table~\ref{tab:cvmatrix} presents the results of the computation of growth rates
and oscillation frequencies of the fiducial model ${ (\alpha , q ) \!=\! (1 , -6) }$ when
we vary these control parameters.
\begin{table}
\centering
\renewcommand{\arraystretch}{1.0}
\begin{tabular}{l c c c c}
 \hline
  Parameter & Interval & Standard & $\eta$ & $\omega_0$ \\ 
  \hline
  & Standard & & 0.045  &  0.54 \\ \hline
  $r_{\min}$ & 0.01 - 0.001 & 0.01 & 0.045  &  0.54 \\
  $r_{\max}$ & 5 - 30 & 10 & 0.045  &  0.54 \\
  $\Delta u$ & 2 - 0.15 & 1 & 0.027 - 0.045  &  0.44 - 0.54 \\
  $\pb$ & 3 - 8 & 4 & 0.054 - 0.037  &  0.54 - 0.49 \\
  $\Rb$ & 3 - 6 & 5 & 0.024 - 0.045  &  0.44 - 0.54 \\
[1ex]
  $n_{1}^{\mathrm{max}}$ & 1 - 5 & 2 & 0.030 - 0.060  &  0.49 - 0.54 \\
  $n_{\mathrm{max}}$ & 5 - 10 & 5 & 0.045 - 0.054 & 0.54 - 0.60 \\
  $\ell_{\mathrm{max}}$ & 2 - 7 & 2 & 0.045 - 0.081  &  0.54 - 0.60 \\
[1ex]
  & Fiducial & & 0.054 & 0.60 \\
 \hline
\end{tabular}
\caption{Growth rates and oscillation frequencies computed for various values of the parameters
that control the precision of the matrix method calculations. Having chosen a particular set of
parameters as a standard case, we  present the computed growth
rate and oscillation frequency when we vary each control parameter independently.
The second column presents the interval of variation of each parameter,
while the third column gives the standard value of each parameter.
The last two  columns show the interval of variation of the growth rate and
oscillation frequency in each case. When no significant variation is measured,
a single value is given. The first line gives the values of $\eta$ and $\omega_{0}$
for the standard case. We note that this standard case differs from the fiducial case of Section~\ref{sec:OneMode},
which had ${n_{\mathrm{max}}=10}$ and ${\pb=2}$, and is shown in the last line of the table.
}
\label{tab:cvmatrix}
\end{table}
As expected (see also, e.g.,~\cite{Hamilton2018}), the results are mildly dependent on some of these parameters.
Here, we show that we reached a sufficient level of convergence to confirm the existence of unstable modes,
as well as the value of their growth rates and oscillation frequencies.

\begin{figure}
\centering
\includegraphics[width=0.45\textwidth]{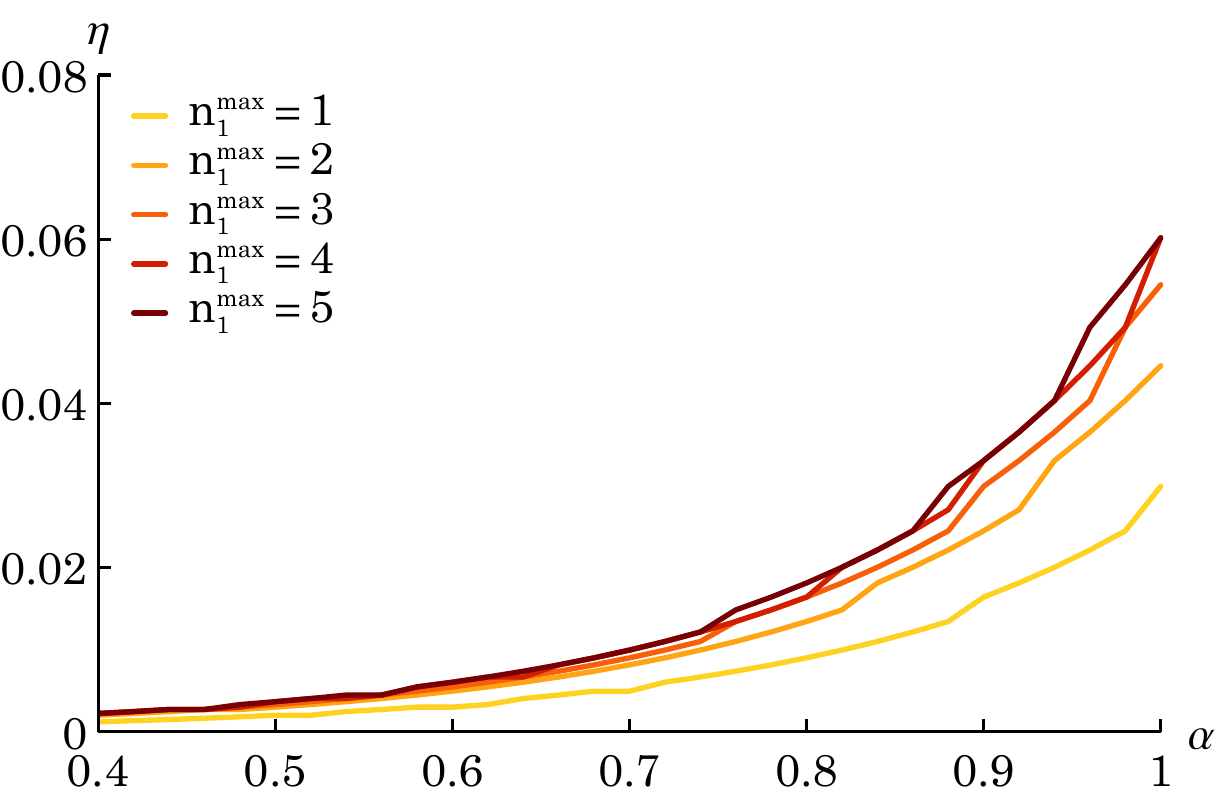}
\includegraphics[width=0.45\textwidth]{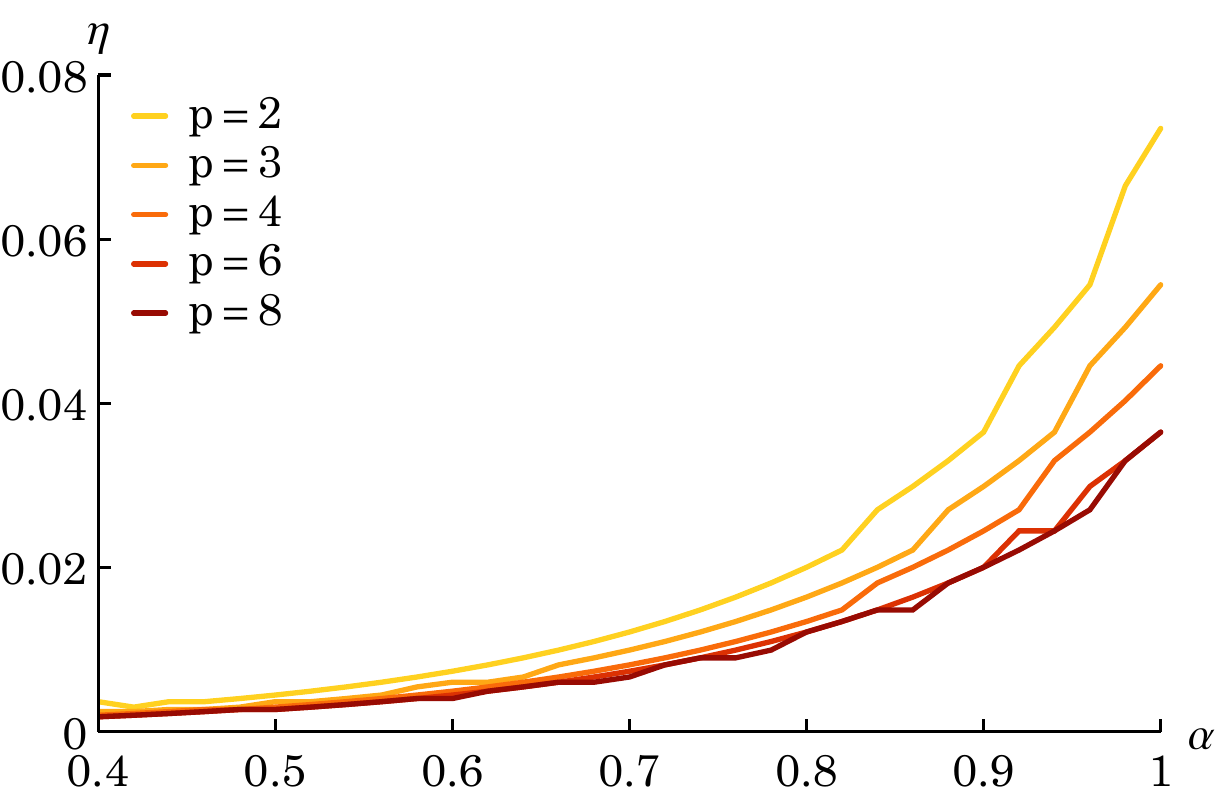}
\caption{
Illustration of the convergence of the matrix method for the measurement of the growth rate $\eta$ for different rotation parameters $\alpha$.
\textit{Top panel}: Dependence of the measured growth on the maximum resonance number, $n_{1}^{\mathrm{max}}$.
\textit{Bottom panel}: Dependence of the measured growth rate with the parameter $p_{\mathrm{b}}$ controlling the density of the ${(\rperi , \rapo)}$-grid, see equation~\eqref{mapping_uv}.
}
\label{fig:convergence-matrix}
\end{figure}

Figure~\ref{fig:convergence-matrix} shows typical  converging ${\eta(\alpha)}$
curves when we increase the level of precision of two of the parameters,
here $\pb$ and $n_1^{\max}$.
In closing, our choice of control parameters ensures that for ${ \eta \!>\! 0.01 }$
the stability manifolds are properly mapped.

\subsection{Convergence study of the $N$-body simulations}
\label{appendix:ConvergenceNbody}

The main control parameters for the force calculation in the \texttt{gyrfalcON} code are $\veps$ (softening parameter), $\theta$ (tolerance parameter) and the softening kernel. The softening kernel used has a density proportional to ${ (1 + x^2)^{-7/2} }$ which, as shown by~\cite{Dehnen2001}, results in much less of a force bias than Plummer softening. The parameter $\theta$ controls the accuracy of the force calculation (and is related to the opening angle in traditional tree codes). The default value of ${0.6}$ gives relative force errors of the order of ${0.001}$ in near-spherical systems~\citep{Dehnen2001}. Besides $N$, the softening length $\veps$ is the main parameter of interest since, if its value is too large, the perturbation which drives the growth of instability might not be resolved. 

\begin{figure}
\centering
\includegraphics[width=0.45\textwidth]{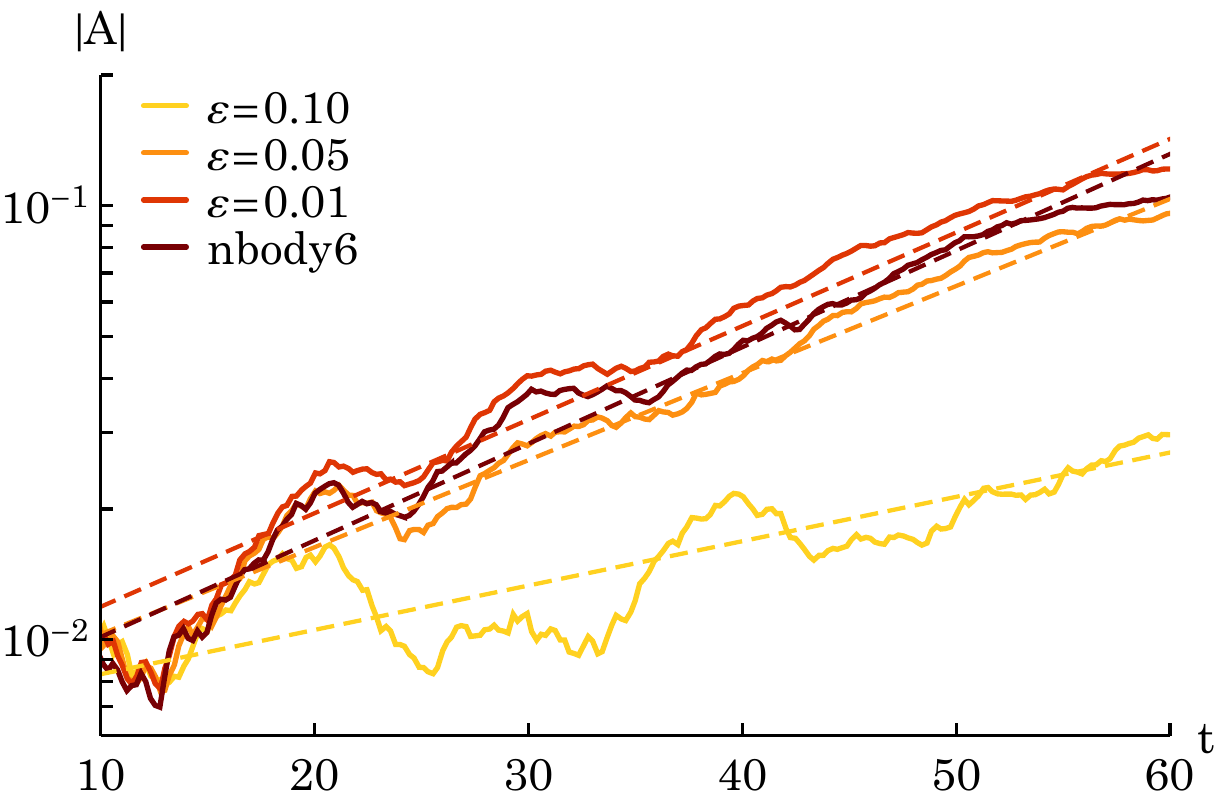}
\includegraphics[width=0.45\textwidth]{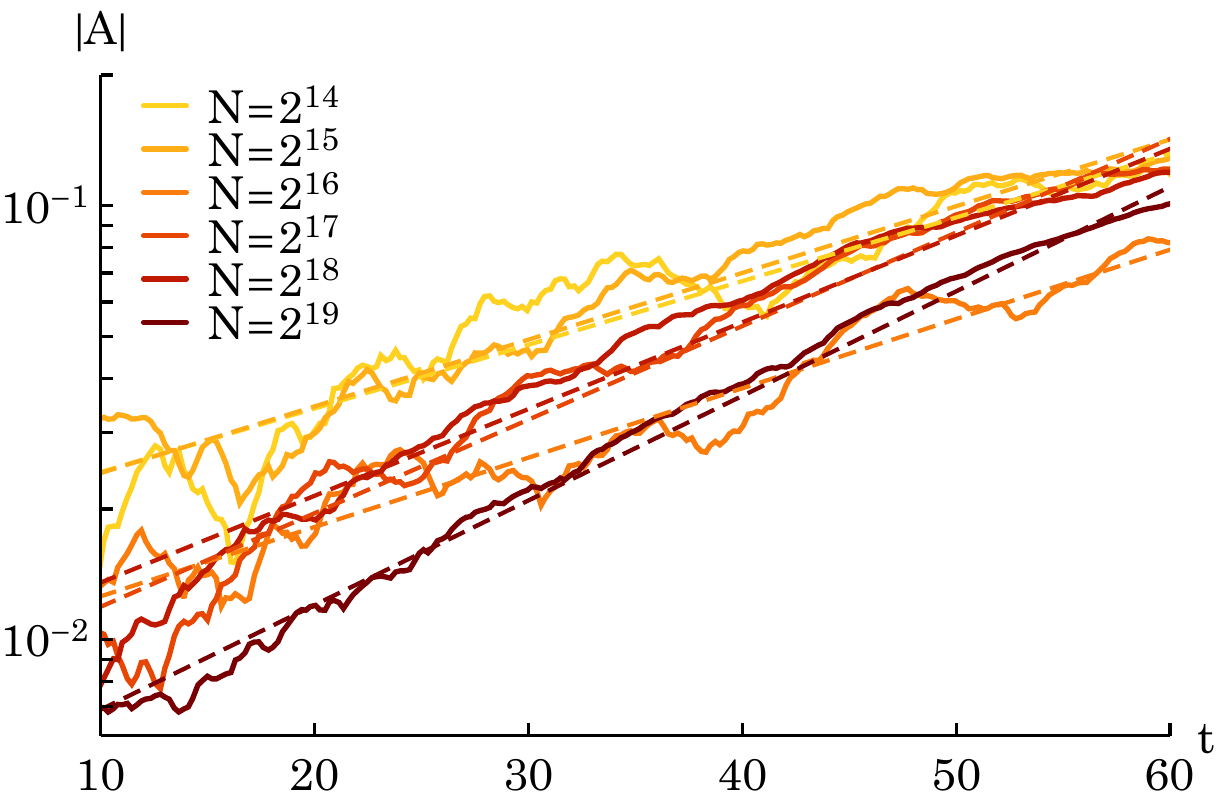}
\caption{ Illustration of the convergence of the $N$-body measurements for the fiducial cluster
${ (\alpha , q) = (1 , -6) }$, as illustrated by the time evolution of the amplitude
${ |A| }$ (continuous lines) and the associated exponential fit (dashed lines).
\textit{Top panel}: Dependence w.r.t.\ the softening length $\veps$.
This panel also includes a fully collisional run.
All runs were performed with ${ N \!=\! 2^{17} }$ particles.
\textit{Bottom panel}: Dependence w.r.t.\ the number of particles $N$.
All runs were performed with the same softening length ${ \veps \!=\! 0.01 }$.}
\label{fig:convergence-nb}
\end{figure}

In Figure~\ref{fig:convergence-nb}, we explore the effect of varying $\veps$ (top panel) and $N$ (bottom panel).
The default value of ${ \veps = 0.05 }$ was found to give good agreement with results using lower values of $\veps$, and with a direct $N$-body simulation using \texttt{NBODY6}~\citep{Nitadori2012}, which in effect has no softening, i.e. ${ \veps = 0 }$. The reference number of particles ${ N = 2^{17} }$ was also found to give good agreement with larger $N$-body simulations: no obvious dependence of the growth rate with increasing $N$ was found.

\end{document}